\documentclass[useAMS,usenatbib,usegraphicx,fleqn]{mn2e}

\usepackage{aas_macros}
\usepackage{multirow}
\usepackage{placeins}
\usepackage{amsmath}
\usepackage{url}

\title{Testing pre-main sequence models: the power of a Bayesian approach.}
\author[M. Gennaro, P.G. Prada Moroni and E. Tognelli]{M. Gennaro$^{1}$\thanks{E-mail:
gennaro@mpia.de (MG)}\thanks{Member of the International Max Planck Research School for Astronomy and Cosmic Physics at the University of Heidelberg, IMPRS-HD, Germany}, P.G. Prada Moroni$^{2,3}$, E.Tognelli$^{2,3}$\\
$^{1}$Max-Planck-Institut f{\"u}r Astronomie, K\"{o}nigstuhl 17, D-69117, Heidelberg, Germany\\
$^{2}$Universit\`{a} di Pisa, Dipartimento di Fisica ''E. Fermi'', Largo B. Pontecorvo, 3, I-56127, Pisa, Italy\\
$^{3}$Istituto Nazionale di Fisica Nucleare - Sezione di Pisa, Largo B. Pontecorvo, 3, I-56127, Pisa, Italy
}

\begin{document}

\date{Submitted to MNRAS 1848 February 21. Accepted 1945 April 25.}

\pagerange{\pageref{firstpage}--\pageref{lastpage}} \pubyear{2011}

\maketitle

\begin{abstract}
Pre-main sequence (PMS) models provide invaluable tools for the study of star forming regions as they allow to assign masses and ages to young stars. Thus it is of primary importance to test the models against observations of PMS stars with dynamically determined mass.
We developed a Bayesian method for testing the present generation of PMS models which allows for a quantitative comparison with observations, largely superseding the widely used isochrones and tracks qualitative superposition. 

Using the available PMS data we tested the newest PISA PMS models establishing their good agreement with the observations. The data cover a mass range from $\sim0.3$ to $\sim3.1 M_{\sun}$, temperatures from $\sim3\times10^3$ to $\sim1.2\times10^4$ K and luminosities from $\sim3\times 10^{-2}$ to $\sim60 L_{\sun}$. Masses are correctly predicted within 20\% of the observed values in most of the cases and for some of them the difference is as small as 5\%. Nevertheless some discrepancies are also observed and critically discussed.

By means of simulations, using typical observational errors, we evaluated the spread of $\log \tau_{sim} - \log \tau_{rec}$, i.e. simulated $-$ recovered ages distribution of the single objects.  We also found that stars in binary systems simulated as coeval might be recovered as non coeval, due to observational errors. The actual fraction of fake non coevality is a complex function of the simulated ages, masses and mass ratios. 
We demonstrated that it is possible to recover the systems' ages with better precision than for single stars using the composite age-probability distribution, i.e. the product of the components' age distributions. Using this valuable tool we estimated the ages of the presently observed PMS binary systems.

\end{abstract}


\maketitle
%

\label{firstpage}

\section{Introduction}

The current understanding of star formation processes largely relies on the ability of assigning ages and masses to young stars using pre-main sequence (PMS) models. The observed luminosity and effective temperature of stars in their early evolutionary stages can be translated into mass and age only by the comparison with PMS stellar tracks. Unfortunately the early evolutionary stages of the stellar life are among those less tightly constrained by observations and most uncertain from the theoretical point of view. This situation becomes progressively worse for stellar mass below $ \approx 1.2$ M$_{\sun} $. This is mainly a consequence of the poor treatment of superadiabatic convection. Moreover there are still large uncertainties on the main input physics describing the cold and dense matter typical of low-mass stars interiors adopted in modern evolutionary codes. This theoretical uncertainty is testified by the large discrepancy still present between different sets of low-mass PMS models \citep[see e.g.][]{SD00,baraffe2002,tognelli11}.  

An ever growing amount of detailed information is becoming available for both star-forming regions in the Milky Way \citep[see][]{2008hsf1.book.....R,2008hsf2.book.....R} and in the Magellanic Clouds \citep[][]{cignoni09,cignoni10,2010A&A...515A..56G}, prompted by the remarkable improvement in the observational techniques over the last decade. The aforementioned theoretical uncertainties imply that many of the properties inferred for these regions, such as the initial mass function and the star formation history, strongly depend on the adopted PMS models, particularly for stars less massive than $\approx 1.2$ M$_{\sun} $.  

The importance of these kind of studies urges an empirical calibration of PMS tracks and isochrones based on a statistically significant sample of young stars with precisely determined parameters (mass, temperature, radius, luminosity and chemical abundances).  
Most useful in this respect are the detached, double lined, eclipsing systems which directly provide stellar masses, temperatures and radii, yielding also a distance independent luminosity through the Stephan law. 
Unfortunately, the currently available sample of PMS stars in eclipsing binaries amounts only to 10 objects in 6 systems (see Sect.~\ref{sec:data} and Table \ref{Tab:data}).
Other important observational constraints to PMS models are provided by astrometric measurement of binary systems that can be resolved thanks to interferometric observations. For these systems the radii can not be determined though. Three astrometric binary systems with both stars in the PMS phase are currently known and studied (Table \ref{Tab:data}).
A third technique providing mass values for young stars is based on spectroscopic observations of circumstellar disks. From the keplerian velocities masses are inferred, although these measurements require an independent estimate of the distance to determine the linear value of the radius at which velocities are measured. Also the sample of known objects in this category is quite small with only 9 stars (see again Table \ref{Tab:data})

As already shown in the early attempts to test PMS stellar tracks against observations of eclipsing binaries \citep[see e.g.][]{palla01,hillenbrand04,stassun04,alecian07,boden07,mathieu07}, the mass values inferred from theoretical models are in reasonable agreement with the dynamical ones for intermediate mass stars, whereas for low-mass stars theoretical values tend to underestimate the stellar mass \citep[see Fig.~3 in][]{mathieu07}. 

From these studies it is also clear that the usefulness of such tests in constraining the theoretical PMS models is severely limited by the still scarce accuracy of the current empirical measurements of the other stellar parameters, i.e. the luminosity, the chemical abundances and above all the effective temperature \citep[see e.g.][]{hillenbrand04,mathieu07}. In the next future both the size and the quality of the observed sample of test objects to calibrate PMS stellar tracks and isochrones are bound to increase. 

In the present paper we apply an objective Bayesian method to compare theoretical predictions with observations, obtaining robust uncertainties for the output values and assessing the overall quality of the comparison. Since the method, which is detailed in Sect.~\ref{Sec:Baymeth}, allows for the use of stellar tracks for a large and very fine grid of metallicity, mass and age values, we tested only our own Pisa-PMS models. These are calculated using the newest version of the \texttt{FRANEC} evolutionary code \citep[see e.g.][]{tognelli11}. The main characteristics of the models are described in Sect.~\ref{sec:models}. In Sect.~\ref{sec:synttest} we asses the ability of the method to retrieve the stellar properties by means of synthetic tests. In Sect.~\ref{sec:data} we describe the observational data set, which includes all the currently available low-mass PMS data. The complete data set is analysed in Sect.~\ref{sec:theovsdyn}, where theoretical masses derived from our standard set of models are compared to the dynamical measurements. In Sect.~\ref{sec:compmod} we compare the results for multiple sets of models.
Section \ref{sec:binanalysis} is dedicated to detailed study of each binary system while the stars in the Taurus-Auriga association are analysed in Sect.~\ref{sec:TauAu}.
A summary with concluding remarks is presented in Sect.~\ref{sec:concl}.

\section{The Bayesian method}
\label{Sec:Baymeth}

The general question we try to answer can be described as the problem of determining certain parameters (the age and mass of a star) by comparing models' predictions with empirical evidence (effective temperatures, luminosities, radii, dynamical masses of stars in binary systems). In order to do so we used a Bayesian approach,  which allows to fully exploit the data.

One of the main advantages of the Bayesian approach over the frequentist one is the possibility of using the available information about the model parameters --the \emph{prior probability}. Thanks to Bayes' Theorem this information is naturally included in the calculation of the new parameters' probability after additional evidence is collected --the \emph{posterior probability}. In this way it is possible to further constrain the models' parameter space in an iterative process of refinement.
On the other hand the main disadvantage is that often the whole space of possible models is not accessible. This means that the normalizing factor appearing on the r.h.s of Bayes' Theorem --eq. (\ref{eq:BayTheo}) below--, can not always be evaluated. In such cases it is impossible to rigorously compute the normalized probability for a model to be correct given the empirical evidence.
Nevertheless it is still possible to compare and choose between two different models, by taking the ratio of the posterior probabilities thus removing the normalization factor.

The notation we will adopt is the same as in \citet{2005A&A...436..127J} --hereafter JL05. The method described in JL05 has been successfully used to provide stellar ages and masses for the stars in the Geneva-Copenhagen survey \citep[see][]{2004A&A...418..989N,2007A&A...475..519H,2009A&A...501..941H}. It is a general method that can be applied to many other astrophysical cases, such ours.
We retain most of the formalism of JL05, even though we customized the method for use with PMS objects. One major difference between the present work and JL05 is the use of the full covariance matrix when dealing with correlated variables such as luminosities and temperatures as determined for stars in eclipsing binary systems (see Sect.~\ref{subsec:lklcov}). An other important difference is the adoption of prior distributions that are appropriate for our observed sample. The flexibility of the Bayesian approach introduced by JL05 lies also in the opportunity of choosing the prior distributions that are best suited for the particular problem. For example our objects have well-determined masses, hence we adopt these values as priors (see Sect.~\ref{subsec:prdstr}).

Let $\bmath{q}$ be a set of observational quantities (or any combination of them), for example temperature and luminosity or gravity and temperature. Let $\bmath{p}$ be a set of model parameters and $\bmath{\Xi}$ a set of meta-parameters identifying a class of models.
We introduce this distinction between $\bmath{p}$ and $\bmath{\Xi}$ for practical reasons. The parameters $\bmath{p}$ are the triple $(\tau,\mu,\zeta)$, i.e. age, mass and metallicity of the model. The meta-parameters $\bmath{\Xi}$ are instead the mixing-length parameter, $\alpha$, the primordial helium abundance, $Y_{\rmn{P}}$, and the helium-to-metals enrichment ratio $\Delta Y/\Delta Z$. These three meta-parameters are chosen on the basis of some considerations and can be regarded as fixed inputs for the evolutionary models library as a whole, which gives them a different status compared to the $\bmath{p}$ set. 
The $\alpha$ parameter is usually calibrated on a solar model \citep[see][and references therein]{2008PhR...457..217B}. The $Y_{\rmn{P}}$ value is constrained by big-bang nucleosynthesis and observation of metal poor H{\sevensize II} regions \citep[][]{2007ApJ...662...15I,2007ASPC..374...81P,2009ApJS..180..306D,2010JCAP...04..029S}. $\Delta Y/\Delta Z$ is constrained by chemical evolution models of the Galaxy \citep[][]{2005A&A...430..491R,2008RMxAA..44..341C} or by comparing the absolute magnitude of unevolved nearby dwarf stars with stellar models \citep[][]{jimenez03,2007MNRAS.382.1516C,2010A&A...518A..13G}. 

The $\bmath{\Xi}$ triple is usually fixed for any set of stellar tracks or isochrones available in the literature. Nevertheless there is no strong reason to assume that the solar calibration of $\alpha$ has to be suitable also for PMS stars of any mass \citep[see e.g. the discussion in][]{montalban04,tognelli11}. Also $Y_{\rmn{P}}$ and $\Delta Y/\Delta Z$ are known with some uncertainty, which is quite large specially for the latter. Hence, having the opportunity to calculate our own stellar models, we allowed for variations of the meta-parameters, calculating stellar models libraries for a total of nine combinations of them (see Sect.~\ref{sec:models} for a detailed description). In principle other meta-parameters exist, such as the relative distribution of metals (the mixture), the opacity tables, the equation of state. Indeed any \emph{choice} of physical inputs identifies a class of models, but we will not explore the effects of changes in the microphysics.

In the following we will drop the $\bmath{\Xi}$ term and, if not explicitly stated, we will refer only to one particular class of models, i.e. one fixed choice of $\bmath{\Xi}_j = (\alpha_j,{Y_{\rmn{P}}}_j,{\Delta Y/\Delta Z}_j)$. We will come back to the comparison of different classes in Sects. \ref{subsec:evidence} and \ref{sec:compmod}.

\subsection{Definition}
Bayes' Theorem states that the posterior probability of the parameters $\bmath{p}$, given the observations $\bmath{q}$, is:
\begin{equation}
\label{eq:BayTheo}
 f( \bmath{p} | \bmath{q} ) = \frac{ f(\bmath{q}|\bmath{p}) f(\bmath{p}) }{ f(\bmath{q}) } \; .
\end{equation}
The probability $f(\bmath{q}|\bmath{p})$ of observing the quantities $\bmath{q}$ given the parameters $\bmath{p}$ is proportional to $\mathcal{L}(\bmath{p}|\bmath{q})$, the \emph{Likelihood} of the parameters $\bmath{p}$ given the evidence $\bmath{q}$.
The quantity $f(\bmath{p})$ is the prior distribution of the parameters, which incorporates the information already available about them. The normalizing factor, $f(\bmath{q}) = \int f(\bmath{q}|\bmath{p})  f(\bmath{p}) \, \rmn{d}\bmath{p} $ is called marginal distribution; it represents the probability of observing new evidence $\bmath{q}$ under a complete set of mutually exclusive hypothesis, i.e. under all possible values for $\bmath{p}$.

To calculate the integral, we should have access to all possible models, for all possible sets of parameters (and meta-parameters), or at least the subset of all plausible models, i.e. models for which $f(\bmath{p})$ is not negligible. Even though the integration is not possible here, this is not a problem. As long as we are interested only in comparing different classes of models or estimating the most probable set of parameters within a single class, this can be accomplished by taking probability ratios, hence removing the normalization. Having considered that, we can then rewrite the posterior probability as:
\begin{equation}
\label{eq:postprob}
 f( \bmath{p} | \bmath{q} ) \propto \mathcal{L}(\bmath{p}|\bmath{q}) f(\bmath{p}) \, .
\end{equation}

\subsection{The Likelihood function and the 2-variables covariance matrix}
\label{subsec:lklcov}
Equation (\ref{eq:postprob}) is identical to eq. (3) in JL05, but in our work we extend the definition of Likelihood to the case of pairs of observables with non-zero covariance. This is particularly important --and often neglected-- when the observables used to determine the stellar parameters are luminosity and temperature of stars in eclipsing binaries. Because of the way the two quantities are derived, they are strongly correlated \citep[see][]{mathieu07}.
Let the vector of observables be a 2D vector: $\bmath{q} = (x,y)$; the definition of Likelihood in the general case is:

\begin{eqnarray}
 \mathcal{L}(\bmath{p}|\bmath{q}) & = & \frac{1}{2\,\pi\,\sigma_x\,\sigma_y \,\sqrt{1-\rho^2}}\times \exp \left\{ -\frac{1}{2\,(1-\rho^2)} \times \right. \nonumber \\
				    & \times & \left[ \frac{ \left[ x(\bmath{p})-\hat{x} \right]^2}{\sigma^2_x} + \frac{ \left[y(\bmath{p})-\hat{y} \right]^2}{\sigma^2_y} + \right. \nonumber \\
				    & -      & \left. \left. \frac{2\rho\left[x(\bmath{p})-\hat{x}\right]\left[y(\bmath{p})-\hat{y} \right]}{\sigma_x \sigma_y} \right] \right\} 
\end{eqnarray}

Here $\hat{x}$ and $\sigma_x$ are the measured value for the observable $x$ and its uncertainty, respectively (the same for $\hat{y}$ and $\sigma_y$). The quantity $\displaystyle \rho = \frac{\rmn{Cov}(x,y)}{\sigma_x \sigma_y} $ is the \emph{correlation coefficient} of $x$ and $y$. The quantities $x(\bmath{p})$ and $y(\bmath{p})$ are the values predicted by the model for the parameter values $\bmath{p}$.

\subsection{The value of the covariance between luminosity and temperature}
In the case of eclipsing binaries the quantities that can be determined from the light curve are the effective temperatures ratio between the primary and secondary, $T_{\rmn{eff},1}/T_{\rmn{eff},2}$, and the radii, $R_1$ and $R_2$. The temperature of the primary has to be inferred by other indicators such as some temperature-sensitive lines in the spectrum and a subsequent spectral type - temperature conversion. Luminosities are not directly measured, but derived using Stephan's Law: $L=4 \pi \sigma_{\rmn{SB}} R^2 {T_{\rmn{eff}}}^4 $. 

On the other hand, since the most used tool of stellar evolution is the Hertzsprung-Russel diagram (HR diagram), where $\log L$ and $\log T_{\rmn{eff}}$ are displayed, most of the analysis of binary systems is done in the HR diagram. So it is useful to have a proper treatment of the covariance matrix for luminosity and temperature.
Given Stephan's law, the standard deviation of $\log L$ is calculated as:
\begin{equation}
\label{eq:varL}
\sigma_{\log L}  = \sqrt{ \frac{1}{\ln 10} \left( 2\frac{\sigma_{R}}{R} + 4\frac{\sigma_{T_{\rmn{eff}}} }{ T_{\rmn{eff}}}\right) }
\end{equation}
where $\sigma_R$ and $\sigma_{T_{\rmn{eff}}}$ are the uncertainties on radius and temperature\footnote{Note that $\ln$ is the base-e logarithm and $\log$ the base-10}.
The covariance between $\log L$ and $\log T_{\rmn{eff}}$ is given by:
\begin{eqnarray}
\label{eq:covarLT}
\rmn{Cov}(\log L, \log T_{\rmn{eff}}) & = & 2\, \rmn{Cov}(\log R, \log T_{\rmn{eff}}) + 4 \,\rmn{Var}(\log T_{\rmn{eff}}) \nonumber \\
				   & \approx & 4\,\rmn{Var}(\log T_{\rmn{eff}})
\end{eqnarray}
where $\rmn{Var}(\log T_{\rmn{eff}}) \equiv \sigma^2_{\log T_{\rmn{eff}}}$.

In both eqs. (\ref{eq:varL}) and (\ref{eq:covarLT}) we assume that temperatures and radii have vanishing covariance. This is not necessarily true for eclipsing binaries, for which they are derived from the same light curve using fitting algorithms. However we are forced to neglect the corresponding term in the total luminosity-temperature covariance since we do not have access to the covariance matrix for radii and temperatures. Nevertheless we expect this covariance to be small specially when radii and temperatures are derived by multiple fitting of light-curves obtained independently in several photometric bands.

\subsection{Prior distributions}
\label{subsec:prdstr}
\begin{figure}
 \centering
 \resizebox{\hsize}{!}{\includegraphics{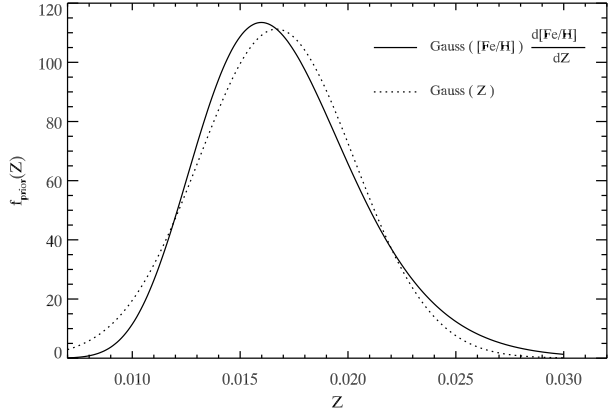}}
 \caption{Comparison between a Gaussian prior in Z and the prior that is derived as in eq. (\ref{eq:gprior}). The assumed starting [Fe/H] distribution is a Gaussian with  $\mu=0.1$ dex and $\sigma=0.1$ dex.}
 \label{fig:comp2prior}
\end{figure}

We will make use of different types of prior distributions for the parameters. Here is a brief description of each of them:
\begin{description}
\item {\bf Mass:} for most of the systems in our sample the dynamical mass is the observable that is known with the best precision. For this reason we will often use a Gaussian mass prior $f(\bmath{p}) \propto \exp \left\{ -\frac{1}{2}\left[ (\mu - m_{\rmn{ob}})/{\sigma_{m_{\rmn{ob}}}} \right]^2 \right\}$. Apart from better constraining the mass values, this particular kind of prior is very informative and helps constraining also the stellar ages (see e.g. Sect.~\ref{sec:synttest}).

\item {\bf Metallicity:} for most of the systems measurements of [Fe/H] are available, too. These can also be included as priors after converting [Fe/H]$_{\rmn{ob}}$ into the corresponding Z$_{\rmn{ob}}$ value. The value of $Z_\rmn{ob}$ depends on [Fe/H]$_{\rmn{ob}}$ but also on $\Delta Y/\Delta Z$, $Y_P$ and $(Z/X)_{\sun}$ \citep[for details see equations (1) and (2) in][]{2010A&A...518A..13G}. As a consequence we use different $Z_{\rmn{ob}}$ values for a given [Fe/H]$_{\rmn{ob}}$ when we compare actual data to different classes of models $\bmath{\Xi}_j$. 

If we assume that [Fe/H] errors are distributed as Gaussians, the corresponding error distribution in $Z$ is calculated as:
\begin{equation}
\label{eq:gprior}
f_{\rmn{prior}}(Z) = \mathcal{G}( \phi(Z)) \frac{\rmn{d\phi(Z)}}{\rmn{d}Z} \, ,
\end{equation}
where we introduced $\phi (Z) \equiv \rmn{[Fe/H]}$ and [Fe/H] is regarded as a function of $Z$; $\mathcal{G}$ is a Gaussian function.
In the general case, the derived $f_{\rmn{prior}}(Z)$ can be asymmetric and, in particular, very different from a Gaussian. Nevertheless, given the typical errors in our data set (i.e. $\sigma(\rmn{[Fe/H]}) \sim 0.1$ dex), the departure from Gaussianity is very small and can be neglected as can be seen in Fig.~\ref{fig:comp2prior}, where the Gaussian distribution in $Z$ and that obtained as in Eq. (\ref{eq:gprior}) are shown. In both cases, we start from an observed [Fe/H] distributed as a Gaussian and with $\mu=0.1$ dex and $\sigma=0.1$ dex. The mean value and variance of the $\mathcal{G}(Z)$ are determined using eq. (1) and (2) of \cite{2010A&A...518A..13G} and simple error propagation rules.

Moreover we can not always be sure that the error distribution on [Fe/H] is itself Gaussian. [Fe/H] errors certainly have a random, Gaussian component that can, however, be smaller than the overall uncertainty due to the poorly constrained systematics. With this in mind, we opted for a Gaussian functional form for the prior in $Z$.


\item{\bf Age:} in the case of two stars in the same system, coevality might be considered as an additional prior, assuming that these two stars formed at the same time. In this case, we simply impose that both stars are coeval by multiplying their age marginal distributions (see below) hence getting a system age distribution.

\end{description}

\subsection{Marginal distributions, best values, uncertainties and relative precision}
\label{sec:md_bv_rp}
As in JL05, the integration of eq.(\ref{eq:postprob}) with respect to all parameters but one, $p_i$, yields the marginal distribution for ${p_i}$. From this distribution it is possible to determine the most probable value for $p_i$ and its confidence interval.
The two parameters we are interested to determine are the stellar age $\tau$, and mass $\mu$.
We will use the same symbol as in JL05 for the age marginal distribution, $G(\tau)$, and analogously define the mass marginal distribution, $H(\mu)$. By writing explicitly the triple of parameters, we define:

\begin{subequations}
\begin{eqnarray}
\label{eq:Gtau}
 G(\tau) & = & \int \mathcal{L}(\tau,\mu,\zeta|\bmath{q})  f(\tau,\mu,\zeta)\, \rmn{d}\mu \,\rmn{d}\zeta \\
\label{eq:Hmu}
H(\mu)  & = & \int \mathcal{L}(\tau,\mu,\zeta|\bmath{q})  f(\tau,\mu,\zeta)\, \rmn{d}\tau \, \rmn{d}\zeta
\end{eqnarray}
\end{subequations}

JL05 demonstrated that the mode of the marginal distribution is a more robust indicator than the mean for estimating stellar ages. This is particularly true for strongly asymmetric distributions or distributions showing multiple peaks. We also adopted the mode as the best value estimator but changed the definition of the uncertainty interval with respect to JL05. 
If $\mathcal{A}$ is the total area under the distribution curve, $F(x)$, we define the confidence interval  $[x_{\rmn{min}},x_{\rmn{max}}]$:
\begin{equation}
 \int_{x_{\rmn{l}}}^{x_{\rmn{min}}} F(x) \,\rmn{d}x = \int_{x_{\rmn{max}}}^{x_{\rmn{u}}} F(x) \,\rmn{d}x = 0.16 \mathcal{A}
\end{equation}
where we assume that the variable $x$ is defined in the interval $[x_{\rmn{l}},x_{\rmn{u}}]$. 
In this way 16\% of the total probability is rejected on each side of the confidence interval.
This definition coincides with that of a $1\sigma$ interval in the case of a Gaussian distribution.
We followed again JL05 in the definition of the relative precision, $\epsilon$:
\begin{equation}
\label{eq:relprec}
 \epsilon = \sqrt{x_{\rmn{max}} / x_{\rmn{min}} } - 1
\end{equation}
Using this definition it is possible to compare the \emph{quality} of different age and mass determinations. The worst relative precision is attained when the marginal distribution is flat. In this case, assuming again $x \in [x_{\rmn{l}},x_{\rmn{u}}]$ we have:
\[
  \frac{ x_{\rmn{max}} }{ x_{\rmn{min}} } = \frac{ x_{\rmn{u}} -\frac{16}{100}  (x_{\rmn{u}} - x_{\rmn{l}}) }{x_{\rmn{l}} + \frac{16}{100}  (x_{\rmn{u}} - x_{\rmn{l}})} 
\]
We calculated models in the age interval $[0.5,100] \, \rmn{Myr}$ and mass interval $[0.2,3.6]\, M_{\sun}$, hence the worst relative precisions attainable are $\epsilon (\tau) \approx 1.26 $ and $\epsilon (\mu) \approx 1.03$.

\subsection{Comparison of different classes of models}
\label{subsec:evidence}

Comparison of two classes of models is possible by calculating the Bayes Factor, i.e. the ratio of the \emph{evidences} for the two classes. The evidence itself is defined as the integral of the likelihood marginalized over the model parameters prior distributions.
Hence the Bayes Factor for the i-th and j-th class of models is:
\begin{equation} 
\label{eq:BF}
BF_{ij} = \frac{f(\bmath{q}| \bmath{\Xi}_i)}{f(\bmath{q}| \bmath{\Xi}_j)} \, ,
\end{equation}
where the evidence for each class is defined as:
\begin{eqnarray}
\label{eq:singleev}
 f(\bmath{q}| \bmath{\Xi}) & = & \int f(\bmath{q},\bmath{p}|\bmath{\Xi}) \,\rmn{d}\bmath{p} \nonumber \\
		       & = & \int f(\bmath{q} | \bmath{p}, \bmath{\Xi}) f(\bmath{p} | \bmath{\Xi}) \,\rmn{d}\bmath{p}
\end{eqnarray}

The Bayes Factor tells us nothing about the best values of the parameters $\bmath{p}$, but it can be used to estimate which class of models --which set of meta-parameters $\bmath{\Xi}$-- gives an overall best-fit to the data. Strong deviations of $BF_{ij}$ from one indicate that one class of models is a significantly better choice than the others.

\section{The set of models}
\label{sec:models}
In the present analysis we used the very recent PMS tracks from the Pisa database\footnote{The database is available at the URL: \\ \url{http://astro.df.unipi.it/stellar-models/}} which contains a very fine grid of models for 19 metallicity values between $Z = 0.0002$ and $Z = 0.03$, three different initial helium abundances and three values of the mixing-length parameter for each metallicity. The models have been computed using an updated version of the \texttt{FRANEC} evolutionary code which takes into account the state-of-art of all the input physics \citep[see][for a detailed description]{tognelli11}. Here we briefly summarize the main characteristics of the code that are relevant for the present work and deeply affect both the morphology and the position of the PMS tracks in the HR diagram.

We adopted the equation of state (EOS) released in 2006 by the OPAL group \citep[see e.g.,][]{rogers02}, the OPAL high-temperature radiative opacity released by the same group in 2005 \citep[see e.g.,][]{iglesias96} for $\log T\rm{[K]} > 4.5$, and the \citet{ferguson05} low-temperature radiative opacities for $\log T\rm{[K]} \leq 4.5$. The radiative opacity tables, both for low and high temperatures, are computed assuming the solar-scaled heavy-element mixture by \citet{asplund05}. 

The outer boundary conditions, required to integrate the stellar structure equations, have been taken from the detailed atmosphere models computed by \citet{brott} for $T_{\rm{eff}} ~\leq~10\,000$ K, and by \citet{castelli03} for higher temperatures.

Convection is treated according to the Mixing Length Theory \citep{bohm58}, following the formalism described in \citet{cox}. We used the classical Schwarzschild criterion to evaluate the borders of convectively unstable regions.  

The hydrogen burning reaction rates are from the NACRE compilation \citep{nacre}, with the exception of the $^{14}$N(p,$\gamma$)$^{15}$O from the LUNA collaboration \citep{imbriani05}. The code explicitly follows the chemical evolution of the light elements (D, $^3$He, Li, Be and B) from the early phases at the beginning of the Hayashi track. The models are evolved starting from a completely formed and fully convective structure, neglecting accretion.

We extracted from the Pisa PMS database tracks for 12 metallicities, namely $Z$ = 0.007, 0.008, 0.009, 0.01, 0.0125, 0.015, 0.0175, 0.02, 0.0225, 0.025, 0.0275, 0.03. The purpose was to cover the full range of metallicities for the observed sample of stars. For models in this range of metallicities we adopted an initial deuterium abundance $X_{\rmn{D}} = 2\cdot 10^{-5}$, suitable for pop. I stars \citep[see e.g.,][]{vidal98,linsky06,steigman07}. 

For each value of $Z$ the initial helium abundance, $Y$, has been obtained by the linear relation,
\begin{equation}
Y = Y_{\rmn{P}} + Z\frac{\Delta Y}{\Delta Z}
\end{equation}
where $Y_{\rmn{P}}$ is the primordial helium abundance, and $\Delta Y/\Delta Z$ is the helium-to-metals enrichment ratio. For $Y_{\rmn{P}}$ we adopted both the recent WMAP estimation $Y_{\rmn{P}}=0.2485$ \citep[see e.g.,][]{cyburt04,steigman06} and a lower value $Y_{\rmn{P}} = 0.230$ \citep{lequeux79,pagel89,olive91}. In the first case we used both $\Delta Y/\Delta Z = 2 $ as commonly adopted in literature \citep[see e.g.,][]{pagel98,jimenez03,flynn04,2007MNRAS.382.1516C} and $\Delta Y/\Delta Z = 5$ that is the extreme value suggested by recent analysis \citep[see e.g.,][and references therein]{2010A&A...518A..13G}, while for $Y_{\rm{P}} = 0.230$ we fixed $\Delta Y/\Delta Z = 2$. Hence, for each value of $Z$, we computed models with three initial helium abundances.

The efficiency of superadiabatic convection is parametrized by the $\alpha$ parameter where the mixing length $\ell$ is given by $\ell=\alpha H_{\rmn{P}}$ and $H_P$ is the pressure scale-height. Following the usual procedure of calibrating the mixing length efficiency using the Solar observables, we obtained $\alpha = 1.68$ for our reference set of models. However there is no strong reason to adopt this value for stars in different evolutionary phases compared to the Sun. Recent analysis of PMS stars in binary systems \citep[see e.g.,][]{2000ApJ...545.1034S,steffen01,stassun04} and studies of lithium depletion in young clusters \citep{ventura98,dantona03} suggest a sub-solar efficiency of the superadiabatic convection in low-mass PMS stars. Therefore we decided to adopt tracks for three $\alpha$ values, i.e. $\alpha=1.2$ (low efficiency), $\alpha =1.68$ (our solar calibrated), $\alpha = 1.9$ (high efficiency). 

Taking into account the three $Y$ values and the three $\alpha$ values, we computed models for each metallicity using a total of nine combinations of the $\mathbf{\Xi}$ triples of meta-parameters.

The tracks have been computed for a very fine grid of masses, with a spacing of 0.05 $M_{\sun}$ in the range $M = 0.2 - 1.0 \,M_{\sun}$, of $0.1 M_{\sun}$ for  the range $M = 1.0 - 2.0\, M_{\sun}$ and of $0.2 M_{\sun}$ for the range $M = 2.0 - 3.0 M_{\sun}$. The tracks have been further interpolated on a finer mass-grid with a spacing of 0.01 M$_{\sun}$ and in age with a spacing of 0.05 Myr in the full mass range. This was done in order to achieve a very high precision in the determination of both the mass and age for the observed stars.

\section{Synthetic data sets: testing the method}
\label{sec:synttest}
\begin{figure*}
 \centering
 \resizebox{0.48\hsize}{!}{\includegraphics{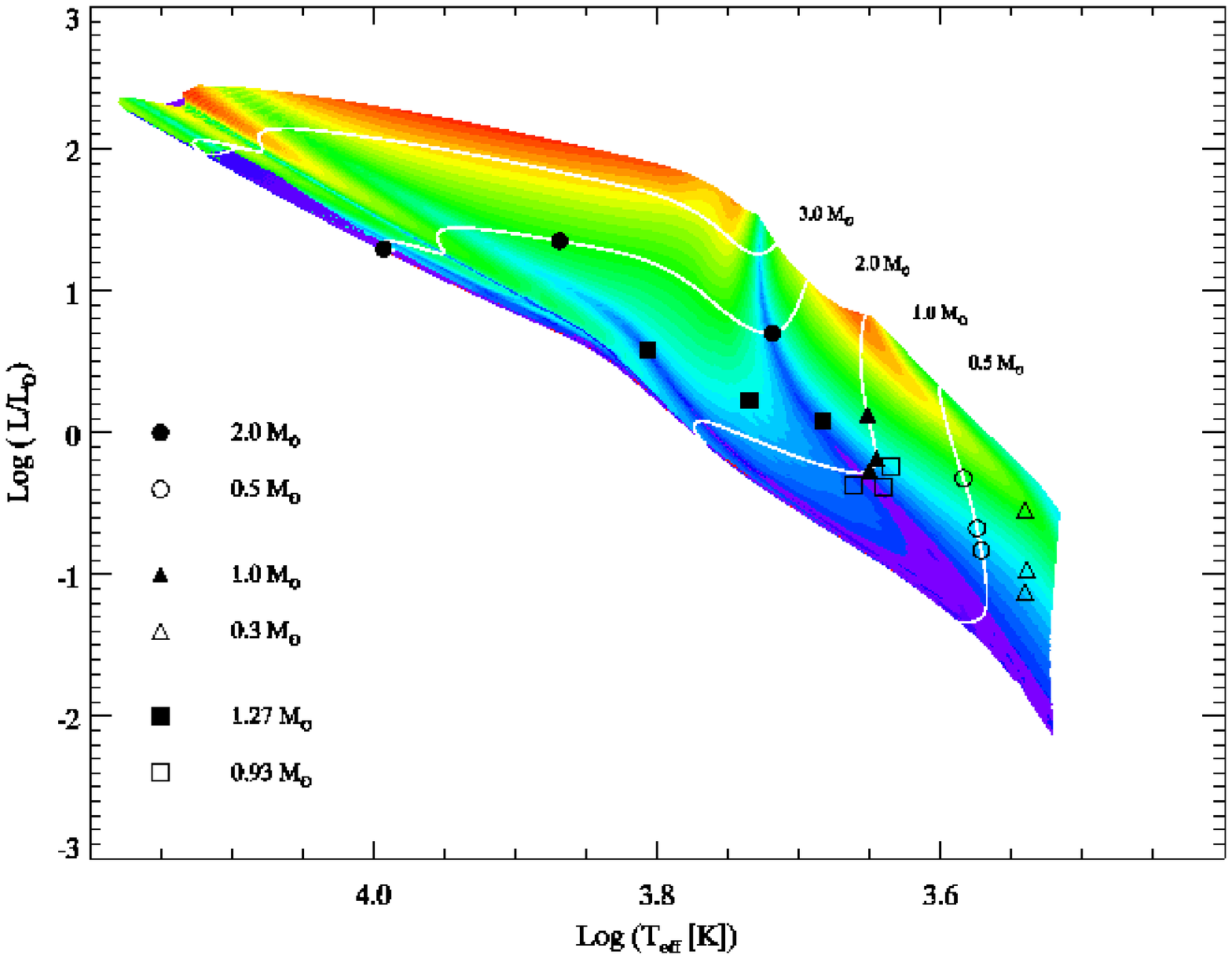}}
 \resizebox{0.48\hsize}{!}{\includegraphics{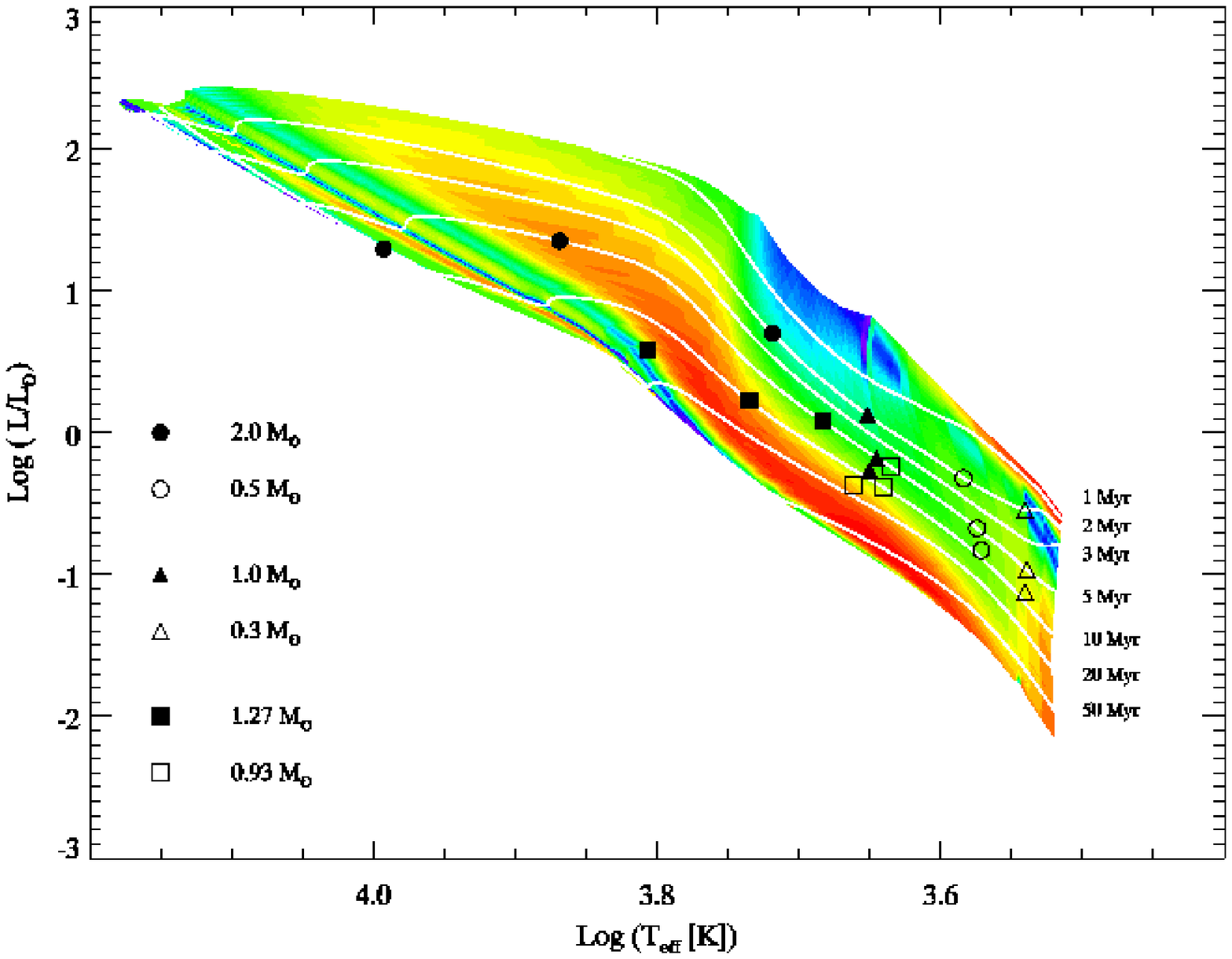}}
 \caption{\emph{Left:} evolutionary speed along stellar tracks in the HR diagram. The age-gradient is inversely proportional to this quantity. The colour coding is a scale from red-orange (fast evolution, small age-gradient, good age determination) to blue-purple (slow evolution, large age-gradient, bad age determination). \emph{Right:} mass-gradient calculated along isochrones. Red-orange regions are regions of low mass-gradient (high precision in mass determination), blue-purple regions are regions of high mass-gradient where masses are determined with worse precision. In both panels symbols indicate the positions of the simulated stars before the random errors are added. Superimposed in white are some reference tracks and isochrones. }
 \label{fig:evspeed}
\end{figure*}

In order to check the accuracy of our method we first tested it against simulated data.
As already demonstrated by JL05, the precision of inferred ages and masses is related to the detailed morphology of isochrones and tracks in the HR diagram.
Depending on the mass and the age of a star, its evolution might be faster in some parts of the diagram than in others. 
The \emph{evolutionary speed} of a star along its track together with the positional uncertainty in the HR diagram determine the absolute precision of the method. Although the PMS is globally a very fast evolutionary phase compared to the MS, there are some stages --i.e. the descent of the Hayashi tracks-- slower than others --i.e. the approach towards MS along Heyney tracks when radiative cores are developed. Another general rule is that more massive stars evolve faster, leading to progressively larger spacing between isochrones, and consequently a better relative precision in age determinations for a given evolutionary phase, as the mass increases. 

The evolutionary speed is inversely proportional to the age-gradient calculated along one evolutionary track. In the regions of the HR diagram where the evolutionary speed is large, the age-gradient is small and the precision in estimating stellar ages is good. To explain this let us define an effective temperature-luminosity error box, as given by observational uncertainties. If we move it across the HR diagram, in regions with small age-gradient we will encircle models with similar ages, leading to a precise age determination, while in regions of high age-gradient (low evolutionary speed) within the same box there will be models with very different ages, leading to a less precise age estimate.
In Fig.~\ref{fig:evspeed} (left panel) we show the evolutionary speed. Moving from blue-purple towards red-orange regions the evolutionary speed increases allowing for progressively more precise age determinations.

In analogy to the age-gradient along a track we can calculate the mass-gradient along an isochrone. In this case, regions of small mass-gradient are regions where the stellar masses can be estimated with better precision. The mass-gradient along isochrones is shown in Fig.~\ref{fig:evspeed}, right panel. In this case regions with lower mass-gradient (better precision) are in red.

To explore how the position in the HR diagram affects age and mass determinations, we generated a sample of synthetic eclipsing binaries (EBs) for different masses and ages (see Table \ref{tab:simstars}).

The simulations of EBs have been done by selecting models from the PISA stellar library with $\alpha = 1.68, \, Y_{\rmn{P}}=0.2485$ and $\Delta Y/\Delta Z = 2$. We fixed the $Z$ value to $Z=0.0125$, similar to our solar model ($Z_{\sun} = 0.0137$). For each combination of masses and ages we generated 100 systems. 
We added random Gaussian uncertainties to the quantities predicted by the models using standard deviations values equal to the typical errors in our data set ($\sigma_{M} = 0.015-0.020 \,M_{\sun}$, $\sigma_{\log T_{\rmn{eff}}} = 0.015 \,\rmn{dex}$, $\sigma_{R} = 0.05 R_{\sun}$ and $\sigma_{\log L} = 0.1 \,\rmn{dex}$).
To simulate the observed errors behaviour of EBs, we allowed for random errors in the primary star temperature, $T_{\rmn{eff},1}$, keeping the ratio of primary-to-secondary effective temperatures, $T_{\rmn{eff},1}/T_{\rmn{eff},2}$, fixed. We allowed for independent errors in the radii, $R_1$ and $R_2$. Luminosities are calculated from the temperatures and radii after the errors have been added.
\setcounter{table}{0}
\begin{table}
 \caption{Mass and ages combination for the simulated EBs. Systems C.x have the same masses of the binary system RXJ 0529.4+0041 A.}
 \label{tab:simstars}
 \centering
 \begin{center}
  \begin{tabular}{lccc}
\hline
\hline
\multirow{2}{*}{System} & Primary Mass    & Secondary Mass     &  Age  \\
			& $[M_{\sun}]$    & $M[M_{\sun}]$      &  [Myr]\\
\hline
A.2                     & 2.0             & 0.5                & 2     \\
A.5                     & 2.0             & 0.5                & 5     \\
A.8                     & 2.0             & 0.5                & 8     \\
B.2                     & 1.0             & 0.3                & 2     \\
B.5                     & 1.0             & 0.3                & 5     \\
B.8                     & 1.0             & 0.3                & 8     \\
C.5                     & 1.27            & 0.93               & 5     \\
C.10                    & 1.27            & 0.93               & 10    \\
C.15                    & 1.27            & 0.93               & 15    \\
\hline
 \end{tabular}
 \end{center}
\end{table}

For each simulated system, we applied our Bayesian method to recover the best ages and masses. 
Since we have chosen to fix $Z=0.0125$, we fixed it also in the recovery method, which is equivalent to using a prior $f(\zeta) = \delta(\zeta-0.0125)$ in equations (\ref{eq:Gtau}) and (\ref{eq:Hmu}). We ran the method both with a flat prior defined over the whole mass interval and applying a Gaussian prior on the simulated mass. The Gaussian prior is centred on the simulated value of the mass and its $\sigma$ is of the order of the typical error for the dynamical masses available in the literature (few \%). We obtained stellar ages for the single stars and also for the systems. In this last case coevality is imposed by considering $G_{\rmn{C}}(\tau) = G_{\rmn{P}}(\tau) \times G_{\rmn{S}}(\tau)$, i.e. the product of primary and secondary marginal age distributions.

\setcounter{table}{1}
\begin{table}
 \caption{Percentage of cases in which simulated ages and masses are recovered within the 68\% uncertainty interval. P indicates the primary stars in the systems, S the secondary, C stands for Coeval, indicating the cases in which the product of the marginal age distributions of the primary and secondary star $G_{\rmn{P}}(\tau) \times G_{\rmn{S}}(\tau)$ is used to infer the age of the whole system. Flat and Gaussian are the adopted mass priors.}
 \label{tab:simres}
 \centering
 \begin{center}
\renewcommand\tabcolsep{4.5pt}
  \begin{tabular}{@{}l|ccc|ccc|ccc|@{}}
\hline
\hline
 & & & & & & & & & \\[-2.5mm]
System & \multicolumn{3}{c|}{A.2} & \multicolumn{3}{c|}{A.5} & \multicolumn{3}{c|}{A.8} \\
\hline
 & & & & & & & & & \\[-2.5mm]
		& P & S & C & P & S & C & P & S & C \\
 & & & & & & & & & \\[-4.5mm]
Flat	& & & & & & & & & \\	
 & & & & & & & & & \\ [-2.5mm]
Ages 	 	& 87 & 92 & 83 & 100 & 74 & 100 & 0  & 79 & 54 \\
Masses  	& 91 & 68 &  - &  98 & 72 &  -  & 78 & 79 &  - \\
 & & & & & & & & & \\ [-1.5mm]
Gaussian	& & & & & & & & & \\			
 & & & & & & & & & \\ [-2.5mm]		
Ages 	 	& 83 & 98 & 100 & 97 & 94 & 97 & 0  & 82 & 77 \\
Masses  	& 100 & 100 &  - &  100 & 100 &  -  & 100 & 100 &  - \\
\hline
 & & & & & & & & & \\[-2.5mm]
System & \multicolumn{3}{c|}{B.2} & \multicolumn{3}{c|}{B.5} & \multicolumn{3}{c|}{B.8} \\
\hline
 & & & & & & & & & \\[-2.5mm]
		& P & S & C & P & S & C & P & S & C \\
 & & & & & & & & & \\[-4.5mm]
Flat	& & & & & & & & & \\			
 & & & & & & & & & \\ [-2.5mm]		
Ages 	 	& 94 & 98 & 97 & 96 & 72 & 84 & 95  & 62 & 81 \\
Masses  	& 77 & 76 &  - &  77 & 71 &  -  & 68 & 73 &  - \\
 & & & & & & & & & \\
Gaussian	& & & & & & & & & \\			
 & & & & & & & & & \\ [-2.5mm]		
Ages 	 	& 100 & 100 & 100 & 94 & 94 & 96 & 92  & 83 & 83 \\
Masses  	& 100 & 100 &  -  &  100 & 100 &  -  & 100 & 100 &  - \\
\hline
 & & & & & & & & & \\[-2.5mm]
System & \multicolumn{3}{c|}{C.5} & \multicolumn{3}{c|}{C.10} & \multicolumn{3}{c|}{C.15} \\
\hline
 & & & & & & & & & \\[-2.5mm]
		& P & S & C & P & S & C & P & S & C \\
 & & & & & & & & & \\[-4.5mm]
Flat	& & & & & & & & & \\			
 & & & & & & & & & \\ [-2.5mm]
Ages 	 	& 95 & 98 & 89 & 83 & 93 & 78 & 87  & 86 & 94 \\
Masses  	& 82 & 74 &  - &  100 & 68 &  -  & 86 & 84 &  - \\
 & & & & & & & & & \\
Gaussian	& & & & & & & & & \\			
 & & & & & & & & & \\ [-2.5mm]
Ages 	 	& 89 & 94 & 93 & 72 & 88 & 58 & 71  & 77 & 78 \\
Masses  	& 100 & 100 &  - &  100 & 98 &  -  & 100 & 100 &  - \\
\hline
 \end{tabular}
\renewcommand\tabcolsep{6pt}
 \end{center}
\end{table}

The results for the complete set of simulations are shown in Table \ref{tab:simres} where we report the percentage of cases in which the simulated age and mass fall within the confidence interval. From the table it is clear that the method is very successful in recovering the simulated values. When no systematic errors are present (in both the model and the data) and if the random error estimates are reliable, we can expect the results of the method for real data to be very robust. Unfortunately this ideal situation is seldom realized in reality, but it is worth noticing that the method is intrinsically able to give a good fit for almost all the regions of the HR diagram.  

The fraction of \emph{good recoveries} or \emph{success rate} can be related to the position of the stars in Fig.~\ref{fig:evspeed}. For example the secondary star of the B.x systems --a 0.3 $M_{\sun}$ star-- is moving towards slower phases of its Hayashi tracks and consequently the fraction of good age recoveries is decreasing with increasing age. Several other things are worth noticing about the recovery fractions in Table \ref{tab:simres}:

\begin{itemize}
\item  First, we note that the actual number of good recoveries is a complex function of the stellar position in the HR diagram. It is true that the mass- and age-gradient visualization of Fig.~\ref{fig:evspeed} can help understanding this function. On the other hand we have to warn the reader that what we indicated as mass- and age-gradients are partial derivatives calculated along isochrones and tracks respectively. 
Therefore, they do not fully represent the real gradient. As a consequence there are regions of the HR diagram where the recovery 
fraction is different from what one might naively expect by looking at Fig.~\ref{fig:evspeed} alone.
For example, the mass recovery fraction of the primary star of the B.8 case is lower than the B.5 case, even though the B.8 case is in a zone of lower mass-gradient (better mass resolution).

\item Secondly, we want to mention the power of the coevality prior. 
Fig.~\ref{fig:delta_age_all} shows on the left the distribution of the difference between the logarithm of the best fit age and the logarithm of the simulated age for each star. The standard deviation of this distribution is $\sigma=0.185$ dex.
The inset plot shows the recovered ages for those stars giving a bad fit, meaning that the simulated age is outside the 68\% confidence interval. The fraction of these bad-fit cases is 19\% of the total simulated stars.

In the central panel the stars in each system are paired together and the age of the system is evaluated from the composite age distribution $G_{\rmn{C}}(\tau)$. The differences between the resulting best ages and the simulated ages show a much narrower distribution, with $\sigma=0.062$, almost 3 times smaller than the $\sigma$ for the single stellar ages. Hence using the coevality prior strongly reduces the error in the best age estimate.

The rightmost panel of Fig.~\ref{fig:delta_age_all} shows the difference between the logarithms of the primary and secondary components' ages for each pair. This distribution has a $\sigma = 0.257$. 
We note that in 13\% of the cases the two components are found to be non coeval, meaning that the two uncertainty intervals do not overlap; these systems are shown in the inset diagram. 
The fraction of non coeval recoveries varies strongly among the simulated systems and the results are summarized in Table \ref{tab:frnc}.
The simulated binary that causes most of the non coeval fake detections is the system A.8. Out of 100 simulated cases, 95 give non coeval results.
This is due to the fact that the primary star of the 100 simulated A.8 systems is very close to the main sequence. Therefore $G(\tau)$ for such a star is a very flat function, with a broad uncertainty interval. Moreover the position of the peak of the distribution strongly depends on where the simulated system is scattered once that the observational errors are added.
On the other hand, the age of the secondary is well recovered, with a narrow uncertainty interval. As a consequence, in the 95\% of the cases, the two intervals become disjoint and the two stars are considered as non coeval.

The fraction of non coeval recoveries is a complex function of the position in the HR diagram of both stars. For an observed sample of binaries, the fraction of total fake non coeval systems to be expected depends on the actual distribution of relative positions of primaries and secondaries in the HR diagram.

Nevertheless we point out that the tail of non coeval systems disappears in the central diagram of Fig.~\ref{fig:delta_age_all}, i.e. no system age is found to be in disagreement with the simulated age, even when the two components were formally non coeval. 
In our simulated cases, the non coevality is mostly due to one of the stars being in a region of very large age-gradient, e.g. close to the main sequence. On the other hand the companions are generally in a region of smaller gradient. Therefore, while the age of one star is badly determined, the other component still has a very informative $G(\tau)$, which drives the composite distribution towards the simulated value.
This fact can be used to determine the age of main sequence stars in systems where the companion is still on the PMS, for example.

In addition, for the full sample of simulated cases, the use of the composite age distribution allows to find a system age that is in better agreement with the simulated one. The $\sigma$ of the distribution of recovered minus simulated age is reduced when the coevality prior is imposed.

These results are very important. It is sometimes noted in the literature that evolutionary models are not able to fit binary data for the same age and ad hoc solutions are invoked to reconcile the models and the observation. Here, we demonstrated that not being able to reproduce both components in a binary system with the same isochrone does not necessarily imply that the stars are not coeval or that they are coeval but the models are not able to reproduce this coevality.  On the contrary, an age difference or even an age mismatch can simply be a consequence of the observational errors in the HR diagram. It is indeed possible that the random scatter of the positions of coeval stars acts in opposite directions for the two components, making one look older and the other younger, to the point that they might be considered as non coeval. The actual expected artificial age difference depends on the region of the HR diagram where the two stars are located, and on their errors. 


As an example we consider the work by \cite{2009ApJ...704..531K} who analyse the binary population in the Taurus-Auriga association. They found that, in general, stars in physical pairs are more coeval than the association as a whole, with significantly smaller intra-binary age spread than for randomly paired stars selected among the association's members. Nevertheless the authors also find that some of the binaries show an intrinsic age spread larger than what observed for the bulk of the pairs and not consistent --within the errors-- with the hypothesis of coeval pairs. They suggest that these outliers can be multiple system with unrecognized companions or stars seen in scattered light or also stars with disk contamination. While this can certainly be the case, we want to point out that the observational errors themselves --even when one might think that they are completely under control as in our simulations-- can be partially responsible for an artificially large age-spread (or non coevality) within a binary system.
The fraction of objects for which this fake non coevality might be expected is a complex function of the actual distribution of observed stars in the HR diagram, i.e. of their mass, mass-ratios and ages.

\setcounter{table}{2}
\begin{table}
 \caption{Fraction of non coeval fake recoveries for the simulated systems.}
 \label{tab:frnc}
 \centering
 \begin{center}
  \begin{tabular}{lcc}
\hline
\hline
System & Simulations    & Non coeval recoveries   \\
\hline
A.2                     & 100            & 0  \%     \\
A.5                     & 100            & 10 \%        \\
A.8                     & 100            & 95 \%        \\
B.2                     & 100            & 0 \%        \\
B.5                     & 100            & 0 \%        \\
B.8                     & 100            & 8 \%       \\
C.5                     & 100            & 0 \%        \\
C.10                    & 100            & 0 \%        \\
C.15                    & 100            & 2 \%        \\
\hline
Total                   & 900           & 13 \%   \\
\hline
 \end{tabular}
 \end{center}
\end{table}

\begin{figure*}
 \centering
  \resizebox{\hsize}{!}{\includegraphics{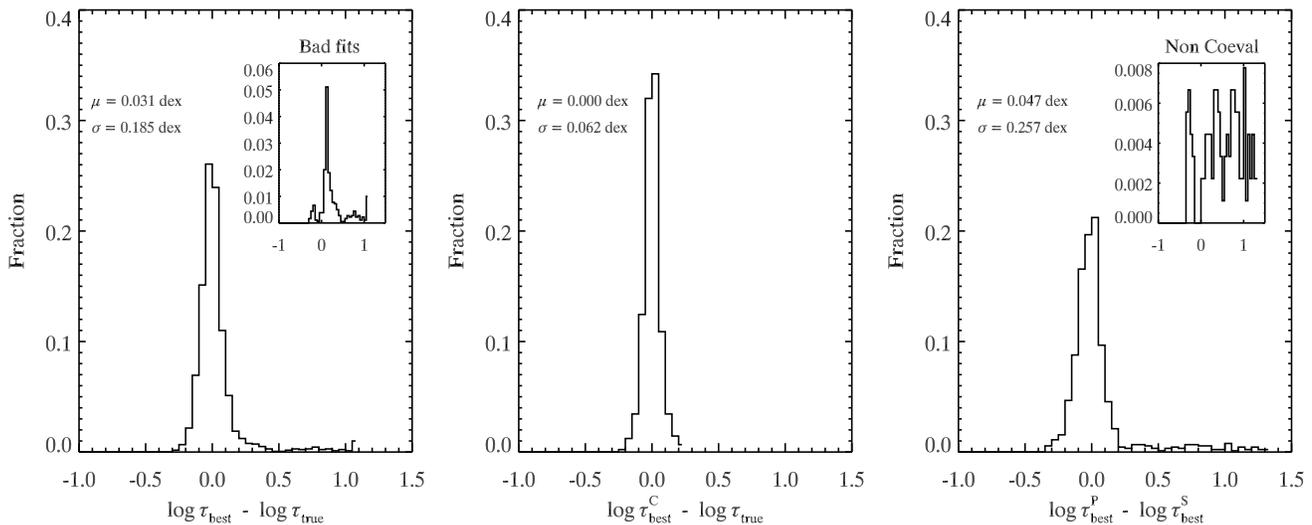}}
 \caption{\emph{Left:} distribution of the difference between the logarithm of the recovered age for the single stars and the logarithm of the simulated age. The inset shows the distribution of age differences only for the stars for which the simulated age is outside of the 68\% confidence interval \emph{Centre:} same as left, but the recovered age is obtained from the composite age distribution for each simulated binary system. \emph{Right:} Distribution of the differences between the logarithm of the primary's recovered age and the logarithm of the secondary's recovered age for each system. The inset shows the distribution of age differences for those systems whose primary and secondary components have disjoint 68\% age confidence interval and are considered as non coeval.}
\label{fig:delta_age_all}
\end{figure*}

\item As a third remark we want to emphasize the power of the Gaussian mass prior, i.e. its large informative value for the recovery of stellar masses. In almost all cases where the Gaussian mass prior is imposed, the recovery fraction raises to 100\%. This might look obvious but we have to recall that dynamical masses are usually the most reliable data available for a binary system. The impact of the Gaussian prior is less strong when the results for stellar ages are compared. Among the 27 cases displayed in Table \ref{tab:simres}, 14 favor the flat prior, while in 12 cases the Gaussian prior gives better results. The primary star of the A.8 system gives equal fractions of 0\% recoveries for Gaussian and flat prior (the star is on the MS). 

We investigated this behaviour and observed that the impact of the Gaussian prior actually depends on the detailed morphology of the 2D posterior probability in the $\mu-\tau$ space.
As an example, consider the primary star for the C.10 case. The success-ratio in the age recovery decreases from 83\% to 72\% for this star when the Gaussian mass prior is applied. 

We show in Fig.~\ref{fig:dropout} one of these drop-out cases. The reason why the age for this particular star is not recovered anymore when the prior is included can be easily understood.
The Gaussian prior is causing an increase in the relative precision for the marginal distribution, $G_{\rmn{Gauss}}(\tau)$. This can be seen in the right panel of the figure  where the $G_{\rmn{Gauss}}(\tau)$ distribution has a clearly larger mode and is narrower than $G_{\rmn{Flat}}(\tau)$. One can notice that the right border of the confidence interval for $G_{\rmn{Flat}}(\tau)$ is already quite close to the simulated age value of 10 Myr, which is barely within the 68\% confidence interval. 
Hence it is the shrinking of the confidence interval --when the Gaussian prior is applied-- that causes the simulated age value to drop out of the 68\% confidence interval. 
Also the actual number of dropouts is related to the detailed structure of the 2D mass- and age-gradients. 



\end{itemize}

\begin{figure*}
 \centering
  \resizebox{\hsize}{!}{\includegraphics{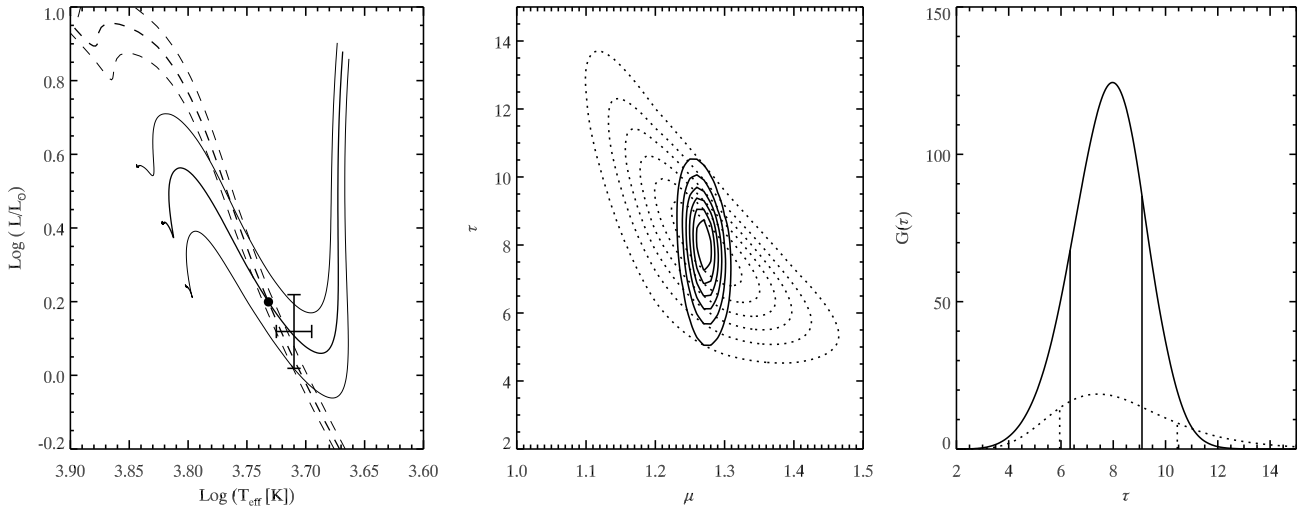}}
 \caption{A particular realization for the primary star of the  C.10 system. In this particular case the age is well recovered in the flat mass prior case and not recovered  when the Gaussian mass prior is imposed. \emph{Left:} the HR diagram position for the star with superimposed some reference isochrones (9,10,11 Myr) and tracks (1.17,1.27 and 1.37 $M_{\sun}$). The cross indicates the position after errors are added, the filled circle represents the original position of a $1.27 M_{\sun}$, 10 Myr star. \emph{Centre:} 2D posterior probability contours. Dotted line is for the case with flat mass prior and solid line for the case with Gaussian mass prior. \emph{Right:} Marginal age distributions. Vertical lines indicate the confidence interval. Dotted and solid line as for the central panel.}
\label{fig:dropout}
\end{figure*}



\section{The data set}
\label{sec:data}
The number of PMS stars with direct mass measurements amounts nowadays to about 30 objects. The sample we used consists of 25 PMS and 2 MS stars, whose properties are summarized in Table \ref{Tab:data}.

Among the 27 objects 10 are PMS stars found in eclipsing binary systems (EB). We included the MS $3.1\, M_{\sun}$ star TYCrA A and $2.1\, M_{\sun}$ star EK Cep A in our sample as well, since their companions are PMS stars in EB systems and we tried to fit both components of binary systems when possible (see Sect.~\ref{sec:binanalysis}).
For EBs the masses, radii, and effective temperature ratios of the two components can be accurately inferred. However the absolute values of the effective temperatures, that rely on the determination of the primary's spectral type and some spectral type-$T_{\rm{eff}}$ relationship, can be affected by systematic errors and constitute a severe source of uncertainty. As an example \cite{hillenbrand04} pointed out that the main sequence empirical scales used for deriving temperatures from the spectral types might cause systematic temperature offsets when applied to PMS stars, due to the different values of surface gravity for a given spectral type \citep[see also][]{luhman97}.

Of the 27 stars in the sample, 6 are found in astrometric/spectroscopic systems (AS), i.e. systems in which the components can be resolved as separate point sources using interferometry. Combining astrometry and line-of-sight velocity measurements the masses of the components are determined in a distance-independent way. With this technique the radii of the components are not measurable, though.

The last 9 objects have masses measured using their circumstellar disks keplerian velocities obtained by means of spectroscopy. The mass for the central star can be determined only if the linear value of the orbital radius at which the velocity is measured is known. Hence stellar masses are in this case distance-dependent. The two stars in the UZ Tau E system form a binary, and their masses are separated using combined spectroscopic measurements for the circumstellar disk and the stellar velocities (DKS).

A very similar sample was already studied in \cite{mathieu07} and \cite{stassun08} and we refer the reader to the first of these papers for a detailed description of the different observational techniques and the different kind of uncertainties affecting them. Compared to \cite{mathieu07} there are some distinct objects in our sample though. The 2M0535-5 brown dwarf EB \citep{2006Natur.440..311S} was excluded because the stellar dynamical masses are smaller than those currently present in the Pisa database. The recently discovered PMS EB ASAS J052821+0338.5 \citep{2008A&A...481..747S} has been added to the sample. We also included the AS binary HD 113449 \citep{2010RMxAC..38...34C} for which we have slightly different parameters from an updated analysis (Cusano, \emph{private communication}).

\begin{figure}
 \centering
   \resizebox{\hsize}{!}{\includegraphics{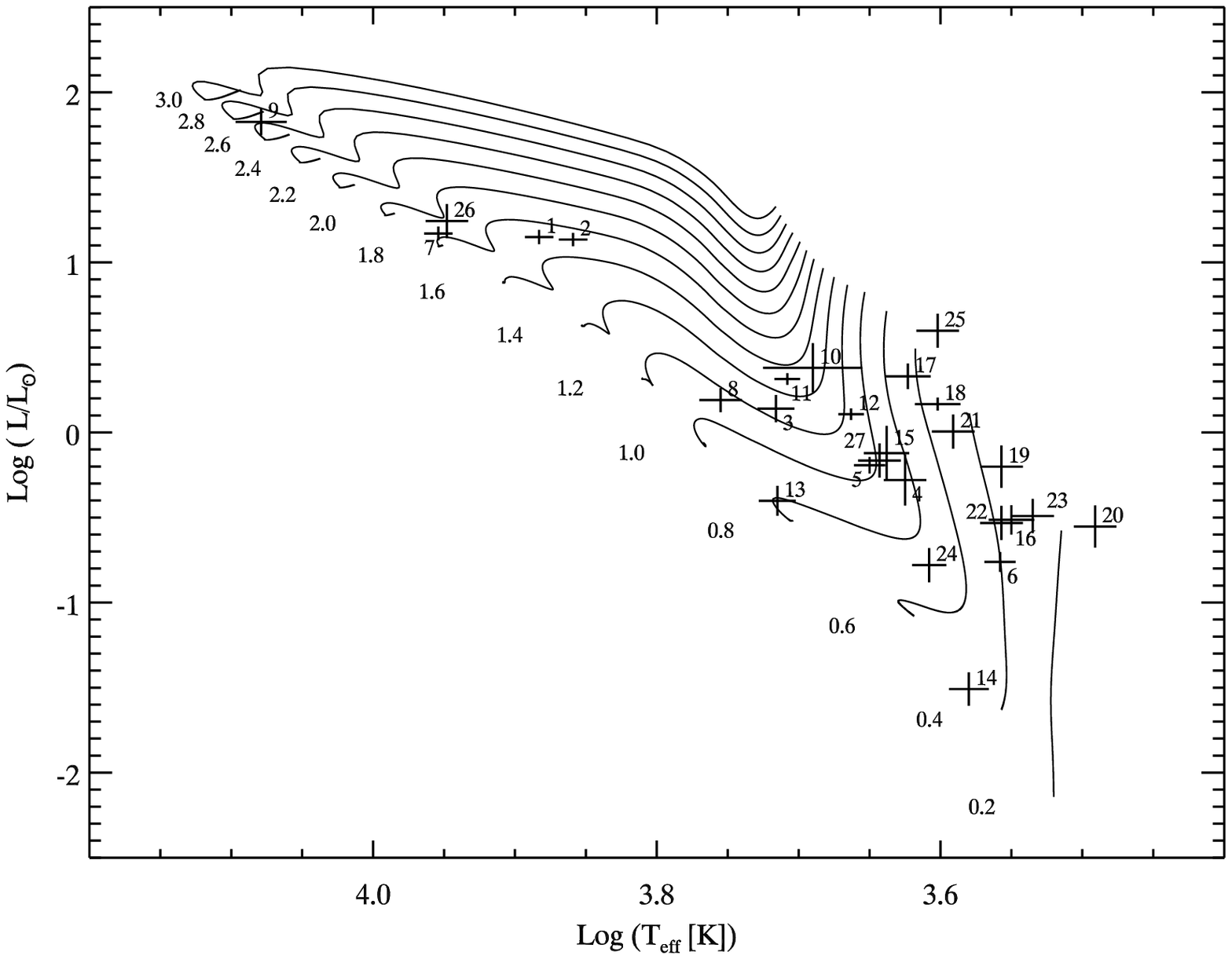}}
 \caption{The HR diagram for our data set. The labels correspond to the ID column in Table \ref{Tab:data}. Superimposed are stellar tracks calculated with $\alpha=1.68, \Delta Y/\Delta Z = 2$ and $Y_{\rmn{P}} = 0.2485$ and $Z=0.0125$. The values of the mass --in solar units-- are displayed on the left of the corresponding track.}
 \label{fig:HRdata}
\end{figure}

The luminosities and effective temperatures for our complete data set are displayed in Fig.~\ref{fig:HRdata}. Overplotted are stellar tracks for $\alpha=1.68$, $\Delta Y/\Delta Z = 2$ and $Y_{\rmn{P}}=0.2485$. We will refer to this set of parameters $\bmath{\Xi}$ as our standard or reference set. In the case of Fig.~\ref{fig:HRdata} the tracks are calculated for a value of $Z =  0.0125$, similar to the metallicity of our standard solar model (i.e., $Z_{\sun} = 0.0137$). 

For several systems, [Fe/H] values are available from direct spectroscopic measurements. We used the [Fe/H] determinations by \cite{2009A&A...501..973D} and by \cite{2011A&A...526A.103D} for the systems in Orion and Taurus-Auriga, respectively. Only 4 stars are left without a [Fe/H] measurement. In order to convert the observed [Fe/H] into the global metallicity $Z_{\rm{ob}}$, we followed eq. (2) in \citet{2010A&A...518A..13G}, adopting $(Z/X)_{\sun} = 0.0181$ by \citet{2009ARA&A..47..481A}. Although the Pisa PMS models have been computed adopting the \citet{asplund05} heavy elements solar mixture, the seeming inconsistency is inconsequential since models computed with \citet{asplund05} and \citet{2009ARA&A..47..481A} but with the same total metallicity $Z$ are essentially indistinguishable \citep[see detailed discussion in ][]{tognelli11}.

A point worth mentioning is that among the objects listed in Table \ref{Tab:data} there are a few cases that appear peculiar. Their location in the HR diagram is indeed incompatible with that of stars of similar mass. 
By looking at the available dynamical masses, luminosities and temperatures we identified four of these peculiar objects, namely NTT 045251 A, UZ Tau Ea, BP Tau, and MWC 480 (see Table \ref{Tab:data}).
\begin{itemize}
 \item NTT 045251 A is close to RXJ 0529.4 Ab and V1174 Ori A in the HR diagram, with approximately the same luminosity and $T_{\rmn{eff}}$. However NTT 045251 A is more massive than  the other two stars by 0.4-0.5 $M_{\sun}$. We checked that this discrepancy can not be reconciled even by assuming that the metallicity of NTT 045251 A is 0.4 dex larger than that of RXJ 0529.4 Ab or V1174 Ori A. 
 \item UZ Tau Ea, $M = 1.016 M_{\sun}$, has a $T_{\rmn{eff}}$ similar to V1174 Ori B, DM Tau, CY Tau, and NTT 045251 B which have lower masses, between 0.55-0.8 $M_{\sun}$. Moreover the star is colder than RXJ 0529.4 Ab and V1174 Ori A by about 850 K in spite of their similar masses, luminosities, and [Fe/H]. 
 \item BP Tau, $M = 1.320 M_{\sun}$, is significantly colder ($\Delta T_{\rmn{eff}} \ga 1000$ K) and fainter ($\Delta \log L/L_{\sun} \approx 0.9$) than RXJ 0529.4 Aa and EK Cep B, although it is slightly more massive and the metallicities are similar. In addition this star is fainter and colder than the 0.96 $M_{\sun}$ HD 113449, which has a similar metallicity. This is not easy to explain because the minimum luminosity of a 1.3 M$_{\sun}$ model is always larger than that achieved by a 0.9-1.0 M$_{\sun}$ star approaching the ZAMS, as in the case of HD 113449 A.
 \item MWC 480 ($M = 1.65 M_{\sun}$ and [Fe/H] = -0.01) is located in the HR diagram between EK Cep A ($M = 2.02 M_{\sun}$ and [Fe/H] = 0.07) and the RS Cha system ($M = 1.87, 1.89 M_{\sun}$ and [Fe/H] = 0.17). We checked with our models that the difference in [Fe/H] can justify neither the similar luminosities of MWC 480 and EK Cep nor the higher luminosity of MCW 480 with respect to the RS Cha stars. 
\end{itemize}

\setcounter{table}{3}
\begin{table*}
    \begin{minipage}{175mm}
 \caption[]{List of stellar properties.}
\label{Tab:data}
\renewcommand\tabcolsep{4.pt}
\begin{tabular}{lllcccccc}
 \hline 
\hline
ID &Name      &Type & Mass              & Radius            & $\log T_{\rmn{eff}}$ & $\log L$          & [Fe/H]           &Ref. \\
   &          &     & $[M_{\sun}]$      & $[R_{\sun}]$      & [K]                     & $[L_{\sun}]$      &                  &  \\
\hline
   &                      &    &                 &                &                 &                  &                &  \\[-2mm]
01 &RS Cha A              & EB & $1.890\pm0.010$ &$2.150\pm0.060$ & $3.883\pm0.010$ & $\phantom{-}1.149 \pm0.041$ & $\phantom{-}0.17\pm0.01$ & And91, Rib00, Ale05 \\
02 &RS Cha B              & EB & $1.870\pm0.010$ &$2.360\pm0.060$ & $3.859\pm0.010$ & $\phantom{-}1.136 \pm0.039$ & $\phantom{-}0.17\pm0.01$ & And91, Rib00, Ale05\\   
03 &RXJ 0529.4$^1$\footnotetext{$^1$ Short form for RXJ 0529.4+0041} Aa     & EB & $1.270\pm0.010$ &$1.440\pm0.050$ & $3.716\pm0.013$ & $\phantom{-}0.140\pm0.080$ & $-0.01\pm0.04$ & Cov04, Dor09 \\
04 &RXJ 0529.4$^1$ Ab     & EB & $0.930\pm0.010$ &$1.350\pm0.050$ & $3.625\pm0.015$ & $-0.280\pm0.150$ & $-0.01\pm0.04$ & Cov04, Dor09 \\
05 &V1174 Ori A           & EB & $1.009\pm0.015$ &$1.339\pm0.015$ & $3.650\pm0.011$ & $-0.193\pm0.048$ & $-0.01\pm0.04$ & Sta04, Dor09 \\
06 &V1174 Ori B           & EB & $0.731\pm0.008$ &$1.065\pm0.011$ & $3.558\pm0.011$ & $-0.761\pm0.058$ & $-0.01\pm0.04$ & Sta04, Dor09 \\
07 &EK Cep A              & EB & $2.020\pm0.010$ &$1.580\pm0.015$ & $3.954\pm0.010$ & $\phantom{-}1.170 \pm0.040$ & $\phantom{-}0.07\pm0.05$ & Pop87, Mar93 \\
08 &EK Cep B              & EB & $1.124\pm0.012$ &$1.320\pm0.015$ & $3.755\pm0.015$ & $\phantom{-}0.190\pm0.070$ & $\phantom{-}0.07\pm0.05$ & Pop87, Mar93 \\
09 &TY CrA A              & EB & $3.160\pm0.020$ &$1.800\pm0.100$ & $4.079\pm0.018$ & $\phantom{-}1.826\pm0.078$ &     -           & Cas98 \\
10 &TY CrA B              & EB & $1.640\pm0.010$ &$2.080\pm0.140$ & $3.690\pm0.035$ & $\phantom{-}0.380\pm0.145$ &     -           & Cas98 \\
11 &ASAS 052821$^2$\footnotetext{$^2$ Short form for ASAS J052821+0338.5} A & EB & $1.387\pm0.017$ &$1.840\pm0.010$ & $3.708\pm0.009$ & $\phantom{-}0.314 \pm0.034$ & $-0.15\pm0.20$ & Ste08 \\
12 &ASAS 052821$^2$ B & EB & $1.331\pm0.011$ &$1.780\pm0.010$ & $3.663\pm0.009$ & $\phantom{-}0.107 \pm0.034$ & $-0.15\pm0.20$ &  Ste08 \\
13 &HD 113449 A           & AS & $0.960\pm0.087$ & -              & $3.715\pm0.013$ & $-0.402\pm0.088$ & $-0.03\pm0.10$ & Pau06, Cus10 \\
14 &HD 113449 B           & AS & $0.557\pm0.050$ & -              & $3.580\pm0.014$ & $-1.509\pm0.098$ & $-0.03\pm0.10$ & Pau06, Cus10 \\
15 &NTT 045251$^3$\footnotetext{$^3$ Short form for NTT 045251+3016} A     & AS & $1.450\pm0.190$ & -              & $3.638\pm0.016$ & $-0.122\pm0.160$ &   -            & Ste01 \\
16 &NTT 045251$^3$ B     & AS & $0.810\pm0.090$ & -              & $3.550\pm0.016$ & $-0.514\pm0.086$ &   -            & Ste01 \\
17 &HD 98800 Ba           & AS & $0.699\pm0.064$ & -              & $3.623\pm0.016$ & $\phantom{-}0.330 \pm0.075$ & $-0.20\pm0.10$              & Bod05, Las09\\
18 &HD 98800 Bb           & AS & $0.582\pm0.051$ & -              & $3.602\pm0.016$ & $\phantom{-}0.167\pm0.038$ & $-0.20\pm0.10$      & Bod05, Las09\\
19 &UZ Tau Ea$^a$\footnotetext{$^a$ The error on the mass does not include the uncertainty on the distance.}             & DKS& $1.016\pm0.065$ & -              & $3.557\pm0.015$ & $-0.201\pm0.124$ & $-0.01\pm0.05$ & Pra02, Dor11 \\
20 &UZ Tau Eb$^a$             & DKS& $0.294\pm0.027$ & -              & $3.491\pm0.015$ & $-0.553\pm0.124$ & $-0.01\pm0.05$ & Pra02, Dor11 \\
21 &DL Tau$^a$                & DK & $0.720\pm0.110$ & -              & $3.591\pm0.015$ & $\phantom{-}0.005 \pm0.100$ & $-0.01\pm0.05$ & Sim00, HW04, Dor11 \\
22 &DM Tau$^a$                & DK & $0.550\pm0.030$ & -              & $3.557\pm0.015$ & $-0.532\pm0.100$ & $-0.01\pm0.05$ &  Sim00, HW04, Dor11\\
23 &CY Tau$^a$                & DK & $0.550\pm0.330$ & -              & $3.535\pm0.015$ & $-0.491\pm0.100$ & $-0.01\pm0.05$ &  Sim00, HW04, Dor11\\
24 &BP Tau$^a$                 & DK & $1.320\pm0.200$ & -              & $3.608\pm0.012$ & $-0.780\pm0.100$ & $-0.01\pm0.05$ &  Joh99, Dut03, Dor11\\
25 &GM Aur$^a$               & DK & $0.840\pm0.050$ & -              & $3.602\pm0.015$ & $\phantom{-}0.598 \pm0.100$ &               $-0.01\pm0.05$ & Sim00, HW04, Dor11\\
26 &MWC 480                & DK & $1.650\pm0.070$ & -              & $3.948\pm0.015$ & $\phantom{-}1.243\pm0.100$ &   $-0.01\pm0.05$             &  Sim00, HW04, Dor11\\
27 &LkCa 15$^a$                & DK & $0.970\pm0.030$ & -              & $3.643\pm0.015$ & $-0.165\pm0.100$ &    $-0.01\pm0.05$           &  Sim00, HW04, Dor11\\
\hline
\end{tabular}
	  And91~=~\citet{1991A&ARv...3...91A};
	  Rib00~=~\citet{2000MNRAS.318L..55R};
	  Ale05~=~\citet{2005A&A...442..993A};
	  Cov04~=~\citet{covino04};
	  Dor09~=~\citet{2009A&A...501..973D}
	  Sta04~=~\citet{stassun04};
	  Pop87~=~\citet{1987ApJ...313L..81P};
	  Mar93~=~\citet{1993A&A...274..274M};
	  Cas98~=~\citet{1998AJ....115.1617C};
	  Ste08~=~\citet{2008A&A...481..747S};
	  Pau06~=~\citet{2006PASP..118..706P};
	  Cus10~=~\citet{2010RMxAC..38...34C};
	  Ste01~=~\citet{steffen01};
	  Bod05~=~\citet{2005ApJ...635..442B};
	  Las09~=~\citet{2009ApJ...698..660L};
	  Pra02~=~\citet{2002ApJ...579L..99P};
	  Sim00~=~\citet{2000ApJ...545.1034S};
	  HW04~=~\citet{hillenbrand04};
	  Dor11~=~\citet{2011A&A...526A.103D};
	  Joh99~=~\citet{1999ApJ...516..900J};
	  Dut03~=~\citet{2003A&A...402.1003D} 
\renewcommand\tabcolsep{6pt}
\end{minipage}
\end{table*}

\section{Theoretical vs. dynamical masses -- the standard set of models}
\label{sec:theovsdyn}
In this section we show the comparison of the whole data set with our evolutionary models. We limited the analysis only to the standard set of models, i.e the $\bmath{\Xi}$ class with $\alpha=1.68, \Delta Y/\Delta Z = 2$ and $Y_{\rmn{P}} = 0.2485$. Section \ref{sec:compmod} is dedicated to the comparison of models with different meta-parameters, $\mathbf{\Xi}$. Since radii are not available for each star in the data set, the comparison was done in the HR diagram. With the notation of Sect.~\ref{Sec:Baymeth}, this means $\bmath{q} = (\log T_{\rmn{eff}}, \log L/L_{\sun}  )$. In the case of stars with available [Fe/H] measurements, a Gaussian prior on the metallicity was applied after converting the [Fe/H] values and their errors into $Z$ values with corresponding errors $\sigma_{Z}$:
\[
f(\zeta) = \frac{1}{\sqrt{2\,\pi \sigma^2_Z } } \times \exp \left[ -\frac{ \left( \zeta - Z \right)^2 }{2\, \sigma^2_Z  } \right] \, .
\]
In the other cases a flat prior for $Z \in [0.007, 0.03]$ was used. 

\begin{figure*}
 \centering
   \resizebox{\hsize}{!}{\includegraphics{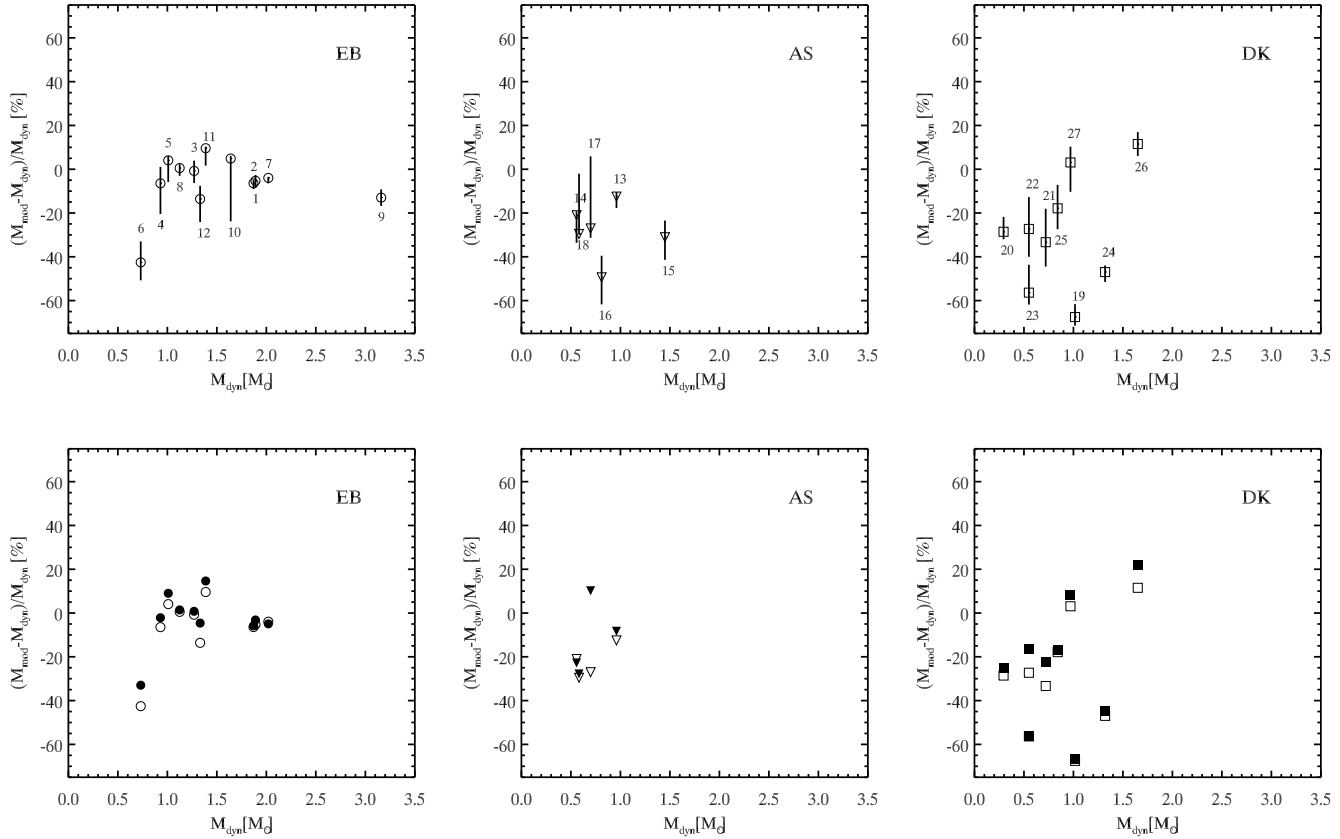}}
 \caption{Inferred masses from our standard set of isochrones compared with the dynamical masses from literature. Numbers correspond to the ID column of Table \ref{Tab:data}. \emph{From left to right}: stars are divided in subset of eclipsing binaries (EB), astrometric/spectroscopic binaries (AS) and stars with masses inferred from disk kinematics (DK). \emph{Upper panels:} best values are indicated by empty symbols. The asymmetric error bars represent the 68\% confidence interval, as defined in Sect.~\ref{Sec:Baymeth}. In these cases a Gaussian metallicity prior was applied when [Fe/H] measurements were available. \emph{Lower panels:} Comparison of the best masses inferred when the metallicity Gaussian prior is imposed (empty symbols, same as upper panels) and when we use a flat prior even for stars with [Fe/H] measurements (filled symbols).}
 \label{fig:6pan}
\end{figure*}

The outcomes of the full data set comparison are summarized in Fig.~\ref{fig:6pan}. The stars have been divided into subgroups: eclipsing binaries (EB), astrometric/spectroscopic binaries (AS) or disk kinematic stars (DK). The stars in the UZ Tau E system (DKS, i.e. disk kinematics plus spectroscopy to disentangle the components) are included in the DK sample for simplicity.
The three subgroups are displayed from left to right.
Each panel shows a comparison of the relative difference between model-inferred mass --$M_{\rmn{mod}}$-- and measured dynamical mass --$M_{\rmn{dyn}}$. 
In the upper panels $M_{\rmn{mod}}$ is derived by applying the Gaussian metallicity prior, when available. The lower panels show a comparison between the Gaussian $Z$ prior case (empty symbols) and the flat $Z$ prior case (full symbols).
The symbols indicate the mode of the posterior probability; the asymmetric error bars indicate the 68\% confidence interval as described in Sect.~\ref{Sec:Baymeth}. In the figure the dynamical mass errors are not added to the error budget. This is meant to purely show the precision of the masses estimated from the models given the observational uncertainties. However this is not a bad approximation for the total mass error budget, given that the quoted errors are of the order of 1\% for most of the dynamical masses and up to 10\% only in very few cases (see Table \ref{Tab:data}).

By looking at the figure, it is clear that EB masses are well recovered in almost every case but for V1174 Ori B. We will discuss this particular object more in detail in Sect.~\ref{sec:V1174Ori}. The general agreement becomes progressively worse for AS binaries and DK stars.
The AS binaries have masses that are underestimated by $\sim 20-30 \%$ on average. The worst case is NTT 045251+3016 B, whose mass is underestimated by $\sim 50$\% (see Sect.~\ref{subsec:NTT}). For DK stars the trend becomes more negative with underestimates as low as 70\% for UZ Tau Ea.
Regarding the DK group, we have to point out that for the latter class of objects the uncertainty on the distance is not included in the dynamical mass error estimate. This uncertainty propagates linearly in the mass uncertainty and quadratically in the luminosity uncertainty. The DK stars in our sample stars are part of the Taurus-Auriga star-forming complex which is located at about 150 pc from the Sun and has a radius of about 15 pc \cite[see e.g][]{2009ApJ...698..242T}. Using an average distance to each star instead of its real distance may then cause a systematic error on the mass estimate of about 10\% and on the luminosity of up to 20\%. 
Part of the disagreement might also arise from the fact that the DK objects are T-Tauri stars, intrinsically variable.
The temperatures and luminosities we adopt for them are all derived by \cite{hillenbrand04}. The authors try to minimize the effects of accretion luminosity using the $I_{\rmn{C}}$ band to estimate the stellar luminosity. Nevertheless the estimated stellar luminosities might still be offset from their real values. An other problem might affect the temperature determination. Temperatures are determined from spectral types using relations calibrated on dwarfs. As \cite{hillenbrand04} point out, there might be a systematic temperature underestimate  due to the fact that PMS stars of a given spectral type are generally warmer than dwarf counterparts.
If the real stellar temperatures would be higher part of the discrepancy in our mass estimates would be removed, since larger masses would be needed to reproduce the observed stellar properties.

The lower panels of fig. \ref{fig:6pan} show a quite surprising result. In the cases where we applied the Gaussian prior on the metallicity the final results for $M_{\rmn{mod}}$ are usually in worse agreement with the $M_{\rmn{dyn}}$ values when compared to the flat-prior case.
It appears that applying the metallicity Gaussian priors we obtain, in general, lower masses values. This behaviour suggests that the $Z_{\rmn{ob}}$ values used here might be too low. If we force the metallicity to assume systematically lower values than the real values by means of the Gaussian prior, then we naturally obtain lower values for the best masses. This is due to the fact that the lower the metallicity used in a stellar model, the hotter and more luminous the model is for a given mass.

We speculate here that part of the problem with the metallicity prior might reside in the low value of $(Z/X)_{\sun}$ used in the present work to convert the observed [Fe/H] into $Z$. In the recent years this value has undergone a drastic change, specially after the introduction of non-LTE and 3D hydrodynamical atmospheric models for the analysis of Solar abundances \citep[see][for a review about the topic]{2009ARA&A..47..481A}.
The traditional value of $Z_{\sun}\sim0.02$ has been strongly revised towards much lower values, down to $Z_{\sun}~0.013\div0.014$ \citep[see e.g.,][and references therein]{serenelli09}.
There is still ongoing debate regarding solar heavy elements relative abundances and total metallicity, and the uncertainty on the absolute values is still large. Nevertheless the most recent results are going in the increasing direction for $(Z/X)_{\sun}$, thus reducing the difference with the traditional estimates. A change in $(Z/X)_{\sun}$ will be reflected directly into $Z_{\rmn{ob}}$  since $(Z/X)_{\rmn{ob}} = (Z/X)_{\sun}\times 10^{\rmn{[Fe/H]}}$ and an increase in this quantity would naturally lead to a systematically larger $M_{\rmn{mod}}$.

\section{Analysis of the data using different classes of models}
\label{sec:compmod}

As described in Sect.~\ref{subsec:evidence}, the ratio of the evidences for two classes of models --the Bayes Factor, $BF$-- can be used to quantify which class of model is better in reproducing the data. Significantly better evidence of a model over an other is claimed when $BF < 0.1$ or $BF > 10$, i.e. when the two evidences differ by one order of magnitude or more \citep{kr95}. Since the evidence is calculated by marginalizing the \emph{posterior} distribution over all the parameters of the model, the \emph{prior} distribution of the parameters has to be considered as part of the model as well \citep[see e.g.][for an application of the Bayes Factor to discriminate between distinct models in a different astrophysical context]{2011MNRAS.tmp..993B}.

We calculated the evidence for each star in the 9 different meta-parameters cases and for each set $\bmath{\Xi}$ we considered 4 different combinations of the prior distributions for the masses and the metallicities. The four combinations are: 1) flat mass prior - flat metallicity prior, 2) Gaussian mass prior - flat metallicity prior, 3) flat mass prior - Gaussian metallicity prior, 4) Gaussian mass prior - Gaussian metallicity prior. Hence we actually have $9 \times 4 = 36$ classes of models.
After evidences have been calculated in the 36 cases for each of the 27 stars of the sample, the $BFs$ have all been calculated by dividing each evidence value for a given star by the evidence value for that same star obtained using our standard class of models. The latter is identified by $Y_{\rmn{P}} = 0.2485$, $\Delta Y/ \Delta Z=2 $ and $\alpha=1.68$ for the 1) case.

The numerical values of the $BFs$ are reported in Appendix \ref{Sec:AppTab}. The Tables \ref{tab:bfactNoMetNomass}, \ref{tab:bfactNoMetYesmass}, \ref{tab:bfactYesMetNomass}, and \ref{tab:bfactYesMetYesmass} display the $BF$ values for the 1), 2), 3), and 4) prior cases respectively.
Looking at each table, it is possible to see the change in the evidence among the different $\bmath{\Xi}_j$ sets, within one of the 4 prior cases.
From one table to the other, the corresponding entries are calculated for the same $\bmath{\Xi}$ set but in the 4 different prior cases; by looking at these corresponding entries it is possible to see the role played by the prior choice and, actually, understand whether this choice is leading to an improvement of the overall fit or not.

The best way to compare different classes using the whole data set is to calculate the composite evidence for the full set of data. The natural extension of equation (\ref{eq:singleev}) is that the evidence for the whole data set --represented here by the set of observables $\{\bmath{q}\}$-- can be written as:
\begin{eqnarray}
\label{eq:combev}
   f(\{\bmath{q}\} | \bmath{\Xi}) & = & \prod_k f(\bmath{q}_k| \bmath{\Xi}) \nonumber \\ 
		                & = & \prod_k \int f(\bmath{q}_k,\bmath{p}_k|\bmath{\Xi}) \,\rmn{d}\bmath{p}_k \nonumber \\
                                & = & \prod_k \int f(\bmath{q}_k | \bmath{p}, \bmath{\Xi}) f_k(\bmath{p} | \bmath{\Xi}) \,\rmn{d}\bmath{p}
\end{eqnarray}
The index $k$ runs over the stars in the sample and $f_k$ indicates the specific prior distribution applied for the $k$-th star.

We multiplied the evidences of the 27 stars of the sample to understand which class of model gives the best general result. In addition we restricted the product to stars belonging only to one type of system (EB, AS, DK). When using the whole data set we can compare only the 1) and 2) cases because the Gaussian metallicity prior can not be applied to the 4 stars for which [Fe/H] measurements are not available. To compare also the 3) and 4) cases we additionally restricted the sub-samples to only stars with [Fe/H] estimates.

In almost all the cases, the class of models with $\Delta Y/\Delta Z =2$, $Y_{P}=0.23$ and $\alpha=1.2$ meta-parameters is the one with the largest evidence. The only exception is the case of AS subset when only stars with known [Fe/H] are considered. For this subset the largest evidence is attained for the $\Delta Y/\Delta Z =2$, $Y_{P}=0.23$ and $\alpha=1.9$ meta-parameters values. We have to point out that in this particular case only 4 stars are part of the subset and the evidence is only 1.13 times larger than in the $\Delta Y/\Delta Z =2$, $Y_{P}=0.23, \alpha=1.2$  case.

These results are the counterpart of what we have already observed in Sect.~\ref{sec:theovsdyn} regarding the mass underestimation by standard models. It was clear --specially by looking at the upper panels of Fig.~\ref{fig:6pan}-- that the general trend for the standard set of models is to predict too low masses compared to $M_{\rmn{dyn}}$. 
A similar trend can be observed in Fig.~3 of \cite{mathieu07}; here all the considered sets of models show the same kind of behaviour in predicting too low stellar masses, with a mean difference that can be of the order of 20\% or more. This suggests that the standard tracks are too hot and luminous when displayed in the HR diagram compared to the observed temperatures and luminosities for the given dynamical masses. A natural way to get a better agreement with observations is to use colder and fainter set of models by adopting both a lower helium initial abundance and mixing length parameter $\alpha$ value.  

This explains why the best overall evidence is achieved by the set with $\alpha=1.2$,  $Y_{\rm{p}} = 0.230$, and $\Delta Y /\Delta Z = 2$. Similar low helium content for a given metallicity $Z$ could be obtained also adopting the currently accepted primordial helium value, $Y_{\rm{P}} = 0.2485$, together with a very small helium-to-metals enrichment ratio ($\Delta Y/\Delta Z \la 1$, for $Z\approx 0.01\div 0.02$). However, both choices are quite unlikely. In the former case the $Y_{\rmn{P}} = 0.23$ value is significantly lower than the recent independent results from extragalactic H{\sevensize II} regions and Big Bang Nucleosynthesis theory \citep[][]{2007ApJ...662...15I,2007ASPC..374...81P,2009ApJS..180..306D,2010JCAP...04..029S}. Regarding the latter case, a $\Delta Y/\Delta Z \la 1$ is smaller than the value suggested by both Galactic chemical evolution models \citep[][]{2005A&A...430..491R,2008RMxAA..44..341C} and by nearby dwarf stars analysis \citep[][]{jimenez03,2007MNRAS.382.1516C,2010A&A...518A..13G}.


We point out that when a single star is considered the value of the evidence for different classes of models are mostly of the same order of magnitude (see the Tables in Appendix \ref{Sec:AppTab}). Hence these global results are more sensitive to the few objects for which we observe major changes in the evidence between different classes. Nevertheless a general analysis is still important to understand the overall behaviour of stellar models in comparison to available data.

We also report that, for the whole sample, as for any subset of stars (EB, AS or DK), the largest value of the evidence is reached in cases 2) or 4), i.e. when the Gaussian mass prior is imposed. This is expected because the prior on the dynamical mass is a much more informative prior than a flat one defined across the entire mass range of simulated models (i.e. 0.2-3.6 M$_{\sun}$).
We already pointed out in Sect.~\ref{sec:synttest} that imposing the dynamical mass constraint improves indeed the quality of the fit, specially for the stellar masses. Here we have an other way to look at this, as the evidence is a quantitative measure of the aforementioned fit improvement.
The inclusion of the metallicity prior has a similar effect of increasing the evidence for the whole sample and for the sub-sample of EB stars (both restricted only to stars with [Fe/H] measurements). Nevertheless, it does not make a big difference for the AS and DK subgroups for which the best evidence is still reached in case 2) even when restricting only to stars for which [Fe/H] measurements are available. It is worth reminding that for most of the DK stars the [Fe/H] values we used are the average values for the Taurus-Auriga star-forming complex.

The above analysis suggests that there are still some problems with the current generation of standard PMS models. However the significance of the disagreement between theory and observations is different depending on the subset of objects considered -- 
for example, we remind that for the subset of EBs the average mismatch between observed and predicted masses is lower than 10\%.
Moreover, as it is clear from the tables of Appendix \ref{Sec:AppTab}, for many of the objects of the sample the single-star evidence may be the largest for other values of the $\bmath{\Xi}$ meta-parameters. Nevertheless this global test and the analysis of Sect.~\ref{sec:theovsdyn} both hint to the fact that a threefold effort is probably needed to a) improve the quality of the data specially assessing the systematic errors, b) better constrain the $\Delta Y/\Delta Z$, and $(Z/X)_{\sun}$ values and c) improve the physics of stellar models.

\section{Application to binaries}
\label{sec:binanalysis}
\setcounter{table}{4}
\begin{table*}
 \caption[]{Results from the comparison with the standard set of models. For the missing entries the confidence interval is poorly defined.}
\label{Tab:resbest}
\renewcommand\tabcolsep{5.pt}
\begin{tabular}{lccccccccc}
 \hline 
\hline
             &                  & & \multicolumn{3}{c}{Flat Mass Prior} & & \multicolumn{3}{c}{Gaussian Mass Prior} \\
\multirow{2}{*}{Name} &  $M_{\rmn{dyn}}$ & &  $M_{\rmn{mod}}$ & Age (stars) & Age (system) & & $M_{\rmn{mod}}$ & Age (stars) & Age (system) \\
             &  [$M_{\sun}$]    & &  [$M_{\sun}$]    & [Myr]  & [Myr] & & [$M_{\sun}$] & [Myr] &  [Myr] \\
\hline
& & & & & & & \\[-2mm]
RS Cha A     & $1.890\pm0.010$ & & $1.79_{-0.07}^{+0.07}$ & $8.40_{-0.60}^{+0.65}$  & \multirow{2}{*}{$8.50_{-0.45}^{+0.50}$}   & &$1.85_{-0.02}^{+0.02}$ & $8.00_{-0.30}^{+0.25}$     &     \multirow{2}{*}{$8.00_{-0.25}^{+0.15}$}        \\
& & & & & & & \\[-2mm]
RS Cha B     & $1.870\pm0.010$ & & $1.74_{-0.07}^{+0.06}$ & $8.70_{-0.75}^{+0.75}$  &    & &   $1.82_{-0.03}^{+0.01}$   &  $7.95_{-0.40}^{+0.30}$   &               \\ 
& & & & & & & \\[-1mm]
RXJ 0529.4 A & $1.270\pm0.010$ & & $1.25_{-0.09}^{+0.09}$ & $8.35_{-1.35}^{+3.45}$  & \multirow{2}{*}{$6.25_{-0.70}^{+1.20}$}   &          &   $1.27_{-0.02}^{+0.01}$  & $8.70_{-1.25}^{+1.20}$  &   \multirow{2}{*}{$6.90_{-0.85}^{+1.15}$}           \\  
& & & & & & & \\[-2mm]
RXJ 0529.4 B & $0.930\pm0.010$ & & $0.87_{-0.12}^{+0.10}$ & $5.25_{-0.70}^{+1.35}$  &    &  &  $0.93_{-0.02}^{+0.01}$   &  $5.20_{-0.70}^{+1.30}$   &               \\   
& & & & & & & \\[-1mm]
V1174 Ori A  & $1.009\pm0.015$ & & $1.04_{-0.08}^{+0.06}$ & $5.85_{-0.40}^{+0.50}$  & \multirow{2}{*}{$7.90_{-0.45}^{+0.45}$} &  &$1.01_{-0.02}^{+0.01}$ & $5.85_{-0.40}^{+0.50}$     &     \multirow{2}{*}{$7.40_{-0.35}^{+0.35}$}        \\
& & & & & & & \\[-2mm]
V1174 Ori B  & $0.731\pm0.008$ & & $0.42_{-0.07}^{+0.08}$ & $9.85_{-0.65}^{+0.45}$  &    &  &  $0.73_{-0.02}^{+0.00}$   &  $8.45_{-0.50}^{+0.55}$   &               \\
& & & & & & & \\[-1mm]
EK Cep A     & $2.020\pm0.010$ & & $1.87_{-0.06}^{+0.06}$ & $30.75_{-8.15}^{+47.60}$  & \multirow{2}{*}{$16.00_{-2.55}^{+2.65}$}   & &$2.02_{-0.02}^{+0.00}$ & $26.85_{-6.55}^{+43.90}$     &     \multirow{2}{*}{$18.95_{-2.05}^{+1.05}$}        \\ 
& & & & & & & \\[-2mm]
EK Cep B     & $1.124\pm0.012$ & & $1.17_{-0.03}^{+0.04}$ & $15.80_{-2.60}^{+2.65}$  &   &  &   $1.13_{-0.01}^{+0.01}$   &  $18.90_{-2.00}^{+1.05}$   &               \\
& & & & & & & \\[-1mm]
TY CrA A     & $3.160\pm0.020$ & & $2.61_{-0.18}^{+0.29}$ & --  & \multirow{2}{*}{$4.25_{-0.40}^{+2.75}$}  & &$3.16_{-0.05}^{+0.01}$ & --     &     \multirow{2}{*}{$3.75_{-0.20}^{+2.65}$}        \\ 
& & & & & & & \\[-2mm]
TY CrA B     & $1.640\pm0.010$ & & $1.52_{-0.35}^{+0.24}$ & $3.10_{-0.40}^{+2.55}$  &   &  &   $1.64_{-0.02}^{+0.01}$   &  $18.90_{-2.00}^{+1.05}$   &               \\ 
& & & & & & & \\[-1mm]
ASAS 052821 A& $1.387\pm0.017$ & & $1.54_{-0.09}^{+0.08}$ & $3.50_{-0.25}^{+0.50}$  & \multirow{2}{*}{$3.50_{-0.20}^{+0.15}$}  & & $1.39_{-0.02}^{+0.01}$ &  $3.25_{-0.20}^{+0.15}$    &     \multirow{2}{*}{$3.45_{-0.15}^{+0.10}$}        \\   
& & & & & & & \\[-2mm]
ASAS 052821 B& $1.331\pm0.011$ & & $1.13_{-0.10}^{+0.10}$ & $3.50_{-0.20}^{+0.15}$  &   &  &   $1.33_{-0.02}^{+0.01}$   &  $3.60_{-0.20}^{+0.10}$   &               \\
& & & & & & & \\[-1mm]
HD 113449 A  & $0.960\pm0.087$ & & $0.84_{-0.05}^{+0.04}$ & $47.60_{-2.40}^{+41.45}$  & \multirow{2}{*}{--}  & & $0.86_{-0.04}^{+0.04}$ &  $48.95_{-1.80}^{+40.90}$    &     \multirow{2}{*}{--}        \\   
& & & & & & & \\[-2mm]
HD 113449 B  & $0.557\pm0.050$ & & $0.44_{-0.06}^{+0.03}$ & --  &   &  &   $0.48_{-0.04}^{+0.02}$   &  --   &               \\ 
& & & & & & & \\[-1mm]
NTT 045251 A & $1.450\pm0.190$ & & $1.00_{-0.14}^{+0.12}$ & $3.60_{-0.65}^{+4.45}$  & \multirow{2}{*}{$2.55_{-0.35}^{+0.65}$}   & &$1.14_{-0.11}^{+0.12}$ & $4.15_{-0.90}^{+3.25}$     &     \multirow{2}{*}{$3.55_{-0.50}^{+0.85}$}        \\ 
& & & & & & & \\[-2mm]
NTT 045251 B & $0.810\pm0.090$ & & $0.41_{-0.10}^{+0.10}$ & $2.40_{-0.35}^{+0.65}$  &   &  &   $0.65_{-0.08}^{+0.07}$   &  $3.40_{-0.50}^{+1.00}$   &               \\ 
& & & & & & & \\[-1mm]
HD 98800 Ba  & $0.699\pm0.064$ & & $0.51_{-0.02}^{+0.25}$ & $0.85_{-0.20}^{+0.00}$  & \multirow{2}{*}{$0.85_{-0.15}^{+0.05}$}   & &$0.68_{-0.07}^{+0.06}$ & $0.85_{-0.10}^{+0.05}$     &     \multirow{2}{*}{$0.90_{-0.10}^{+0.00}$}        \\ 
& & & & & & & \\[-2mm]
HD 98800 Bb  & $0.582\pm0.051$ & & $0.41_{-0.00}^{+0.17}$ & $0.95_{-0.35}^{+0.00}$  &   &  &   $0.56_{-0.06}^{+0.05}$   &  $1.00_{-0.15}^{+0.00}$   &               \\ 
\hline
\end{tabular}
\renewcommand\tabcolsep{6pt}
\end{table*}

In this Section we will analyse the binary systems in our data set. 
For each of them we will check whether the models are able to reproduce the coevality of the two stars. In all the cases where [Fe/H] measurements are available, we will implicitly use the corresponding Gaussian prior on $Z$ in the marginalization of the probability distributions. In the four remaining cases a flat prior will be used in the available range of metallicities for our models set: $Z \in [0.007, 0.03]$.

We will mainly make use of the standard $\mathbf{\Xi}$-set previously introduced. In the case of severe disagreement between the standard models and the observations, we will explore the possibility that non-standard $\mathbf{\Xi}$-set might give a better agreement with the data.

\begin{figure*}
 \centering
  \resizebox{0.98\hsize}{!}{\includegraphics{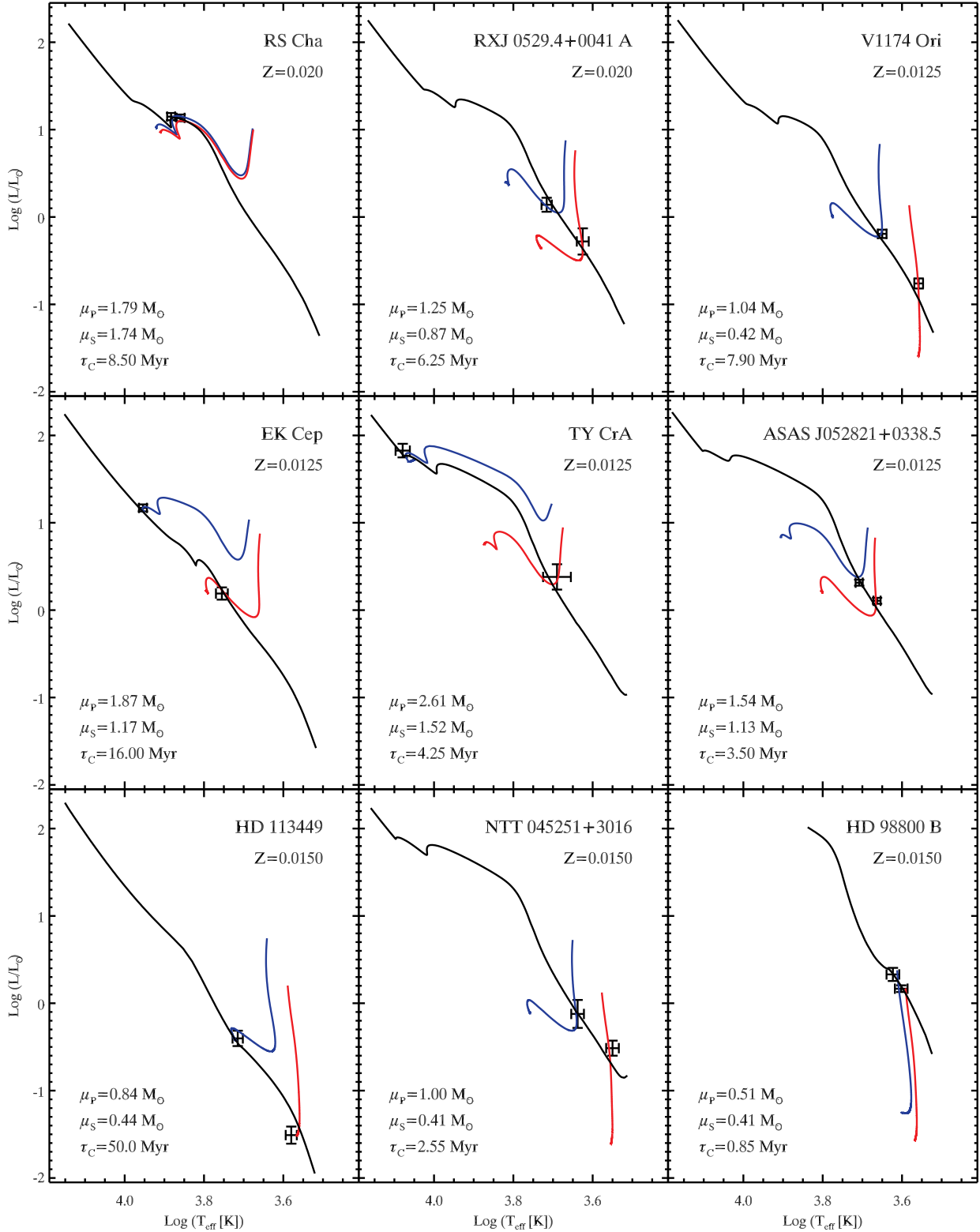}}
 \caption{HR diagrams with best fitting tracks and isochrones for the EB and AS binaries of our sample. Masses and ages are obtained using a flat mass prior and a Gaussian metallicity prior.}
\label{fig:bestfitHRDflat}
\end{figure*}
\begin{figure*}
 \centering
  \resizebox{0.98\hsize}{!}{\includegraphics{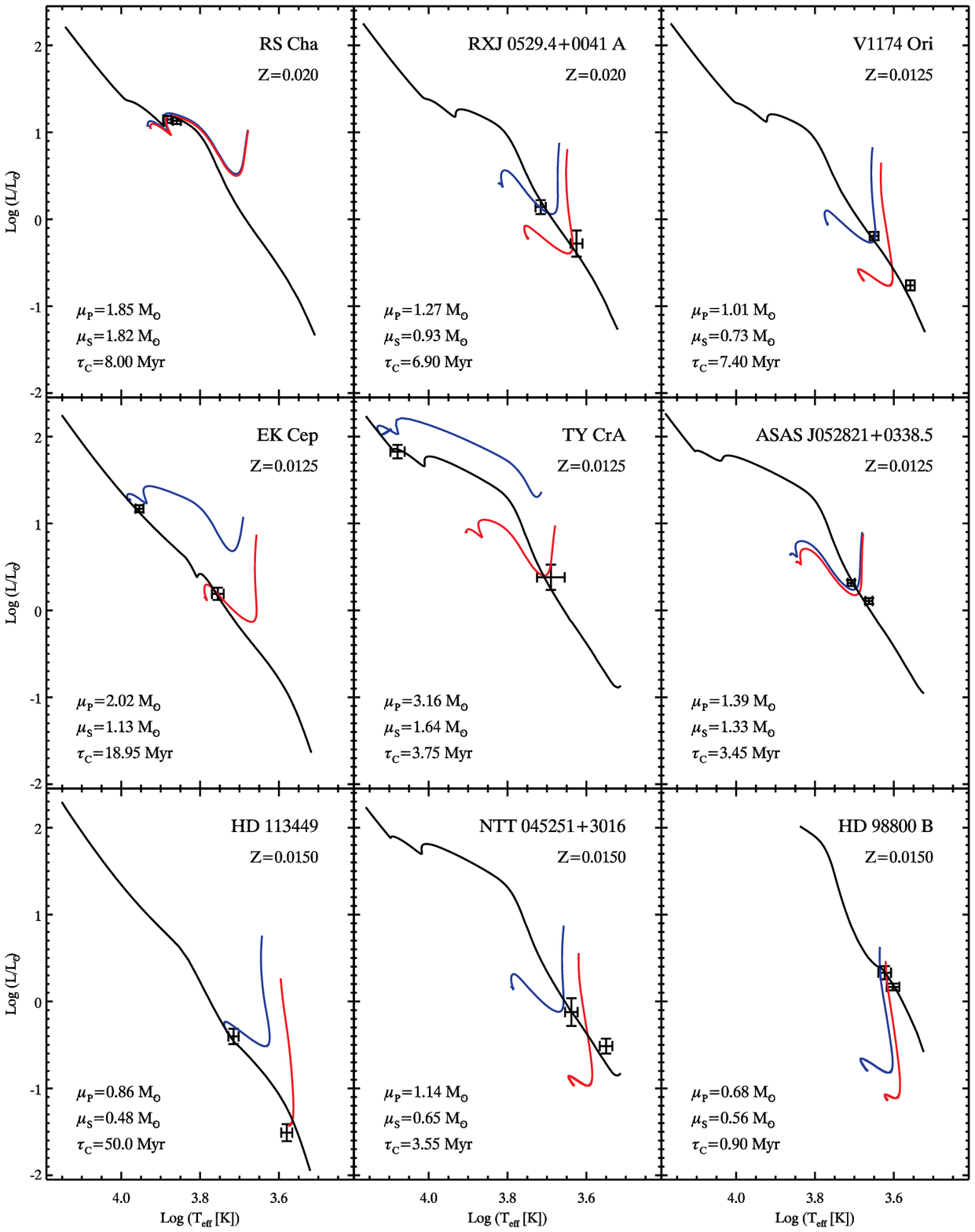}}
 \caption{HR diagrams with best fitting tracks and isochrones for the EB and AS binaries of our sample. Masses and ages are obtained using a Gaussian mass prior and a Gaussian metallicity prior.}
\label{fig:bestfitHRDgauss}
\end{figure*}

Figures \ref{fig:bestfitHRDflat} and \ref{fig:bestfitHRDgauss} show the HR diagrams with the data for each of the 6 EB and 3 AS systems with  the best fitting tracks and isochrones for the standard set of models superimposed. 
The results are also summarized in Table \ref{Tab:resbest}. Some entries are missing in the table, corresponding to the cases in which the confidence intervals are poorly defined. This happens when the posterior probability is a very flat function and its mode falls outside the confidence interval.
The best fit masses and ages are obtained by applying two different priors in mass, namely a flat and a Gaussian one in Figs. \ref{fig:bestfitHRDflat} and \ref{fig:bestfitHRDgauss}, respectively. 
In the case of EBs, for which stellar radii are measured, we have used the surface gravity vs. effective temperature diagram to compare the models with the data. This diagram has the advantage of combining the three measured quantities --mass, radius and temperature-- hence representing the most stringent test for the models. For the AS systems we used the HR diagram.
We display all the best fit models in the HR diagram for homogeneity.

The best values are obtained after marginalization in $Z$ using a Gaussian prior. For displaying purposes only, we used isochrones and tracks with a specific $Z$ value. The values for each system are obtained after transforming the observed [Fe/H] into $Z_{\rmn{obs}}$ using $\Delta Y/\Delta Z =2, Y_{\rmn{P}}=0.2485$ and $(Z/X)_{\sun}=0.0181$. We then took the closest $Z$ available in our models database.
In the case where [Fe/H] is not available we used $Z=0.0125$, the closest to our solar-calibrated $Z$ value.
The overplotted isochrones correspond to the best system composite age, i.e. that obtained by maximizing $G_{\rmn{C}}(\tau) = G_{\rmn{P}}(\tau) \times G_{\rmn{S}}(\tau)$.

Given the large number and size, the figures relative to the each system's subsection are presented in Appendix \ref{Sec:figbinan}. The upper panels show the marginalized mass distributions and the lower ones the age distributions. The left and right panels show the results obtained using a flat and a Gaussian prior on the mass distribution respectively.

\subsection{RS Cha}
\label{sec:RSCha}
This double-lined EB is located in the $\eta$~Cha cluster \citep{2000ApJ...544..356M}. The stellar masses and radii are from \cite{2005A&A...442..993A} who refined the values from the pioneering studies on binary stars by \cite{1975A&A....44..445A,1991A&ARv...3...91A}.
\cite{2005A&A...442..993A} also provide a spectroscopic measurement of [Fe/H]. Temperatures are taken from \cite{2000MNRAS.318L..55R}. Surface gravities are simply calculated as $|g| = GM/R^2$.

Previously thought to be a post-MS system \citep{1969MNSSA..28....5J,1975A&A....44..445A,1991A&ARv...3...91A}, the X-rays emission reported by \cite{1999ApJ...516L..77M} clearly points to the PMS nature of this system. 
The two RS Cha components have very similar masses of $\sim 1.9 M_{\sun}$ and are both close to approaching the Zero-Age MS. Recent literature estimates for the system age range from $6^{+2}_{-1}$ Myr \citep{2004ApJ...609..917L} to $9.13 \pm 0.12$ Myr \citep{2007A&A...473..181A}.

The outcomes of the comparison between our standard set of models and the RS Cha components' gravities and temperatures are shown in Fig.~\ref{fig:RSCha_GH}.  

As already noted in Sect.~\ref{sec:theovsdyn} the standard set of models slightly underestimates the mass values with $\mu_{\rmn{P}} = 1.79_{-0.07}^{+0.07} M_{\sun}$ and $\mu_{\rmn{S}} = 1.74_{-0.07}^{+0.06} M_{\sun}$ for the primary and secondary mass respectively. The dynamical masses fall outside these 68\% confidence intervals, nevertheless the discrepancy is quite small --less than 5\%-- which is a very remarkable agreement.
The relative precision of the mass estimates is strongly increased by the use of the Gaussian prior and also the mode of the mass marginal distributions for both components are more similar to the observed values with $\mu_{\rmn{P}} = 1.83_{-0.02}^{+0.02} M_{\sun}$ and $\mu_{\rmn{S}} = 1.82_{-0.03}^{+0.01} M_{\sun}$.

Even with the slight mass discrepancy, the results on the system's age are very robust. The age estimates of the two components remarkably agree among each other. From the single star's marginal age distribution we obtained the combined system age as $G(\tau)_{\rmn{RS\,Cha\,A}}\times G(\tau)_{\rmn{RS\,Cha\,B}}$. The estimated value for the system age in the case of a flat mass prior is $\tau_{\rmn{C}} = 8.50_{-0.45}^{+0.50}$ Myr which is narrowed down to $\tau_{\rmn{C}} = 8.00_{-0.25}^{+0.15}$ Myr when the Gaussian mass prior is imposed. It is worth noticing that the relative precision of the combined age is improved with respect to the single stellar ages estimates. 

\subsection{RXJ 0529.4+0041 A}
\label{sec:RXJ}

The discovery of this double-lined EB located in the Orion star-forming region was reported by \cite{2000A&A...361L..49C}. The same group refined the system parameters using new photometric observations in \cite{covino04}. We adopt the data from the latter paper and the Orion [Fe/H] from \cite{2009A&A...501..973D}.

Comparing the observed gravity and temperatures with our standard set of models yields stellar masses in agreement with the dynamical measurements with $\mu_{\rmn{P}} = 1.25_{-0.09}^{+0.09} M_{\sun}$ and $\mu_{\rmn{S}} = 0.87_{-0.12}^{+0.10} M_{\sun}$ when a flat mass prior is used. If a Gaussian prior is applied, then the precision improves by a factor of 10 for the estimated masses with $\mu_{\rmn{P}} = 1.27_{-0.02}^{+0.01} M_{\sun}$ and $\mu_{\rmn{S}} = 0.93_{-0.02}^{+0.01} M_{\sun}$.
On the other hand the derived single stellar ages are in slight disagreement. Using a Gaussian mass prior, we obtain respectively $\tau_{\rmn{P}} = 8.70_{-1.25}^{+1.20}$ Myr and $\tau_{\rmn{S}} = 5.20_{-0.70}^{+1.30}$ Myr. Nevertheless, the two ages are both consistent with the composite age of the system, i.e. $\tau_{\rmn{C}} = 6.90_{-0.85}^{+1.15}$ Myr. The results for the standard set of models are summarized in Fig.~\ref{fig:RXJ_GH}.

By looking at the Bayes factors of Table \ref{tab:bfactYesMetYesmass}, it appears that most of the age discrepancy might be ascribed to the poorer fit of the secondary star. For the primary star the standard set of models provides the second highest Bayes Factor value, 18.34, the highest being just 18.35 for the set with $\Delta Y/\Delta Z =2, \alpha=1.68$ but $Y_{\rmn{P}}=0.23$. For the secondary the standard set provides a Bayes Factor that is $\sim 1.5$ times smaller than the one giving the largest evidence, i.e. $\Delta Y/\Delta Z =2, \alpha=1.20$ but $Y_{\rmn{P}}=0.2485$. This is not enough to state that the latter set gives a significantly better agreement with the data, but using the latter set of meta-parameters yields an age of  $6.30_{-0.85}^{+1.65}$ Myr, which is in agreement with the primary star's age within the uncertainty interval.

The fact that the secondary star is better fitted by cooler models (i.e. models with lower $\alpha$) was already reported by \cite{2000ApJ...543L..77D} and confirmed in \cite{covino04}.

\subsection{V1174 Ori}
\label{sec:V1174Ori}
This double-lined EB was discovered by \cite{stassun04}. We adopt stellar parameters from this paper and the average [Fe/H] abundances for Orion \citep{2009A&A...501..973D}. 

As in the case of RXJ 0529.4+0041 A, the primary star of V1174 Ori is moving away from the Hayashi track, while the secondary is still fully convective.
Also in this case the models show some difficulty in reproducing the secondary observables. The standard set of models --with a flat mass prior-- predicts a secondary mass of $\sim0.42 M_{\sun}$, much smaller than the dynamical mass ($\sim0.7 M_{\sun}$). The primary mass is instead well recovered with $\mu_{\rmn{P}} = 1.04_{-0.08}^{+0.07} M_{\sun}$. The situation for the secondary does not improve much even when using the coldest set of models available. 

It has been noted \citep[see e.g.][]{hillenbrand04} that one of the problems in estimating the effective temperatures for PMS stars from the observed spectral type is the adoption of temperature scales that are calibrated on MS stars. For example \cite{stassun04} use the temperature scale for dwarf stars by Schmidt-Kaler in \cite{1982lbg6.conf.....A}. The same authors show how stellar models are not able to reproduce luminosities and temperatures for the secondary star and attribute the discrepancy to the non-adequacy of the dwarf spectral type to $T_{\rmn{eff}}$ conversion when applied to PMS stars.

At a given spectral type PMS stars are in general hotter than the corresponding MS stars. Their surface gravities indicate that a temperature scale intermediate between dwarfs and giants should be adopted. \cite{hillenbrand04} suggest that temperature corrections as high as 100 K could be necessary to compensate for the temperature underestimates. We explored this possibility in the case of V1174 Ori, by artificially increasing the temperature of the primary by 100 K and keeping the effective temperature ratio between primary and secondary constant. The reason why we can not simply increase the effective temperature of the secondary, which is the main responsible of the disagreement with the models, is that in \cite{stassun04} this quantity is not directly and independently measured, but it follows from the determination of the primary effective temperature from the spectral type and the temperature ratio from the light curve. Hence, if any offset is present, it should be reflected in both components.

The results of the comparison of the modified observables with the standard set of models are shown in Fig.~\ref{fig:V1174Ori_mod}. The situation is only slightly improved compared to Fig.~\ref{fig:V1174Ori}. The gain in the secondary mass estimate is that now the best mass is $\sim0.49M_{\sun}$, not yet enough to be in agreement with the dynamical mass. However the primary mass is still recovered within the uncertainty interval. It is clear that a change in $T_{\rmn{eff}}$ has a larger impact on the inferred stellar mass when a star is still in the vertical Hayashi track than when it is located on the almost horizontal Heyney track.

Even with the small improvement achievable by increasing the estimated $T_{\rmn{eff}}$, V1174 Ori remains a challenge for stellar evolution theory. However parallel observational efforts are required to assess the issues related to the effective temperatures determinations. 

\subsection{EK Cep}
This system is known to be an EB for more than 50 years \citep{1959QB837.S75......}. Quite some time after its discovery it was recognized to host a $\sim 1.1 M_{\sun}$ PMS star, together with a $\sim 2.0 M_{\sun}$ primary already on its main sequence \citep{1987ApJ...313L..81P}. We adopted stellar parameters from this paper and the spectroscopic determination of [Fe/H] by \cite{1993A&A...274..274M}.

With our standard set of models the stellar masses for EK Cep are not recovered within the 68\% confidence interval (see Fig.~\ref{fig:EKCep}). If we use a flat mass prior we obtain $\mu_{\rmn{P}} = 1.87_{-0.06}^{+0.06} M_{\sun}$ and $\mu_{\rmn{P}} = 1.17_{-0.03}^{+0.04} M_{\sun}$, while the measured dynamical values are $2.020\pm0.010M_{\sun}$ and $1.124\pm0.012 M_{\sun}$ for primary and secondary, respectively. Hence the primary mass is slightly underestimated and the secondary mass slightly overestimated. It has to be noted that the absolute difference between model-predicted and dynamical masses are of the order of 7\% and 4\% hence quite small. Nevertheless, the discrepancy are significant according to our definition of the confidence interval.

Given the Bayes Factors of Table \ref{tab:bfactYesMetYesmass}, we have compared the data with models from the class with $\alpha=1.20, Y_{\rmn{P}} = 0.23$, and $\Delta Y / \Delta Z =2$, i.e. the one that gives the largest composite evidence for the system. The agreement is still not satisfactory. As shown in Fig.~\ref{fig:EKCepNS}, a further improvement of the fit is achieved by assuming a higher metallicity (i.e. $Z=0.0193$ rather than $Z=0.0157$) as if the old $(Z/X)_{\sun} = 0.0231$ by \cite{1998SSRv...85..161G} was used instead of the recent one by \cite{2009ARA&A..47..481A}. The inferred masses are in this case $\mu_{\rmn{P}} = 1.97_{-0.06}^{+0.07} M_{\sun}$ and $\mu_{\rmn{P}} = 1.12_{-0.02}^{+0.14} M_{\sun}$. This results are in very good agreement with those of \cite{2006A&A...445.1061C}, who using similar values, namely $Z=0.0175$ and $\alpha=1.3$, is able to reproduce the system observables.

This test shows how the success of a set of models in reproducing the observations might be severely affected by the current uncertainties on the meta-parameters. Paradoxically, in the case of EK Cep system models calculated with out-of-date meta-parameters seem to give a better agreement with the data than the state-of-the-art ones.

One interesting thing that this system shows about our method is the power of the combined system age marginal distribution, $G(\tau)_{\rmn{EK\,Cep\,A}}\times G(\tau)_{\rmn{EK\,Cep\,B}}$. Since the primary star is already on the Main Sequence, its evolution is very slow, resulting in a very flat $G(\tau)$ and, consequently a very poor precision in the age determination. Nevertheless the age of the system is very well determined --given the choice of the model class.
As a consequence also the primary star has a very precise age determination, which is very valuable for MS stars.
In the case of the standard set of model and Gaussian mass prior the system age is $\tau = 18.95_{-2.05}^{+1.05}$ Myr, while for the non-standard set used in this Section, again with a Gaussian mass prior, we obtain $\tau = 26.55_{-1.80}^{+0.85}$Myr.

\subsection{TY CrA}
This double lined EB is part of a hierarchical system with three or possibly four stellar components \citep[see][]{2003A&A...406L..51C}. The fundamental parameters we adopted are taken from \cite{1998AJ....115.1617C}. For this particular system we could not find any spectroscopic determination of [Fe/H]. Therefore, instead of applying a Gaussian prior on the metallicity, the marginalization over $Z$ was made using a flat prior with $Z\in [0.007,0.03]$, i.e. the range of metallicities available in our models grid.

As in the case of EK Cep, the primary star is already on the Main Sequence, while the slower evolving secondary is still on its Hayashi track. Similarly to the EK Cep case, our standard set of models is able to reproduce the secondary mass quite well, while the primary mass is once again underestimated. The values we obtain when a flat mass prior is used are $\mu_{\rmn{P}} = 2.61_{-0.18}^{+0.29} M_{\sun}$ and $\mu_{\rmn{S}} = 1.52_{-0.35}^{+0.24} M_{\sun}$ while the dynamical masses are estimated to be $3.16 \pm 0.02 M_{\sun}$ and $1.64 \pm 0.01 M_{\sun}$ for the primary and secondary, respectively. The low relative precision of these model predictions --compared e.g to the case of EK Cep-- are mainly due to the larger uncertainties on the effective temperatures and radii for the TY CrA system (see Table \ref{Tab:data}).

Also in this case, we tried to see whether the agreement between data and models might be improved by using the coldest set of models with $\alpha=1.20, Y_{\rmn{P}} = 0.23$ and $\Delta Y / \Delta Z =2$. As it is possible to see by comparing Figs. \ref{fig:TYCrA} and \ref{fig:TYCrANS}, there is a slight improvement in the primary mass determination, without losing the good agreement for the secondary mass. The results obtained with a flat mass prior are $\mu_{\rmn{P}} = 2.69_{-0.18}^{+0.30} M_{\sun}$ and $\mu_{\rmn{S}} = 1.49_{-0.21}^{+0.24} M_{\sun}$. The primary mass is still underestimated by about 15\%, not a too bad result --in absolute terms-- but further investigation is needed to explain this partial disagreement.

TY CrA A is on the MS, therefore its age is not well constrained. Nevertheless the mode of the age distribution for the primary is very close to the mode of the secondary, which has a better constrained age determination. From the composite age distribution $G(\tau)_{\rmn{TY\,CrA\,A}}\times G(\tau)_{\rmn{TY\,CrA\,B}}$ in the case of the standard set and applying a Gaussian mass prior we obtain $\tau = 3.75_{-0.20}^{+2.65}$ Myr. For the coldest set of models and still applying a Gaussian mass, prior we obtain a slightly older age of $\tau = 5.20_{-0.70}^{+3.05}$ Myr.
These age values are slightly older than the age found by \cite{1998AJ....115.1617C} who roughly estimate a system age of $\sim 3$ Myr. Also these authors show that the models have problems in consistently predicting the stellar observable for both components. While being able to reproduce the observed secondary properties, they also find that models overestimate the effective temperature of the primary star. This is equivalent to our finding of an underestimated stellar mass for the given $T_{\rmn{eff}}$ and $\log g$.

\subsection{ASAS J052821+0338.5}

This is the most recently discovered double-lined EB in our data set. The stellar parameters are from \cite{2008A&A...481..747S}. The two stars have very similar masses, and the slightly more massive primary ($1.387 \pm 0.017 M_{\sun}$) is just moving away from its Hayashi track while the secondary ($1.331 \pm 0.011 M_{\sun}$) is located just before the end of the fully convective phase.
The [Fe/H] value we used is $-0.15 \pm 0.2$ dex, i.e. the average of the quoted values for the primary ($-0.2 \pm 0.2$ dex) and the secondary ($-0.1 \pm 0.2 $ dex) in  \cite{2008A&A...481..747S}.

The predictions of our standard set of models slightly differ from the measured masses. The values we obtain when a flat mass prior is used are $\mu_{\rmn{P}} = 1.54_{-0.11}^{+0.08} M_{\sun}$ and $\mu_{\rmn{S}} = 1.13_{-0.10}^{+0.10} M_{\sun}$. Therefore the primary mass is overestimated by $\sim 11$\% and the secondary is underestimated by $\sim 15$\%. \cite{2008A&A...481..747S} provide a double solution for the system parameters depending whether stellar spots are taken into account in the light-curve analysis (as in the case reported in Table \ref{Tab:data}) or not. We applied our method using also the measurements for the latter case and the results are shown in Fig.~\ref{fig:ASAS_nospot}. Both of the predicted masses are in better agreement with the data in this case with $\mu_{\rmn{P}} = 1.53_{-0.10}^{+0.08} M_{\sun}$ and $\mu_{\rmn{S}} = 1.24_{-0.09}^{+0.11} M_{\sun}$. Hence for the primary mass the situation is slightly better with an overestimate of 10\%, while the situation is much improved for the secondary which is now predicted to be 7\% less massive than the observed value.

It is clear that the detailed modeling of the light curve plays an important role in determining stellar properties and, as a consequence, in constraining the models' predictions.
Once again the models give an overall satisfactory agreement, being 10 or even 15\% still a quite good error in stellar masses predictions. Nevertheless more work is needed to explain these differences. 

The results for the age of the system are more robust with the two components having ages in good agreement. In the case where the light curve solution including star spots is used, we obtain a system age --using a Gaussian mass prior-- of $3.45_{-0.15}^{+0.10}$ Myr.
When using the light-curve solution without star spots, we instead obtain $3.65_{-0.20}^{+0.10}$ Myr.
In both cases the age is much younger than what is found by \cite{2008A&A...481..747S} who, using solar metallicity models by \cite{1998A&A...337..403B}, found an age of $\sim 10$ Myr for the system. We used the [Fe/H]$\simeq-0.15$ dex quoted in the same paper to derive our Gaussian Z prior. This value is slightly sub-solar, hence part of the difference between our age estimate and that by \cite{2008A&A...481..747S} could be ascribed to that. Still it is quite hard, even using our solar metallicity models, to reproduce a $\sim 1.4 M_{\sun}$ star close to the base of its Hayashi track at such an old age like 10 Myr. 

\subsection{HD 113449}

This system is an AS binary whose orbital parameters have been recently estimated by \cite{2010RMxAC..38...34C}. Here we use slightly different parameters (yet unpublished) kindly provided by the same group after more accurate analysis of the data and the [Fe/H] by \cite{2006PASP..118..706P}.

As already noted by \cite{2010RMxAC..38...34C}, there is a slight disagreement between dynamical and inferred masses by several sets of stellar models. Also the masses predicted by our standard set of stellar tracks are slightly underestimated. The primary mass is found to be $\mu_{\rmn{P}} = 0.84_{-0.05}^{+0.04} M_{\sun}$ while the secondary is $\mu_{\rmn{S}} = 0.44_{-0.06}^{+0.03} M_{\sun}$ (see also Fig.~\ref{fig:HD113449}). From Table \ref{Tab:resbest} we can see the primary dynamical mass of $0.960\pm0.087 M_{\sun}$ is still consistently recovered while for the secondary the dynamical mass of $0.557\pm0.050 M_{\sun}$ is outside the 68\% confidence interval. 

To see whether the discrepancy could be reduced, we used the coldest set of models and, in addition, we derived the $Z$ value using the spectroscopic [Fe/H] and the $(Z/X)_{\sun} = 0.0231$ by \cite{1998SSRv...85..161G}. The results, displayed in Fig.~\ref{fig:HD113449NS}, show a better agreement with the observations.  The improvement is not substantial, though, and the predicted stellar masses are in this case $\mu_{\rmn{P}} = 0.89_{-0.05}^{+0.04} M_{\sun}$ and $\mu_{\rmn{S}} = 0.45_{-0.06}^{+0.03} M_{\sun}$.

The stellar ages in this particular case are not very well determined. The two stars are indeed very close to their MS position, which makes age determination very hard. Nevertheless the primary shows a small peak in its $G(\tau)$ distribution at an age of $\tau \sim 50$ Myr (for both the standard and non-standard set of models). The secondary instead does not show any peak in the stellar age, with a very flat $G(\tau)$ slightly increasing towards the edge of our models age-interval (100 Myr). The system age is poorly defined as well.

\subsection{NTT 045251+3016}
\label{subsec:NTT}

The discovery of this AS binary was first reported by \cite{steffen01} from which we adopted the stellar parameters. In this case no spectroscopic [Fe/H] is available. 
This system is quite young and both the primary and the secondary are found in their fully convective phase along the Hayashi track.
As pointed by \cite{steffen01}, all the stellar models adopted by them predict too low masses for both components. The set of models that gives the best agreement with observations is the one by \cite{1998A&A...337..403B} when a mixing length parameter $\alpha=1.0$ is adopted. This is not surprising given that models with a lower $\alpha$, being intrinsically colder, predict larger masses for given observed luminosities and temperatures.

With our standard set of models and using a flat mass prior, the two masses are found to be $\mu_{\rmn{P}} = 1.00_{-0.14}^{+0.12} M_{\sun}$ and $\mu_{\rmn{S}} = 0.41_{-0.10}^{+0.10} M_{\sun}$, severely lower than the dynamical masses by $\sim 30$\% and $\sim50$ \% respectively (see Fig.~\ref{fig:NTT}).
From Tables \ref{tab:bfactNoMetNomass} and \ref{tab:bfactNoMetYesmass} it is possible to see that for both of the components of NTT 045251+3016 the set of models that provides the largest Bayes Factor is, once again, the coldest set available $(Y_{\rmn{P}} = 0.23,\, \Delta Y/\Delta Z = 2,\, \alpha=1.2)$.
When  using this particular set and a flat mass prior, we obtain slightly larger masses of  $\mu_{\rmn{P}} = 1.13_{-0.13}^{+0.16} M_{\sun}$ and $\mu_{\rmn{S}} = 0.50_{-0.12}^{+0.13} M_{\sun}$ (see Fig.~\ref{fig:NTTNS}). The improvement is not enough to obtain an agreement between predicted and observed mass for the secondary, while the primary mass, though still underestimated, is in agreement within the errors.

We report that using the BASE software (courtesy of Tim Schulze-Hartung, in prep) for analysing the system's astrometric measurements and radial velocities, slightly lower masses are predicted. The primary mass is found to be $\mu_{\rmn{P}} = 1.383 \pm 0.220 M_{\sun}$ (-4.60\%) while the secondary is $\mu_{\rmn{S}} = 0.766 \pm 0.089$   (-5.41\%).

Even with this latter improvements, there is still a larger disagreement in the predicted vs. dynamical mass for NTT 045251+3016 than what we found for the EBs cases or even the other two AS binaries. This suggests that part of the problem might reside in observations as well, and we already noted in Sect.~\ref{sec:data} the peculiar location of this star in the HR diagram given its measured mass. Apart from a theoretical effort, which is certainly needed, this system demands attention also from the observational side to exclude, e.g., higher order multiplicity that to-date interferometric observations are not capable to resolve. 

Concerning the system composite age, the values we obtain when using a Gaussian mass prior are $\tau = 3.55_{-0.50}^{+0.85}$ Myr and $4.65_{-0.65}^{+1.1}$ Myr for the case of standard and coldest set of models, respectively.

\subsection{HD 98800 B}
This AS binary is part of a quadruple system. \cite{2005ApJ...635..442B} reported preliminary visual and physical orbit for the binary subsystem. They derived the components' masses of $0.699\pm0.064 M_{\sun}$ and $0.582\pm0.051 M_{\sun}$ for the primary and the secondary, respectively. We adopted the [Fe/H] value from \cite{2009ApJ...698..660L}.
Both the components of the system are very young and located at the beginning of their Hayashi track.

For this system the standard set of models provides a good fit to both components (see Fig.~\ref{fig:HD98800}). 
The predicted mass values in the case of a flat mass prior are $\mu_{\rmn{P}} = 0.51_{-0.02}^{+0.25} M_{\sun}$ and $\mu_{\rmn{S}} = 0.41_{-0.00}^{+0.17} M_{\sun}$. The location of the two stars in the HR diagram makes their marginal distribution extremely asymmetric. As a consequence, the best values are located quite close to (or exactly at) the boundary of the confidence intervals.
This is the reason why the quoted lower error for the secondary mass is zero.
The best values for both the primary and secondary are slightly smaller than the dynamical mass values but in this case there is consistency within the 68\% confidence intervals.

The inferred ages for the two components are very similar and when a Gaussian mass prior is adopted we obtain $\tau_{\rmn{P}} = 0.85_{-0.10}^{+0.05}$ Myr and $\tau_{\rmn{S}} = 1.00_{-0.15}^{+0.00}$ Myr. Also in this case the marginal distributions are quite asymmetric. The composite system age is found to be $\tau_{\rmn{C}} = 0.90 _{-0.10}^{+0.00}$ Myr, very young indeed.

\section{The stars in Taurus-Auriga}
\label{sec:TauAu}
In this section we will present the results for the 9 stars found in the Taurus-Auriga star-forming region and whose masses are derived using disk kinematics (see Sect.~\ref{sec:data}).
For all of these stars, the Gaussian metallicity prior is applied using the average value of [Fe/H] = $-0.01 \pm 0.05$ dex for the region \citep{2011A&A...526A.103D}.

This sample exactly coincides with the DK+DKS sample of Sect.~\ref{sec:data}. As it is possible to see in the rightmost panels of Fig.~\ref{fig:6pan}, most of the DK+DKS stars have strongly underestimated values of the mass. We discussed some possible reasons for this discrepancy in Sect.~\ref{sec:theovsdyn}

We used the standard set of model to derive the ages of the Taurus-Auriga stars. The predicted values when a Gaussian mass prior is used are reported in Table \ref{Tab:TA}. 
In addition to single stellar ages we also computed the composite age for the DKS system UZ Tau E.
In general these age determinations have a worst precision when compared to the EB and AS sample. Moreover for the star MWC 480 the peak of the distribution is outside the 68\% confidence interval. This is because the $G(\tau)$ is very flat and the corresponding peak is just barely visible. This peak is located in the area corresponding to the leftmost 16\% probability that is excluded according to our definition. This is not strange given our definition of the confidence interval, it is just an indication that the age of this system is very poorly defined.

As we mentioned in Sect.~\ref{sec:data} the stars UZ Tau Ea, BP Tau and MCW 480 have a peculiar location in the HR diagram. 
We excluded the latter two stars, while we kept UZ Tau Ea in the sample to obtain an average age of $2.1 \pm 1.3$ Myr where the quoted uncertainty is the standard deviation of the ages of the remaining stars.
For both the stars in the UZ Tau E system we considered the composite age as the best age estimator.
This average age is in very good agreement with the estimated age for the Taurus-Auriga star forming region of 1-2 Myr \cite[see e.g][and references therein]{2009ApJ...704..531K}.

\setcounter{table}{5}
\begin{table}
\begin{minipage}{70mm}
 \caption{Derived ages for Taurus-Auriga DK stars.}
\label{Tab:TA}
 \centering
 \begin{center}
\renewcommand\tabcolsep{4.5pt}
\begin{tabular}{lcc}
 \hline 
 \hline
      Name       &     Age[Myr]       & Rel. Prec. \\
\hline
     UZ Tau Ea   &     $2.65_{-0.45}^{+1.50}$    & 0.373 \\
& & \\[-2mm]
     UZ Tau Eb   &     $1.45_{-0.70}^{+0.45}$    & 0.592 \\
& & \\[-2mm]
     UZ Tau E$^a$\footnotetext{$^a$ Value for the composite system age} &     $1.85_{-0.25}^{+0.45}$    & 0.199 \\
& & \\[-2mm] 
     DL Tau      &     $1.20_{-0.15}^{+0.30}$    & 0.195 \\
& & \\[-2mm]
     DM Tau      &     $3.05_{-0.50}^{+1.10}$    & 0.276 \\
& & \\[-2mm]
     CY Tau      &     $1.85_{-0.65}^{+0.30}$    & 0.338 \\
& & \\[-2mm]
     BP Tau      &    $17.25_{-3.35}^{+8.85}$    & 0.370 \\
& & \\[-2mm]
     GM Aur      &     $0.50_{-0.00}^{+0.10}$    & 0.095 \\
& & \\[-2mm]
     MWC 480     & $10.30^b$\footnotetext{$^b$ Uncertainty interval poorly defined}& 0.957 \\
& & \\[-2mm]
     LkCa 15     &     $4.45_{-0.75}^{+2.50}$    & 0.371 \\
\hline
\end{tabular}
\end{center}
\end{minipage}
\end{table}

\section{Summary and Conclusions}
\label{sec:concl}
The importance of a stringent test of PMS models against stars with accurately known parameters (i.e. mass, luminosity, radius, effective temperature, [Fe/H]) can hardly be overestimated, as these models represent the main tool to derive masses and ages of stars observed in star forming regions and young stellar clusters. Consequently the inferred star formation histories and mass functions of young stellar groups strongly depend on the answers provided by stellar evolutionary codes.  

In order to constrain PMS models we relied on a data set containing 25 PMS stars of measured mass (plus 2 MS companions in binary systems). This is the full up-to-date sample of known PMS stars with dynamical mass measurements in the range 0.2-3.0 $M_{\sun}$. Among them 10 PMS objects belong to double-lined eclipsing binary systems and 6 to astrometric and spectroscopic binaries; the remaining 9 objects are stars whose masses are derived using the measured orbital velocity of their circumstellar disks.

The main novelties of the paper are both the approach followed for comparing theory with observations and the set of PMS models used in the comparison. 
Regarding the former, we applied for the first time to the whole sample of PMS stars a very general Bayesian method. This approach allows a full exploitation of the available information about the observed objects which is included in the form of prior probability distributions. In addition it provides robust uncertainties for the inferred quantities. 

The models are extracted from the very recent Pisa PMS database \citep{tognelli11}. They include the state-of-art input physics and are available for a large and very fine grid of metallicities, masses and ages and for different primordial helium abundances, $Y_{\rmn{P}}$, helium-to-metals enrichment ratios, $\Delta Y/\Delta Z$, and mixing-length parameter, $ \alpha$, values. 
  
We checked the robustness and accuracy of the method in recovering stellar ages and masses against simulated binary data sets. One interesting result is that even synthetic binary stars  --coeval by construction-- might mimic non coevality as a consequence of the random uncertainty in the effective temperature, radius, and luminosity. 
The actual fraction of fake non coeval recoveries strongly depends on the sample characteristics, since we demonstrated that the ability of recovering the simulated masses and ages is a complex function of the actual position of the star in the HR diagram.
This fraction can be as large as 95\% for systems with one component close to its main sequence position. 

This suggests that the inability to fit both components of a binary system with a single isochrone does not necessarily imply that the two stars are not coeval or that the models present some deficiency.   
We also showed that, even in the simulated systems that are recovered as non coeval, the inferred system age obtained using the composite age distribution is in very good agreement with the simulated age.

When the real data are used, the Pisa PMS models show an overall agreement with the observations. 
With the exception of  V1174 Ori B, the masses of EB stars are well recovered within 10\%.
The agreement progressively worsens for AS binaries and DK stars, but also the observational uncertainties become more severe for the latter objects. 

A slightly worse situation is observed for stellar ages compared to stellar masses. Within our sample 6 binary systems are present for which ages can be derived for both components.  The systems EK Cep and TY CrA  both have a primary on the MS, hence their primaries' ages are not very accurate and we exclude them from the following discussion. HD 113449 is excluded as well, since the age of the secondary is also not well determined.    
Of the remaining 6 systems, 4 are predicted to be coeval --within the uncertainties-- by our models. The systems RXJ 0529.4+0041 A and V1174 Ori are instead not consistent with coevality.
 
We have shown by simulations that the probability of deriving non coeval ages for two stars in a binary system --because of observational uncertainties-- is a complex function of the position of stars in the HR diagram. Nevertheless we have also shown that, for the subset of simulated system for which both components are still far from their main sequence position, this probability is very low, always below 10\% and as small as 0\% for most simulated cases.
The observed non coevality fraction of $~30 \%$ is higher than what expected, given the typical quoted observational errors and the location in the HR diagram of the stars of the 6 aforementioned systems. With the present low number statistics, it is hard to draw robust conclusions, nevertheless these results might suggest that PMS models still have some problems in correctly predicting stellar ages. On the other hand the hypothesis that some systems might really be non coeval can not be ruled out given the current uncertainties on the models. An other possibility is that observational errors, specially the systematic ones are underestimated, leading to a spurious non coevality of the stars in RXJ 0529.4+0041 A and V1174 Ori.

With our Bayesian approach it is possible to evaluate the probability for different sets of models, i.e. the models' evidence. 
We analysed the entire data set using several classes of models computed with different $Y_{\rmn{P}}$, $\Delta Y/\Delta Z$, and  $\alpha$ values. We calculated the evidence for each star using 9 different meta-parameters configurations. Furthermore, four combinations of the prior distributions for mass and metallicity have been used for each meta-parameter choice, for a total of 36 classes. 
We found that adopting a Gaussian rather than a flat mass prior significantly improves the composite evidence for the full data set; the same effect, but to a lesser degree, is obtained imposing a Gaussian metallicity prior, mainly for EBs.
 
Although our standard set of models shows a reasonable general agreement with the data, predicting mass values almost always 
within 20\% of the dynamical ones --and in several case even within 5\%--, the general trend suggests that standard models 
tend to underestimate the stellar mass, confirming previous results \citep[see][and references therein]{mathieu07}. As a consequence, the largest composite evidence is obtained with our coldest set of models, i.e. with the mixing-length parameter $ \alpha$=1.2 and the lowest helium abundance at fixed metallicity.  

Given that the discrepancy between theory and observations increases going from the most precise data set of EBs to the others,
we point out that a twofold effort is needed to achieve a better agreement. From the theoretical point of view a better understanding on the treatment of superadiabatic convection and a better characterization of the models' meta-parameters is desirable.
Moreover, one of the open questions of PMS modeling concerns the role played by accretion in affecting the observable properties of very young stars, i.e. their effective temperatures and luminosities.
Different authors disagree on the impact accretion can have on the stellar ages which are inferred using non-accreting models \citep{2009ApJ...702L..27B,2011ApJ...738..140H}. Unfortunately, a fully consistent treatment of accretion for stars spanning a wide range of parameters is still lacking, therefore preventing a proper and quantitative evaluation of the impact of accretion on the inferred stellar properties.

From the observational side the significance of such a comparison could be improved in the future by a larger sample of well studied and characterized PMS stars and by a better control on the systematic errors affecting AS and DK stars' measurements.

\section*{acknowledgements}
      We would like to thank Felice Cusano (INAF-Osservatorio Astronomico di Capodimonte) for providing refined values for the parameters of the HD 113449 system.
      We would also like to thank Tim Schulze-Hartung for applying his BASE software to the NTT 045251+3016 data set and for providing a new orbital solution.
      Special thanks to Emiliano Gregori (Universit\`{a} di Pisa) for useful discussion and comments and for inspiring part of the analysis in this paper.
      MG would like to thank Wolfgang Brandner for constant advice and for supporting and encouraging this work.
      MG would like to thank Coryn Bailer-Jones (MPIA) for very useful discussion and advice.
      MG would also like to thank the INFN-Pisa for supporting a collaborative visit to the astrophysics group in Pisa, where part of this work was realized.
      ET and PGPM would like to thank MPIA Heidelberg, and specially Wolfgang Brandner, for hosting them and supporting a visit.
      ET and PGPM acknowledge financial support from PRIN-INAF 2008 (P.I. M. Marconi).

\bibliography{biblioBinPMS}{}
\bibliographystyle{mn2e}

\clearpage
\appendix
\section{Tables of Bayes Factors}
\label{Sec:AppTab}

\setcounter{table}{0}
\begin{table*}
\begin{minipage}{160mm}
 \caption{Bayes factor values for the case of flat prior on the mass and flat prior on the metallicity}
 \label{tab:bfactNoMetNomass}
\renewcommand\tabcolsep{4.pt}
  \begin{tabular}{@{}l|lccc|lccc@{}}
\hline
\hline
& & $\alpha=1.20$ & $\alpha=1.68$ & $\alpha=1.90$ & & $\alpha=1.20$ & $\alpha=1.68$ & $\alpha=1.90$ \\ 
\hline
 & RS Cha A & & & & RS Cha B & & &  \\
& & & & & & & & \\[-4.5mm]
$\Delta Y/\Delta Z=2,\, Y_\rmn{P}=0.23$ & & 1.252& 1.252& 1.253& & 1.054&1.077& 1.111\\
$\Delta Y/\Delta Z=2,\, Y_\rmn{P}=0.2485 $& &  1.004& 1.000& 1.009& & 0.975&1.000& 1.033\\
$\Delta Y/\Delta Z=5,\, Y_\rmn{P}=0.2485 $& &  0.556& 0.561& 0.559& & 0.787&0.802& 0.828\\
\hline
 & RXJ Aa & & & & RXJ Ab & & &  \\
& & & & & & & & \\[-4.5mm]
$\Delta Y/\Delta Z=2,\, Y_\rmn{P}=0.23$ & & 1.002& 1.042& 1.035& & 1.377&1.101& 1.003\\
$\Delta Y/\Delta Z=2,\, Y_\rmn{P}=0.2485 $& &  0.955& 1.000& 0.996& & 1.292&1.000& 0.900\\
$\Delta Y/\Delta Z=5,\, Y_\rmn{P}=0.2485 $& &  0.833& 0.889& 0.889& & 1.038&0.716& 0.620\\
\hline
 & V1174 Ori A & & & & V1174 Ori B & & &  \\
& & & & & & & & \\[-4.5mm]
$\Delta Y/\Delta Z=2,\, Y_\rmn{P}=0.23$ & & 1.108& 1.051& 1.009& & 1.589&1.183& 1.086\\
$\Delta Y/\Delta Z=2,\, Y_\rmn{P}=0.2485 $& &  1.073& 1.000& 0.948& & 1.365&1.000& 0.914\\
$\Delta Y/\Delta Z=5,\, Y_\rmn{P}=0.2485 $& &  0.960& 0.824& 0.741& & 0.821&0.579& 0.525\\
\hline
 & EK Cep A & & & & EK Cep B & & &  \\
& & & & & & & & \\[-4.5mm]
$\Delta Y/\Delta Z=2,\, Y_\rmn{P}=0.23$ & & 0.950& 0.955& 0.956& & 5.684&1.181& 0.946\\
$\Delta Y/\Delta Z=2,\, Y_\rmn{P}=0.2485 $& &  0.995& 1.000& 1.005& & 4.447&1.000& 0.866\\
$\Delta Y/\Delta Z=5,\, Y_\rmn{P}=0.2485 $& &  1.451& 1.457& 1.460& & 1.386&0.712& 0.712\\
\hline
 & TY CrA A & & & & TY CrA B & & &  \\
& & & & & & & & \\[-4.5mm]
$\Delta Y/\Delta Z=2,\, Y_\rmn{P}=0.23$ & & 0.962& 0.963& 0.964& & 1.166&1.066& 0.977\\
$\Delta Y/\Delta Z=2,\, Y_\rmn{P}=0.2485 $& &  0.999& 1.000& 1.001& & 1.111&1.000& 0.909\\
$\Delta Y/\Delta Z=5,\, Y_\rmn{P}=0.2485 $& &  1.145& 1.147& 1.147& & 0.957&0.815& 0.720\\
\hline
 & ASAS A & & & & ASAS B & & &  \\
& & & & & & & & \\[-4.5mm]
$\Delta Y/\Delta Z=2,\, Y_\rmn{P}=0.23$ & & 0.922& 1.036& 1.052& & 1.251&1.083& 0.943\\
$\Delta Y/\Delta Z=2,\, Y_\rmn{P}=0.2485 $& &  0.884& 1.000& 1.017& & 1.208&1.000& 0.848\\
$\Delta Y/\Delta Z=5,\, Y_\rmn{P}=0.2485 $& &  0.784& 0.906& 0.913& & 1.071&0.745& 0.580\\
\hline
 & HD 113449 A & & & & HD 113449 B & & &  \\
& & & & & & & & \\[-4.5mm]
$\Delta Y/\Delta Z=2,\, Y_\rmn{P}=0.23$ & & 0.569& 0.971& 1.019& & 0.665&0.832& 0.883\\
$\Delta Y/\Delta Z=2,\, Y_\rmn{P}=0.2485 $& &  0.667& 1.000& 0.998& & 0.806&1.000& 1.060\\
$\Delta Y/\Delta Z=5,\, Y_\rmn{P}=0.2485 $& &  1.003& 0.837& 0.670& & 1.272&1.589& 1.686\\
\hline
 & NTT A & & & & NTT B & & &  \\
& & & & & & & & \\[-4.5mm]
$\Delta Y/\Delta Z=2,\, Y_\rmn{P}=0.23$ & & 1.422& 1.098& 0.976& & 1.776&1.173& 1.043\\
$\Delta Y/\Delta Z=2,\, Y_\rmn{P}=0.2485 $& &  1.340& 1.000& 0.877& & 1.529&1.000& 0.887\\
$\Delta Y/\Delta Z=5,\, Y_\rmn{P}=0.2485 $& &  1.101& 0.723& 0.607& & 0.946&0.612& 0.541\\
\hline
 & HD 98800 Ba & & & & HD 98800 Bb & & &  \\
& & & & & & & & \\[-4.5mm]
$\Delta Y/\Delta Z=2,\, Y_\rmn{P}=0.23$ & & 3.162& 1.120& 0.841& & 2.719&1.148& 0.903\\
$\Delta Y/\Delta Z=2,\, Y_\rmn{P}=0.2485 $& &  2.816& 1.000& 0.756& & 2.322&1.000& 0.811\\
$\Delta Y/\Delta Z=5,\, Y_\rmn{P}=0.2485 $& &  1.937& 0.754& 0.600& & 1.563&0.911& 0.632\\
\hline
 & UZ Tau Ea & & & & UZ Tau Eb & & &  \\
& & & & & & & & \\[-4.5mm]
$\Delta Y/\Delta Z=2,\, Y_\rmn{P}=0.23$ & & 1.731& 1.126& 0.968& & 2.795&1.363& 0.862\\
$\Delta Y/\Delta Z=2,\, Y_\rmn{P}=0.2485 $& &  1.540& 1.000& 0.846& & 2.228&1.000& 0.614\\
$\Delta Y/\Delta Z=5,\, Y_\rmn{P}=0.2485 $& &  1.202& 0.795& 0.623& & 0.897&0.229& 0.123\\
\hline
 & DL Tau & & & & DM Tau & & &  \\
& & & & & & & & \\[-4.5mm]
$\Delta Y/\Delta Z=2,\, Y_\rmn{P}=0.23$ & & 2.465& 1.139& 0.924& & 1.784&1.182& 1.051\\
$\Delta Y/\Delta Z=2,\, Y_\rmn{P}=0.2485 $& &  2.147& 1.000& 0.814& & 1.529&1.000& 0.886\\
$\Delta Y/\Delta Z=5,\, Y_\rmn{P}=0.2485 $& &  1.400& 0.735& 0.616& & 0.929&0.589& 0.518\\
\hline
 & CY Tau & & & & BP Tau & & &  \\
& & & & & & & & \\[-4.5mm]
$\Delta Y/\Delta Z=2,\, Y_\rmn{P}=0.23$ & & 1.551& 1.131& 1.025& & 1.103&1.058& 1.045\\
$\Delta Y/\Delta Z=2,\, Y_\rmn{P}=0.2485 $& &  1.371& 1.000& 0.902& & 1.042&1.000& 0.987\\
$\Delta Y/\Delta Z=5,\, Y_\rmn{P}=0.2485 $& &  0.996& 0.735& 0.646& & 0.890&0.848& 0.832\\
\hline
 & GM Aur & & & & MWC 480 & & &  \\
& & & & & & & & \\[-4.5mm]
$\Delta Y/\Delta Z=2,\, Y_\rmn{P}=0.23$ & & 7.963& 1.329& 0.287& & 1.118&1.128& 1.131\\
$\Delta Y/\Delta Z=2,\, Y_\rmn{P}=0.2485 $& &  6.900& 1.000& 0.173& & 0.991&1.000& 1.002\\
$\Delta Y/\Delta Z=5,\, Y_\rmn{P}=0.2485 $& &  5.211& 0.244& 0.023& & 0.564&0.570& 0.571\\
\hline
 & LkCa 15 & & & & & & & \\
& & & & & & & & \\[-4.5mm]
$\Delta Y/\Delta Z=2,\, Y_\rmn{P}=0.23$ & &  1.259 &  1.080 & 0.992 & & & & \\
$\Delta Y/\Delta Z=2,\, Y_\rmn{P}=0.2485$ & &  1.202 &  1.000 & 0.904 & & & & \\
$\Delta Y/\Delta Z=5,\, Y_\rmn{P}=0.2485$ & &  1.029 &  0.754 & 0.646 & & & & \\
\hline
 \end{tabular}
\end{minipage}
\end{table*}

\setcounter{table}{1}
\begin{table*}
\begin{minipage}{160mm}
  \caption{Bayes factor values for the case with Gaussian prior on the mass and flat prior on the metallicity}
 \label{tab:bfactNoMetYesmass}
\renewcommand\tabcolsep{4.pt}
  \begin{tabular}{@{}l|lccc|lccc@{}}
\hline
\hline
& & $\alpha=1.20$ & $\alpha=1.68$ & $\alpha=1.90$ & & $\alpha=1.20$ & $\alpha=1.68$ & $\alpha=1.90$ \\ 
\hline
 & RS Cha A & & & & RS Cha B & & &  \\
& & & & & & & & \\[-4.5mm]
$\Delta Y/\Delta Z=2,\, Y_\rmn{P}=0.23$ & & 6.968& 7.171& 7.039& & 17.54&17.92& 18.36\\
$\Delta Y/\Delta Z=2,\, Y_\rmn{P}=0.2485 $& &  7.123& 7.150& 7.212& & 12.47&12.64& 12.88\\
$\Delta Y/\Delta Z=5,\, Y_\rmn{P}=0.2485 $& &  0.147& 0.142& 0.143& & 0.003&0.003& 0.003\\
\hline
 & RXJ A & & & & RXJ B & & &  \\
& & & & & & & & \\[-4.5mm]
$\Delta Y/\Delta Z=2,\, Y_\rmn{P}=0.23$ & & 11.72& 16.17& 13.57& & 11.71&13.11& 12.30\\
$\Delta Y/\Delta Z=2,\, Y_\rmn{P}=0.2485 $& &  8.582& 16.72& 14.98& & 12.81&11.93& 10.14\\
$\Delta Y/\Delta Z=5,\, Y_\rmn{P}=0.2485 $& &  2.719& 14.44& 15.56& & 12.08&3.470& 1.650\\
\hline
 & V1174 Ori A & & & & V1174 Ori B  & & &  \\
& & & & & & & & \\[-4.5mm]
$\Delta Y/\Delta Z=2,\, Y_\rmn{P}=0.23$ & & 2.844& 7.566& 9.745& & 6.218&1.236& 0.658\\
$\Delta Y/\Delta Z=2,\, Y_\rmn{P}=0.2485 $& &  4.949& 10.52& 12.47& & 3.032&0.358& 0.158\\
$\Delta Y/\Delta Z=5,\, Y_\rmn{P}=0.2485 $& &  16.15& 12.65& 6.966& & 0.024&$< 10^{-3}$& $< 10^{-3}$\\
\hline
 & EK Cep A & & & & EK Cep B & & &  \\
& & & & & & & & \\[-4.5mm]
$\Delta Y/\Delta Z=2,\, Y_\rmn{P}=0.23$ & & 9.530& 9.597& 9.580& & 18.02&17.25& 9.205\\
$\Delta Y/\Delta Z=2,\, Y_\rmn{P}=0.2485 $& &  2.259& 2.254& 2.256& & 17.48&23.61& 14.84\\
$\Delta Y/\Delta Z=5,\, Y_\rmn{P}=0.2485 $& &  0.001& 0.001& 0.001& & 28.67&14.20& 21.99\\
\hline
 & TY CrA A & & & & TY CrA B & & &  \\
& & & & & & & & \\[-4.5mm]
$\Delta Y/\Delta Z=2,\, Y_\rmn{P}=0.23$ & & 0.409& 0.409& 0.409& & 13.52&6.466& 4.870\\
$\Delta Y/\Delta Z=2,\, Y_\rmn{P}=0.2485 $& &  0.145& 0.145& 0.145& & 13.13&6.728& 4.847\\
$\Delta Y/\Delta Z=5,\, Y_\rmn{P}=0.2485 $& &  0.005& 0.005& 0.005& & 5.550&6.863& 4.718\\
\hline
 & ASAS A & & & & ASAS B & & &  \\
& & & & & & & & \\[-4.5mm]
$\Delta Y/\Delta Z=2,\, Y_\rmn{P}=0.23$ & & 3.446& 0.909& 2.070& & 8.715&12.54& 7.418\\
$\Delta Y/\Delta Z=2,\, Y_\rmn{P}=0.2485 $& &  9.576& 2.131& 3.682& & 11.10&9.666& 3.843\\
$\Delta Y/\Delta Z=5,\, Y_\rmn{P}=0.2485 $& &  29.33& 7.729& 12.42& & 21.51&0.383& 0.023\\
\hline
 & HD 113449 A & & & & HD 113449 B & & &  \\
& & & & & & & & \\[-4.5mm]
$\Delta Y/\Delta Z=2,\, Y_\rmn{P}=0.23$ & & 5.754& 10.44& 10.94& & 2.548&3.270& 3.497\\
$\Delta Y/\Delta Z=2,\, Y_\rmn{P}=0.2485 $& &  5.856& 9.333& 9.159& & 2.548&3.252& 3.476\\
$\Delta Y/\Delta Z=5,\, Y_\rmn{P}=0.2485 $& &  4.563& 2.854& 2.015& & 2.559&3.129& 3.291\\
\hline
 & NTT A & & & & NTT B & & &  \\
& & & & & & & & \\[-4.5mm]
$\Delta Y/\Delta Z=2,\, Y_\rmn{P}=0.23$ & & 3.589& 1.323& 0.830& & 2.417&0.372& 0.192\\
$\Delta Y/\Delta Z=2,\, Y_\rmn{P}=0.2485 $& &  2.837& 0.880& 0.516& & 1.391&0.166& 0.078\\
$\Delta Y/\Delta Z=5,\, Y_\rmn{P}=0.2485 $& &  1.104& 0.173& 0.079& & 0.116&0.005& 0.002\\
\hline
 & HD 98800 Ba & & & & HD 98800 Bb & & &  \\
& & & & & & & & \\[-4.5mm]
$\Delta Y/\Delta Z=2,\, Y_\rmn{P}=0.23$ & & 2.327& 6.120& 5.958& & 4.910&7.595& 7.702\\
$\Delta Y/\Delta Z=2,\, Y_\rmn{P}=0.2485 $& &  2.909& 6.173& 5.419& & 4.354&7.662& 6.594\\
$\Delta Y/\Delta Z=5,\, Y_\rmn{P}=0.2485 $& &  4.946& 5.024& 3.010& & 7.062&6.963& 3.018\\
\hline
 & UZ Tau Ea & & & & UZ Tau Eb & & &  \\
& & & & & & & & \\[-4.5mm]
$\Delta Y/\Delta Z=2,\, Y_\rmn{P}=0.23$ & & 0.077& $< 10^{-3}$& $< 10^{-3}$& & 46.45&16.35& 9.575\\
$\Delta Y/\Delta Z=2,\, Y_\rmn{P}=0.2485 $& &  0.030& $< 10^{-3}$& $< 10^{-3}$& & 33.43&9.941& 5.498\\
$\Delta Y/\Delta Z=5,\, Y_\rmn{P}=0.2485 $& &  $< 10^{-3}$& $< 10^{-3}$& $< 10^{-3}$& & 7.188&0.826& 0.355\\
\hline
 & DL Tau & & & & DM Tau & & &  \\
& & & & & & & & \\[-4.5mm]
$\Delta Y/\Delta Z=2,\, Y_\rmn{P}=0.23$ & & 15.75& 6.815& 4.181& & 19.31&12.15& 9.790\\
$\Delta Y/\Delta Z=2,\, Y_\rmn{P}=0.2485 $& &  14.59& 4.953& 2.782& & 16.83&8.719& 6.555\\
$\Delta Y/\Delta Z=5,\, Y_\rmn{P}=0.2485 $& &  9.141& 1.225& 0.515& & 6.555&1.297& 0.675\\
\hline
 & CY Tau & & & & BP Tau & & &  \\
& & & & & & & & \\[-4.5mm]
$\Delta Y/\Delta Z=2,\, Y_\rmn{P}=0.23$ & & 5.420& 3.667& 3.256& & 0.180&0.155& 0.147\\
$\Delta Y/\Delta Z=2,\, Y_\rmn{P}=0.2485 $& &  4.675& 3.145& 2.776& & 0.138&0.115& 0.107\\
$\Delta Y/\Delta Z=5,\, Y_\rmn{P}=0.2485 $& &  3.058& 2.071& 1.788& & 0.054&0.038& 0.034\\
\hline
 & GM Aur & & & & MWC 480 & & &  \\
& & & & & & & & \\[-4.5mm]
$\Delta Y/\Delta Z=2,\, Y_\rmn{P}=0.23$ & & 59.83& 9.330& 1.688& & 0.161&0.162& 0.163\\
$\Delta Y/\Delta Z=2,\, Y_\rmn{P}=0.2485 $& &  53.20& 6.241& 0.865& & 0.306&0.309& 0.309\\
$\Delta Y/\Delta Z=5,\, Y_\rmn{P}=0.2485 $& &  40.15& 0.817& 0.037& & 2.305&2.325& 2.332\\
\hline
& LkCa 15 & & & & & & & \\
& & & & & & & & \\[-4.5mm]
$\Delta Y/\Delta Z=2,\, Y_\rmn{P}=0.23$ & & 3.908 & 7.764 & 8.856 & & & & \\
$\Delta Y/\Delta Z=2,\, Y_\rmn{P}=0.2485 $ & &  5.393 & 9.119 & 9.496 & & & & \\
$\Delta Y/\Delta Z=5,\, Y_\rmn{P}=0.2485 $ & &  10.91 & 8.083 & 5.268 \\
\hline
 \end{tabular}
 \end{minipage}
\end{table*}

\setcounter{table}{2}
\begin{table*}
\begin{minipage}{160mm}
 \caption{Bayes factor values for the case of flat prior on the mass and Gaussian prior on the metallicity}
 \label{tab:bfactYesMetNomass}
\renewcommand\tabcolsep{4.pt}
  \begin{tabular}{@{}l|lccc|lccc@{}}
\hline
\hline
& & $\alpha=1.20$ & $\alpha=1.68$ & $\alpha=1.90$ & & $\alpha=1.20$ & $\alpha=1.68$ & $\alpha=1.90$ \\ 
\hline
 & RS Cha A & & & & RS Cha B & & &  \\
& & & & & & & & \\[-4.5mm]
$\Delta Y/\Delta Z=2,\, Y_\rmn{P}=0.23$ & & 0.887& 0.897& 0.896& & 1.058&1.089& 1.114\\
$\Delta Y/\Delta Z=2,\, Y_\rmn{P}=0.2485 $& &  0.752& 0.752& 0.764& & 0.978&1.002& 1.044\\
$\Delta Y/\Delta Z=5,\, Y_\rmn{P}=0.2485 $& &  0.530& 0.532& 0.531& & 0.790&0.806& 0.833\\
\hline
 & RXJ A & & & & RXJ B & & &  \\
& & & & & & & & \\[-4.5mm]
$\Delta Y/\Delta Z=2,\, Y_\rmn{P}=0.23$ & & 0.992& 1.016& 1.007& & 1.353&1.071& 0.975\\
$\Delta Y/\Delta Z=2,\, Y_\rmn{P}=0.2485 $& &  0.951& 0.976& 0.972& & 1.264&0.967& 0.871\\
$\Delta Y/\Delta Z=5,\, Y_\rmn{P}=0.2485 $& &  0.873& 0.905& 0.904& & 1.077&0.766& 0.675\\
\hline
 & V1174 Ori A & & & & V1174 Ori B & & &  \\
& & & & & & & & \\[-4.5mm]
$\Delta Y/\Delta Z=2,\, Y_\rmn{P}=0.23$ & & 1.103& 1.034& 0.991& & 1.313&0.999& 0.924\\
$\Delta Y/\Delta Z=2,\, Y_\rmn{P}=0.2485 $& &  1.066& 0.984& 0.929& & 1.110&0.836& 0.772\\
$\Delta Y/\Delta Z=5,\, Y_\rmn{P}=0.2485 $& &  0.992& 0.861& 0.786& & 0.785&0.581& 0.534\\
\hline
 & EK Cep A & & & & EK Cep B & & &  \\
& & & & & & & & \\[-4.5mm]
$\Delta Y/\Delta Z=2,\, Y_\rmn{P}=0.23$ & & 3.545& 3.568& 3.570& & 2.769&0.865& 0.840\\
$\Delta Y/\Delta Z=2,\, Y_\rmn{P}=0.2485 $& &  2.343& 2.367& 2.378& & 1.769&0.795& 0.792\\
$\Delta Y/\Delta Z=5,\, Y_\rmn{P}=0.2485 $& &  0.091& 0.093& 0.093& & 0.783&0.699& 0.710\\
\hline
 & TY CrA A & & & & TY CrA B & & &  \\
& & & & & & & & \\[-4.5mm]
$\Delta Y/\Delta Z=2,\, Y_\rmn{P}=0.23$ & & -& -& -& & -&-& -\\
$\Delta Y/\Delta Z=2,\, Y_\rmn{P}=0.2485 $& &  -& -& -& & -&-& -\\
$\Delta Y/\Delta Z=5,\, Y_\rmn{P}=0.2485 $& &  -& -& -& & -&-& -\\
\hline
 & ASAS A & & & & ASAS B & & &  \\
& & & & & & & & \\[-4.5mm]
$\Delta Y/\Delta Z=2,\, Y_\rmn{P}=0.23$ & & 0.632& 0.701& 0.706& & 0.840&0.680& 0.581\\
$\Delta Y/\Delta Z=2,\, Y_\rmn{P}=0.2485 $& &  0.597& 0.666& 0.667& & 0.788&0.604& 0.502\\
$\Delta Y/\Delta Z=5,\, Y_\rmn{P}=0.2485 $& &  0.544& 0.610& 0.604& & 0.706&0.481& 0.380\\
\hline
 & HD 113449 A & & & & HD 113449 B & & &  \\
& & & & & & & & \\[-4.5mm]
$\Delta Y/\Delta Z=2,\, Y_\rmn{P}=0.23$ & & 0.874& 1.052& 0.975& & 1.028&1.296& 1.375\\
$\Delta Y/\Delta Z=2,\, Y_\rmn{P}=0.2485 $& &  0.986& 0.948& 0.827& & 1.325&1.636& 1.731\\
$\Delta Y/\Delta Z=5,\, Y_\rmn{P}=0.2485 $& &  1.043& 0.631& 0.490& & 1.987&2.383& 2.498\\
\hline
 & NTT A & & & & NTT B & & &  \\
& & & & & & & & \\[-4.5mm]
$\Delta Y/\Delta Z=2,\, Y_\rmn{P}=0.23$ & & -& -& -& & -&-&- \\
$\Delta Y/\Delta Z=2,\, Y_\rmn{P}=0.2485 $& &  -& -& -& & -&-& - \\
$\Delta Y/\Delta Z=5,\, Y_\rmn{P}=0.2485 $& &  -& -& -& & -&-& - \\
\hline
 & HD 98800 Ba & & & & HD 98800 Bb & & &  \\
& & & & & & & & \\[-4.5mm]
$\Delta Y/\Delta Z=2,\, Y_\rmn{P}=0.23$ & & 1.499& 0.676& 0.560& & 1.479&0.790& 0.662\\
$\Delta Y/\Delta Z=2,\, Y_\rmn{P}=0.2485 $& &  1.287& 0.614& 0.501& & 1.075&0.708& 0.538\\
$\Delta Y/\Delta Z=5,\, Y_\rmn{P}=0.2485 $& &  1.046& 0.542& 0.420& & 0.843&0.595& 0.425\\
\hline
 & UZ Tau Ea & & & & UZ Tau Eb & & &  \\
& & & & & & & & \\[-4.5mm]
$\Delta Y/\Delta Z=2,\, Y_\rmn{P}=0.23$ & & 1.629& 1.150& 1.007& & 1.092&0.404& 0.294\\
$\Delta Y/\Delta Z=2,\, Y_\rmn{P}=0.2485 $& &  1.446& 1.006& 0.845& & 0.687&0.252& 0.161\\
$\Delta Y/\Delta Z=5,\, Y_\rmn{P}=0.2485 $& &  1.383& 0.857& 0.692& & 0.311&0.072& 0.044\\
\hline
 & DL Tau & & & & DM Tau & & &  \\
& & & & & & & & \\[-4.5mm]
$\Delta Y/\Delta Z=2,\, Y_\rmn{P}=0.23$ & & 2.058& 1.052& 0.892& & 1.474&1.015& 0.914\\
$\Delta Y/\Delta Z=2,\, Y_\rmn{P}=0.2485 $& &  1.777& 0.941& 0.791& & 1.248&0.853& 0.767\\
$\Delta Y/\Delta Z=5,\, Y_\rmn{P}=0.2485 $& &  1.372& 0.822& 0.708& & 0.893&0.604& 0.542\\
\hline
 & CY Tau & & & & BP Tau & & &  \\
& & & & & & & & \\[-4.5mm]
$\Delta Y/\Delta Z=2,\, Y_\rmn{P}=0.23$ & & 1.448& 1.142& 1.061& & 1.105&1.078& 1.068\\
$\Delta Y/\Delta Z=2,\, Y_\rmn{P}=0.2485 $& &  1.298& 1.040& 0.945& & 1.058&1.029& 1.019\\
$\Delta Y/\Delta Z=5,\, Y_\rmn{P}=0.2485 $& &  1.131& 0.829& 0.731& & 0.965&0.928& 0.914\\
\hline
 & GM Aur & & & & MWC 480 & & &  \\
& & & & & & & & \\[-4.5mm]
$\Delta Y/\Delta Z=2,\, Y_\rmn{P}=0.23$ & & 7.278& 0.368& 0.061& & 0.462&0.466& 0.468\\
$\Delta Y/\Delta Z=2,\, Y_\rmn{P}=0.2485 $& &  5.192& 0.218& 0.027& & 0.313&0.316& 0.317\\
$\Delta Y/\Delta Z=5,\, Y_\rmn{P}=0.2485 $& &  4.421& 0.080& 0.008& & 0.125&0.126& 0.127\\
\hline
& LkCa 15 & & & & & & & \\
& & & & & & & & \\[-4.5mm]
$\Delta Y/\Delta Z=2,\, Y_\rmn{P}=0.23$ & &  1.243 & 1.048 & 0.959 & & & & \\
$\Delta Y/\Delta Z=2,\, Y_\rmn{P}=0.2485 $& &1.186 & 0.965 & 0.869 & & & & \\
$\Delta Y/\Delta Z=5,\, Y_\rmn{P}=0.2485 $& &1.059 & 0.791 & 0.690 & & & & \\
\hline
\end{tabular}
 \end{minipage}
\end{table*}

\setcounter{table}{3}
\begin{table*}
\begin{minipage}{160mm}
 \caption{Bayes factor values for the case with Gaussian prior on both the mass and the metallicity}
 \label{tab:bfactYesMetYesmass}
\renewcommand\tabcolsep{4.pt}
  \begin{tabular}{@{}l|lccc|lccc@{}}
\hline
\hline
& & $\alpha=1.20$ & $\alpha=1.68$ & $\alpha=1.90$ & & $\alpha=1.20$ & $\alpha=1.68$ & $\alpha=1.90$ \\ 
\hline
 & RS Cha A & & & & RS Cha B & & &  \\
& & & & & & & & \\[-4.5mm]
$\Delta Y/\Delta Z=2,\, Y_\rmn{P}=0.23$ & & 17.40& 17.78& 17.51& & 26.67&27.27& 27.57\\
$\Delta Y/\Delta Z=2,\, Y_\rmn{P}=0.2485 $& &  10.23& 10.24& 10.23& & 10.36&10.60& 10.85\\
$\Delta Y/\Delta Z=5,\, Y_\rmn{P}=0.2485 $& &  0.003& 0.003& 0.003& & 0.003&0.003& 0.003\\
\hline
 & RXJ A & & & & RXJ B & & &  \\
& & & & & & & & \\[-4.5mm]
$\Delta Y/\Delta Z=2,\, Y_\rmn{P}=0.23$ & & 10.87& 18.35& 17.02& & 15.62&14.00& 11.73\\
$\Delta Y/\Delta Z=2,\, Y_\rmn{P}=0.2485 $& &  7.781& 18.34& 18.31& & 15.98&10.44& 7.579\\
$\Delta Y/\Delta Z=5,\, Y_\rmn{P}=0.2485 $& &  3.366& 14.13& 15.14& & 12.00&3.298& 1.666\\
\hline
 & V1174 Ori A & & & & V1174 Ori B & & &  \\
& & & & & & & & \\[-4.5mm]
$\Delta Y/\Delta Z=2,\, Y_\rmn{P}=0.23$ & & 2.484& 12.76& 16.35& & 0.794&0.091& 0.043\\
$\Delta Y/\Delta Z=2,\, Y_\rmn{P}=0.2485 $& &  6.836& 18.37& 18.09& & 0.217&0.016& 0.006\\
$\Delta Y/\Delta Z=5,\, Y_\rmn{P}=0.2485 $& &  22.02& 13.08& 7.076& & 0.007&$< 10^{-3}$& $< 10^{-3}$\\
\hline
 & EK Cep A & & & & EK Cep B & & &  \\
& & & & & & & & \\[-4.5mm]
$\Delta Y/\Delta Z=2,\, Y_\rmn{P}=0.23$ & & 37.95& 38.17& 38.06& & 24.89&21.46& 9.717\\
$\Delta Y/\Delta Z=2,\, Y_\rmn{P}=0.2485 $& &  8.021& 8.004& 8.010& & 16.38&28.07& 17.43\\
$\Delta Y/\Delta Z=5,\, Y_\rmn{P}=0.2485 $& &  0.001& 0.001& 0.001& & 4.962&16.00& 22.97\\
\hline
 & TY CrA A & & & & TY CrA B & & &  \\
& & & & & & & & \\[-4.5mm]
$\Delta Y/\Delta Z=2,\, Y_\rmn{P}=0.23$ & & -& -& -& & -&-& -\\
$\Delta Y/\Delta Z=2,\, Y_\rmn{P}=0.2485 $& &  -& -& -& & -&-& -\\
$\Delta Y/\Delta Z=5,\, Y_\rmn{P}=0.2485 $& &  -& -& -& & -&-& -\\
\hline
 & ASAS A & & & & ASAS B & & &  \\
& & & & & & & & \\[-4.5mm]
$\Delta Y/\Delta Z=2,\, Y_\rmn{P}=0.23$ & & 3.527& 1.714& 3.897& & 13.81&4.101& 1.250\\
$\Delta Y/\Delta Z=2,\, Y_\rmn{P}=0.2485 $& &  10.47& 4.188& 6.801& & 13.61&1.690& 0.340\\
$\Delta Y/\Delta Z=5,\, Y_\rmn{P}=0.2485 $& &  26.83& 10.93& 11.13& & 7.133&0.056& 0.003\\
\hline
 & HD 113449 A & & & & HD 113449 B & & &  \\
& & & & & & & & \\[-4.5mm]
$\Delta Y/\Delta Z=2,\, Y_\rmn{P}=0.23$ & & 8.426& 9.087& 7.972& & 4.084&5.100& 5.393\\
$\Delta Y/\Delta Z=2,\, Y_\rmn{P}=0.2485 $& &  7.398& 6.058& 4.900& & 4.029&4.877& 5.124\\
$\Delta Y/\Delta Z=5,\, Y_\rmn{P}=0.2485 $& &  3.911& 1.671& 1.155& & 3.304&3.765& 3.882\\
\hline
 & NTT A & & & & NTT B & & &  \\
& & & & & & & & \\[-4.5mm]
$\Delta Y/\Delta Z=2,\, Y_\rmn{P}=0.23$ & & -& -& -& & -&-& - \\
$\Delta Y/\Delta Z=2,\, Y_\rmn{P}=0.2485 $& &  -& -& -& & -&-& - \\
$\Delta Y/\Delta Z=5,\, Y_\rmn{P}=0.2485 $& &  -& -& -& & -&-& - \\
\hline
 & HD 98800 Ba & & & & HD 98800 Bb & & &  \\
& & & & & & & & \\[-4.5mm]
$\Delta Y/\Delta Z=2,\, Y_\rmn{P}=0.23$ & & 4.598& 4.732& 3.154& & 9.824&5.134& 4.441\\
$\Delta Y/\Delta Z=2,\, Y_\rmn{P}=0.2485 $& &  5.117& 3.838& 2.254& & 5.794&4.017& 2.152\\
$\Delta Y/\Delta Z=5,\, Y_\rmn{P}=0.2485 $& &  5.514& 2.523& 1.264& & 3.389&2.762& 0.925\\
\hline
 & UZ Tau Ea & & & & UZ Tau Eb & & &  \\
& & & & & & & & \\[-4.5mm]
$\Delta Y/\Delta Z=2,\, Y_\rmn{P}=0.23$ & & 0.007& $< 10^{-3}$& $< 10^{-3}$& & 14.31&3.507& 2.202\\
$\Delta Y/\Delta Z=2,\, Y_\rmn{P}=0.2485 $& &  0.002& $< 10^{-3}$& $< 10^{-3}$& & 7.555&1.637& 0.915\\
$\Delta Y/\Delta Z=5,\, Y_\rmn{P}=0.2485 $& &  $< 10^{-3}$& $< 10^{-3}$& $< 10^{-3}$& & 2.198&0.252& 0.123\\
\hline
 & DL Tau & & & & DM Tau & & &  \\
& & & & & & & & \\[-4.5mm]
$\Delta Y/\Delta Z=2,\, Y_\rmn{P}=0.23$ & & 15.38& 4.642& 2.638& & 15.09&7.006& 5.228\\
$\Delta Y/\Delta Z=2,\, Y_\rmn{P}=0.2485 $& &  12.89& 2.985& 1.558& & 10.91&4.091& 2.819\\
$\Delta Y/\Delta Z=5,\, Y_\rmn{P}=0.2485 $& &  7.816& 1.055& 0.472& & 4.461&0.943& 0.529\\
\hline
 & CY Tau & & & & BP Tau & & &  \\
& & & & & & & & \\[-4.5mm]
$\Delta Y/\Delta Z=2,\, Y_\rmn{P}=0.23$ & & 4.679& 3.441& 3.139& & 0.134&0.113& 0.106\\
$\Delta Y/\Delta Z=2,\, Y_\rmn{P}=0.2485 $& &  4.056& 3.024& 2.706& & 0.099&0.080& 0.074\\
$\Delta Y/\Delta Z=5,\, Y_\rmn{P}=0.2485 $& &  3.289& 2.275& 1.980& & 0.051&0.038& 0.034\\
\hline
 & GM Aur & & & & MWC 480 & & &  \\
& & & & & & & & \\[-4.5mm]
$\Delta Y/\Delta Z=2,\, Y_\rmn{P}=0.23$ & & 66.12& 2.143& 0.193& & 0.358&0.361& 0.362\\
$\Delta Y/\Delta Z=2,\, Y_\rmn{P}=0.2485 $& &  47.29& 1.166& 0.090& & 0.734&0.741& 0.743\\
$\Delta Y/\Delta Z=5,\, Y_\rmn{P}=0.2485 $& &  34.52& 0.238& 0.009& & 1.134&1.144& 1.147\\
\hline
& LkCa 15 & & & & & & & \\
& & & & & & & & \\[-4.5mm]
$\Delta Y/\Delta Z=2,\, Y_\rmn{P}=0.23$   & & 5.624 & 10.79 &  11.34 & & & & \\
$\Delta Y/\Delta Z=2,\, Y_\rmn{P}=0.2485 $& & 8.308 & 11.78 & 10.77 & & & & \\
$\Delta Y/\Delta Z=5,\, Y_\rmn{P}=0.2485 $& &12.79  & 8.522 & 5.659 & & & & \\
\hline
\end{tabular}
 \end{minipage}
\end{table*}

\clearpage
\section{Mass and age marginal distributions}
\label{Sec:figbinan}

\begin{figure*}
 \centering
	\includegraphics[width=0.48\hsize,height=5cm]{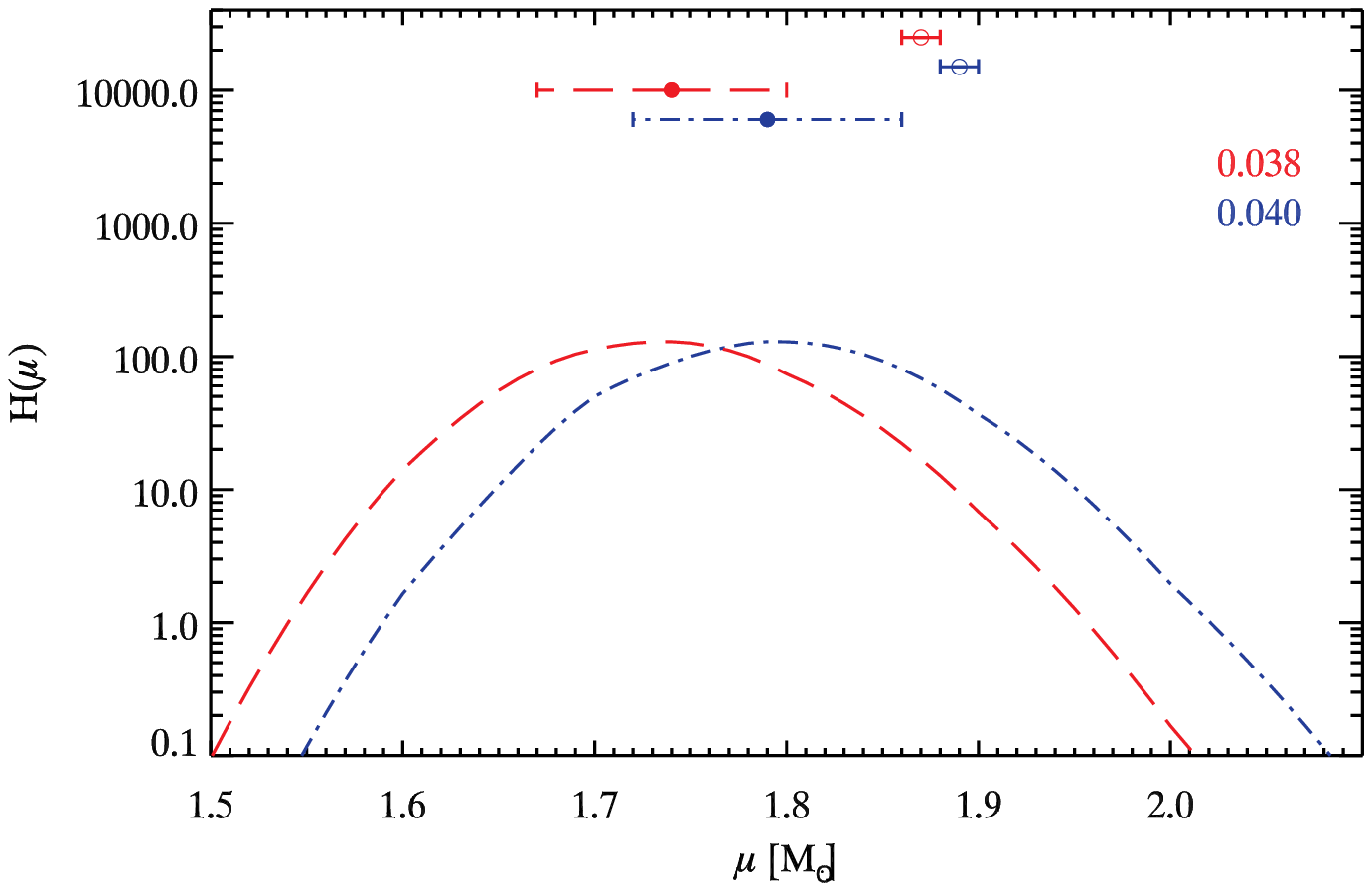}
	\includegraphics[width=0.48\hsize,height=5cm]{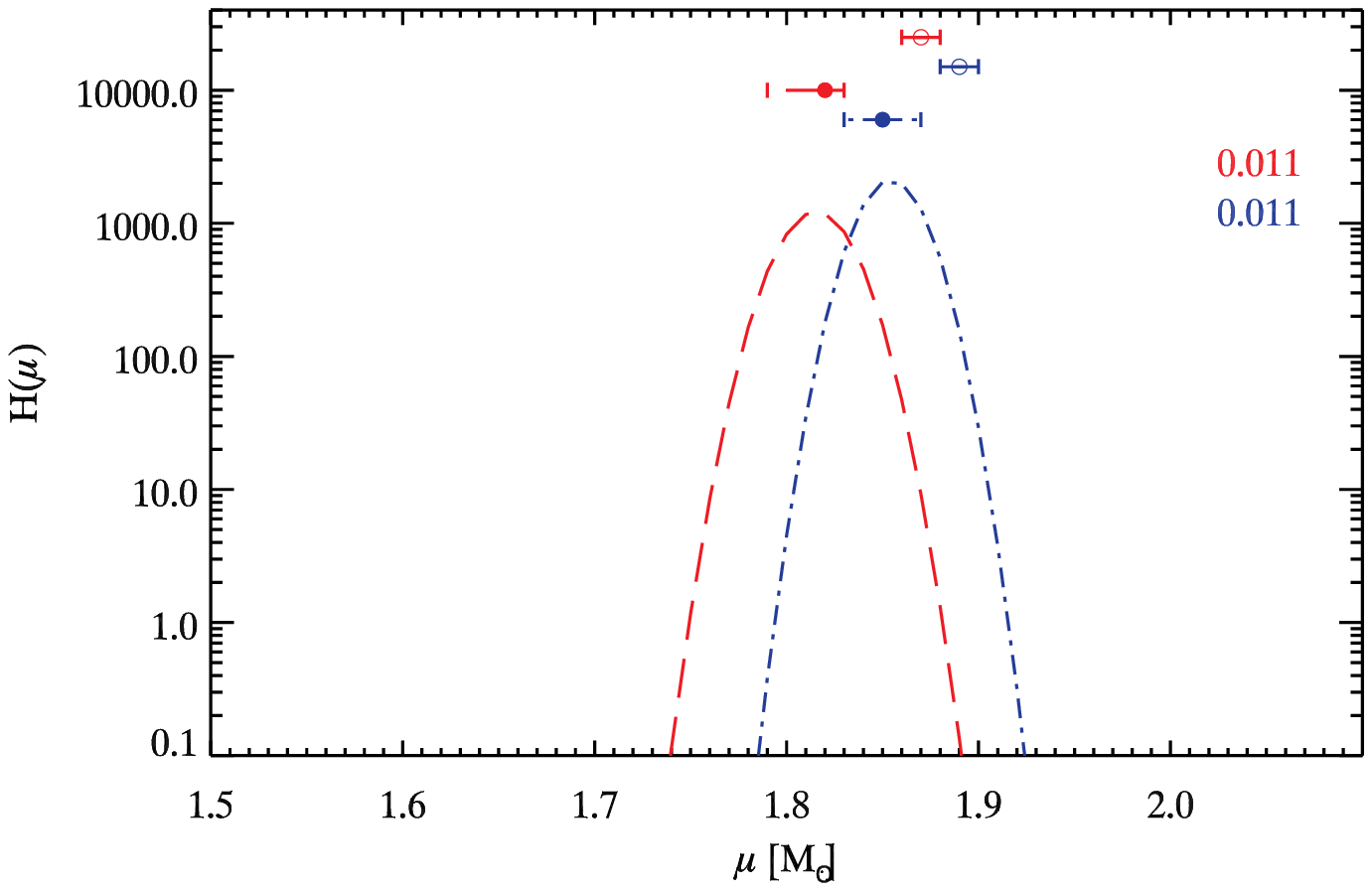}
	\includegraphics[width=0.48\hsize,height=5cm]{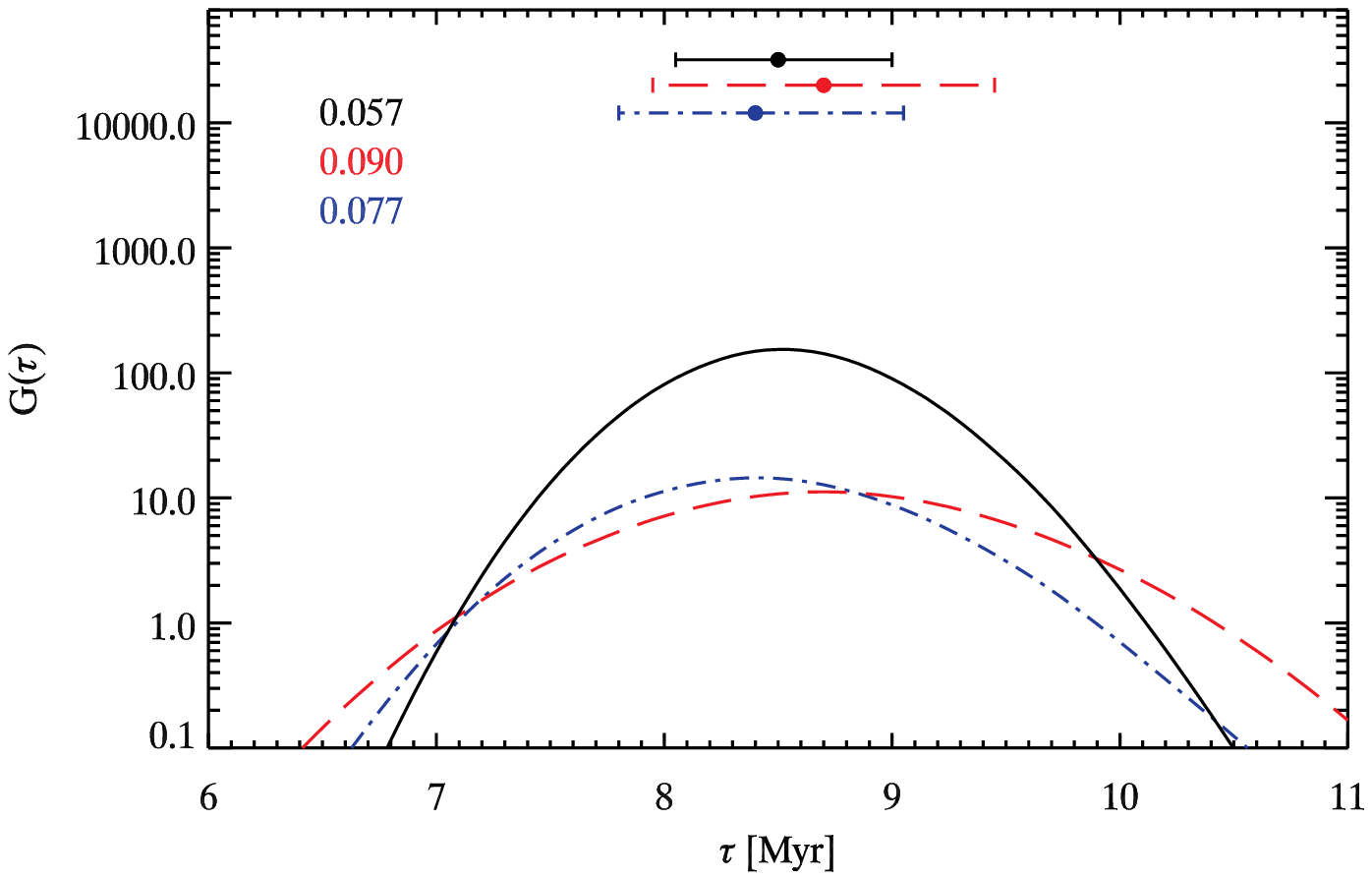}
	\includegraphics[width=0.48\hsize,height=5cm]{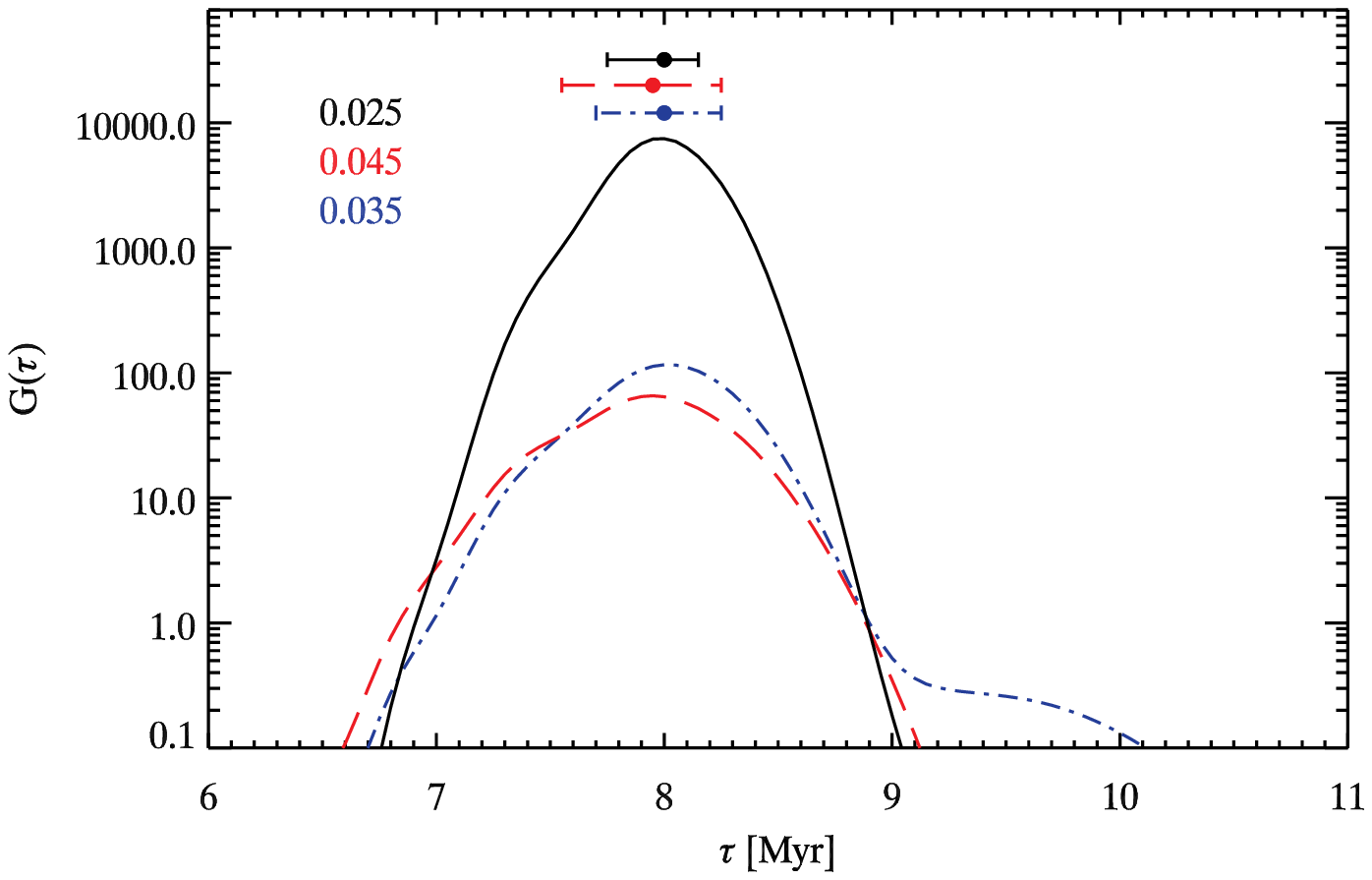}
	\caption{RS Cha components mass and age distribution functions as obtained from the $\log g - \log T_{\rmn{eff}}$ diagram using the standard set of models. \emph{Upper panels}: Marginalized mass distributions. \emph{Lower panels:} Marginalized age distributions. \emph{Left panels:} Marginalization using a flat mass prior. \emph{Right panels:} Marginalization using a Gaussian mass prior. \emph{All panels:} In dot-dashed-blue the primary component, in dashed-red the secondary. Full symbols indicate the mode of the distributions, the bars mark the 68\% confidence interval. The quoted numbers represent the relative precision of the mass or age estimates. In the upper panels the empty symbols and related error bars indicate the dynamical masses and their measurement errors. In the lower panels the solid-black line represents the system age distribution.}
 \label{fig:RSCha_GH}
\end{figure*}
\clearpage
\begin{figure*}
 \centering
        \includegraphics[width=0.48\hsize,height=5cm]{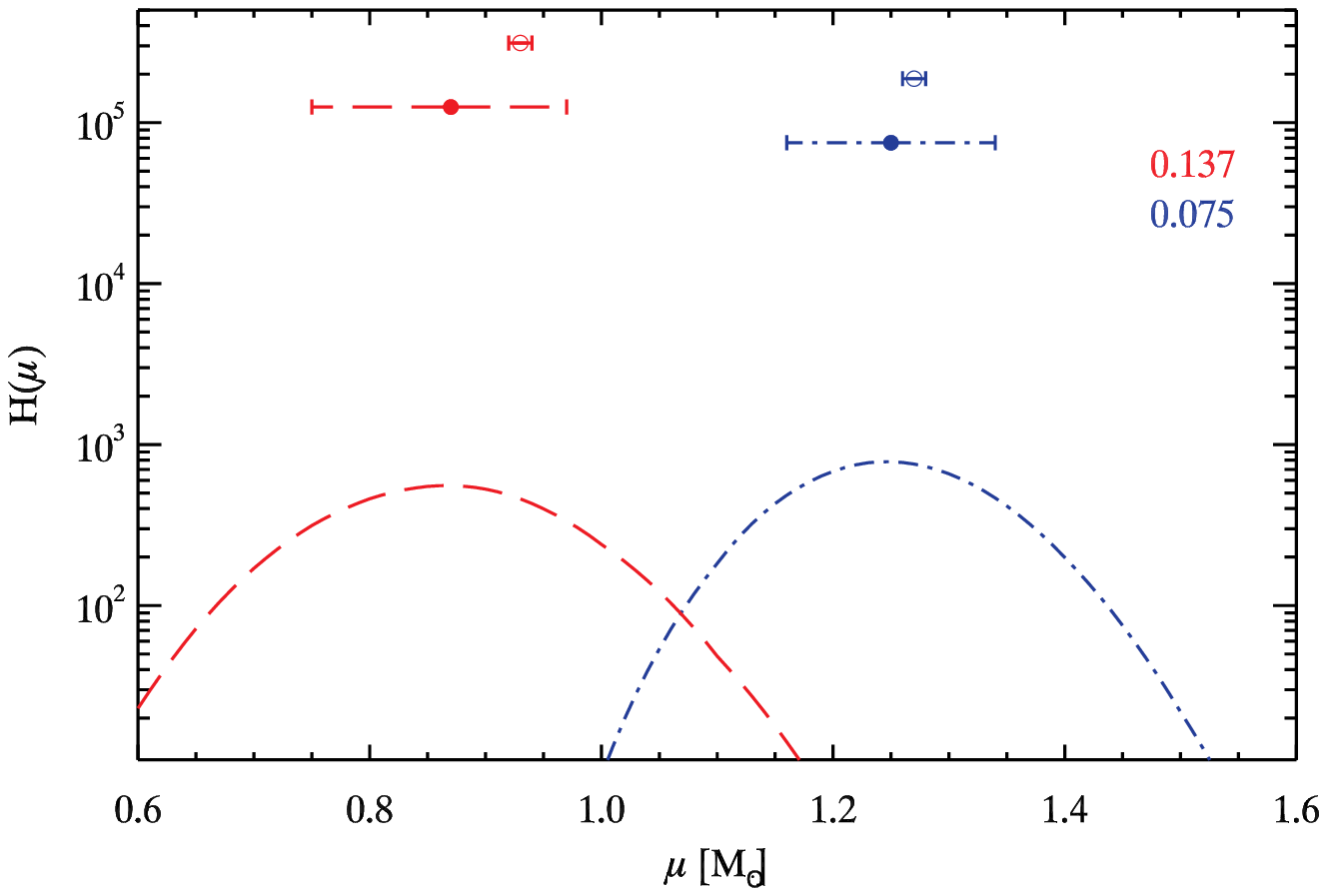}
	\includegraphics[width=0.48\hsize,height=5cm]{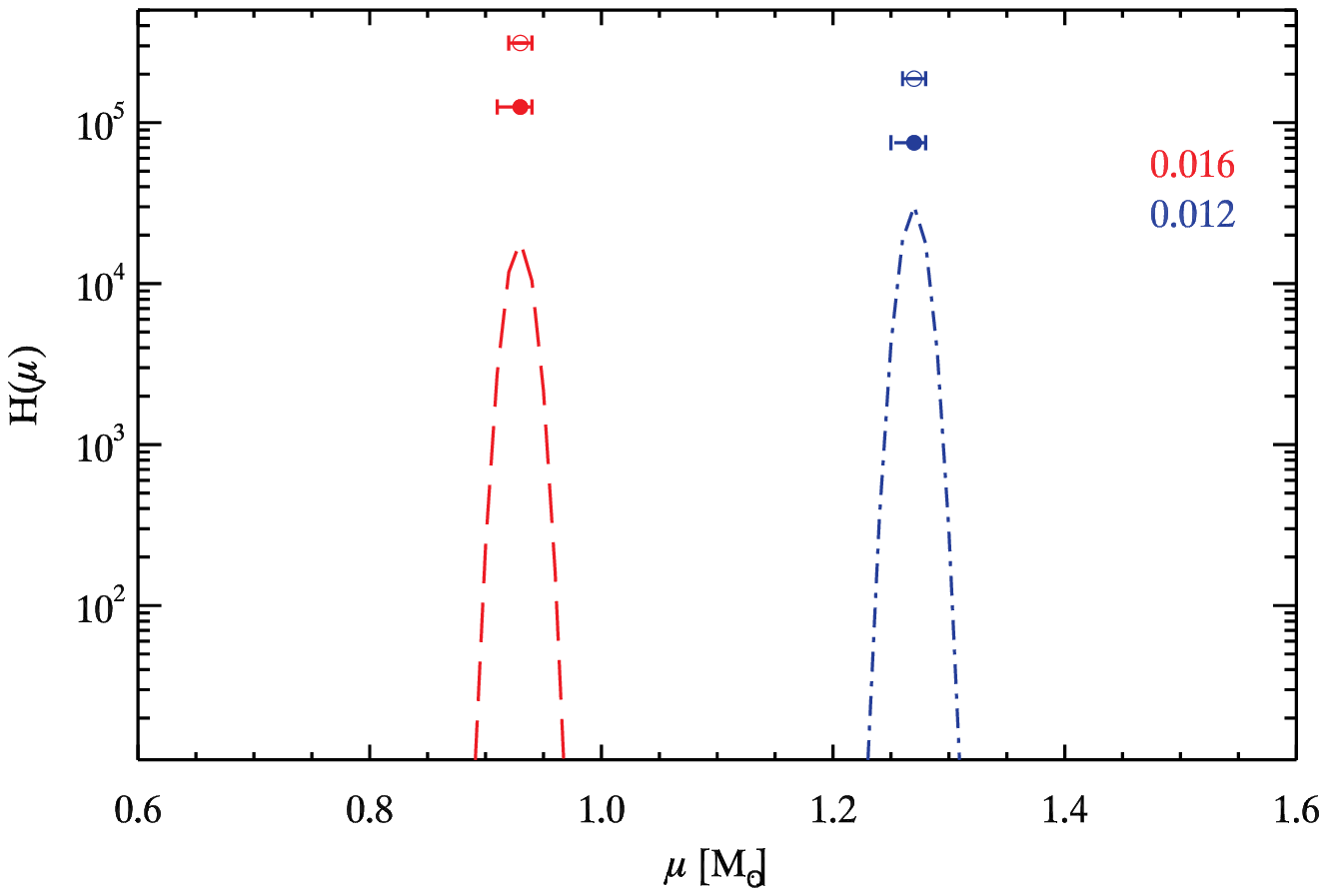}
	\includegraphics[width=0.48\hsize,height=5cm]{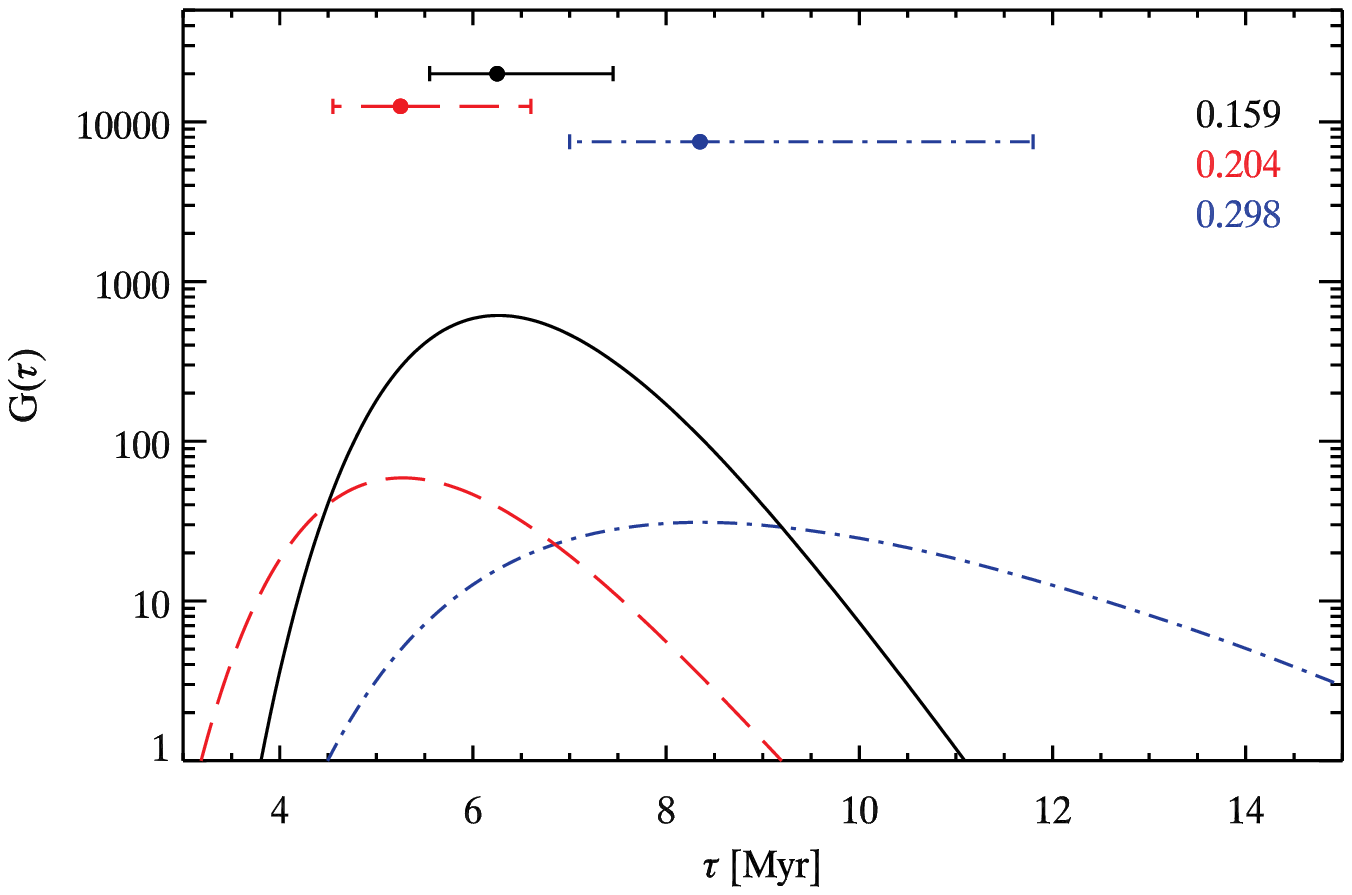}
	\includegraphics[width=0.48\hsize,height=5cm]{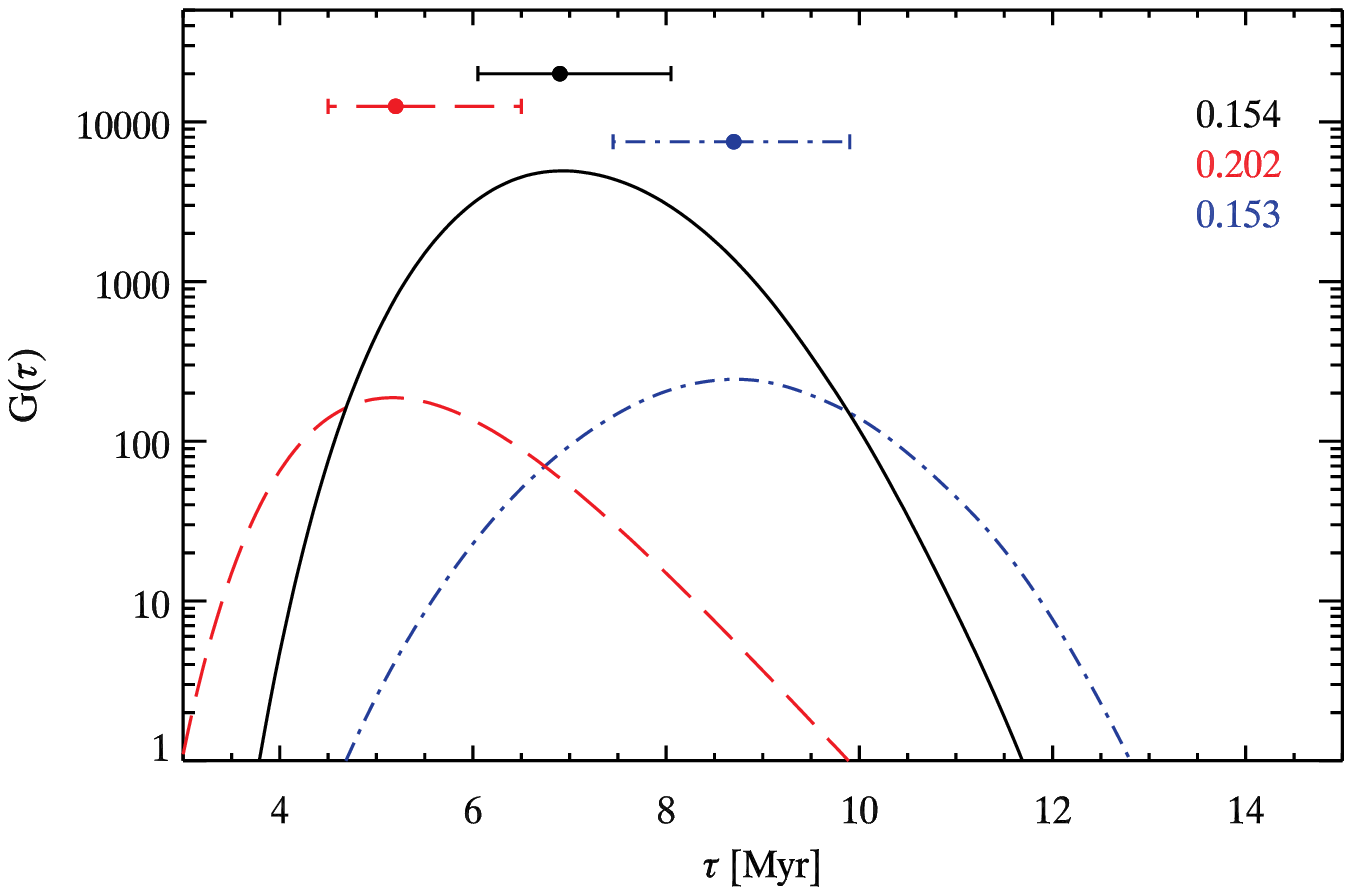}
	\caption{RXJ 0529.4+0041 components mass and age distribution functions from comparison with the standard set of models (see Fig.~\ref{fig:RSCha_GH} for a description).}
 \label{fig:RXJ_GH}
\end{figure*}

\begin{figure*}
 \centering
	\includegraphics[width=0.48\hsize,height=5cm]{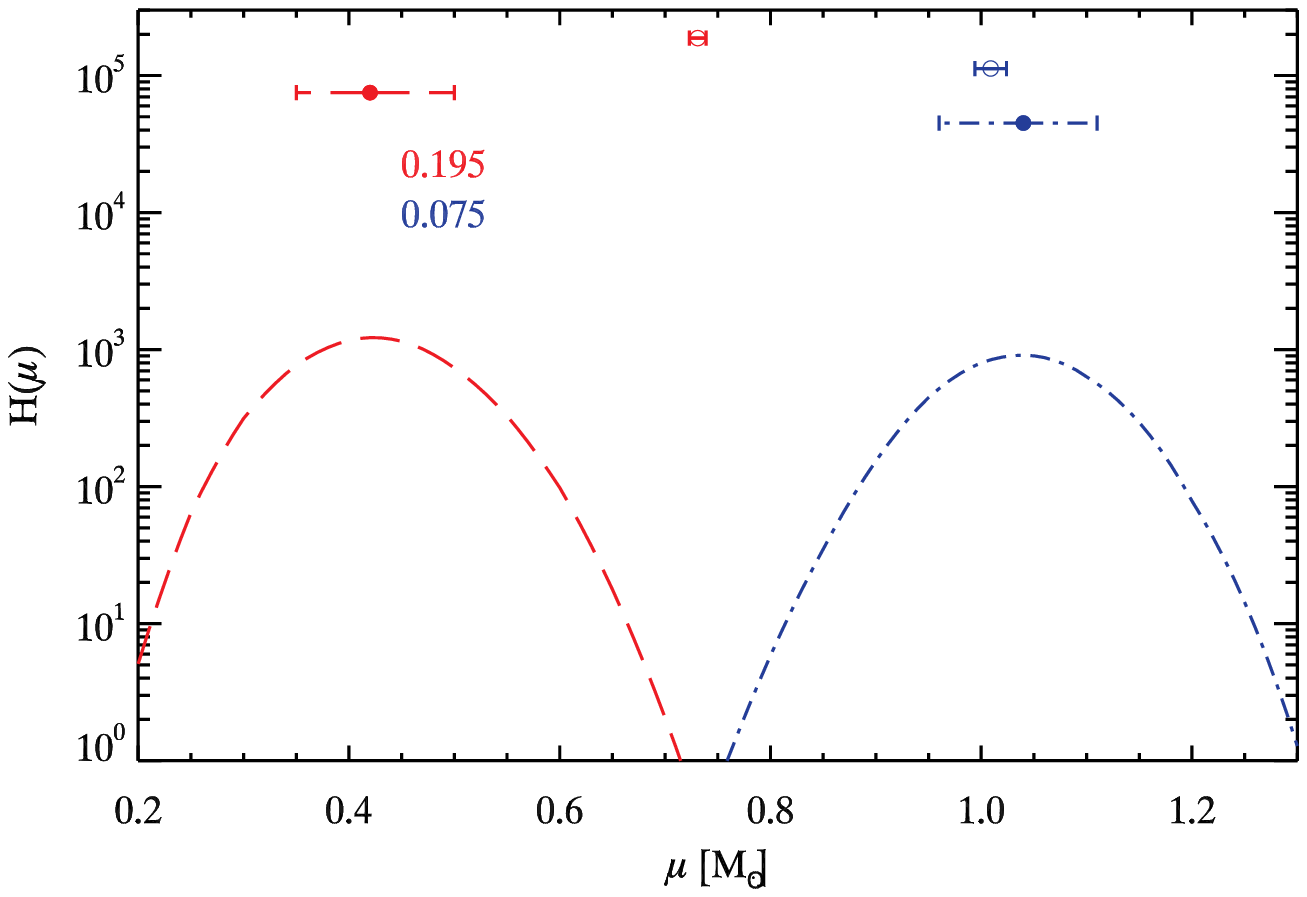}
	\includegraphics[width=0.48\hsize,height=5cm]{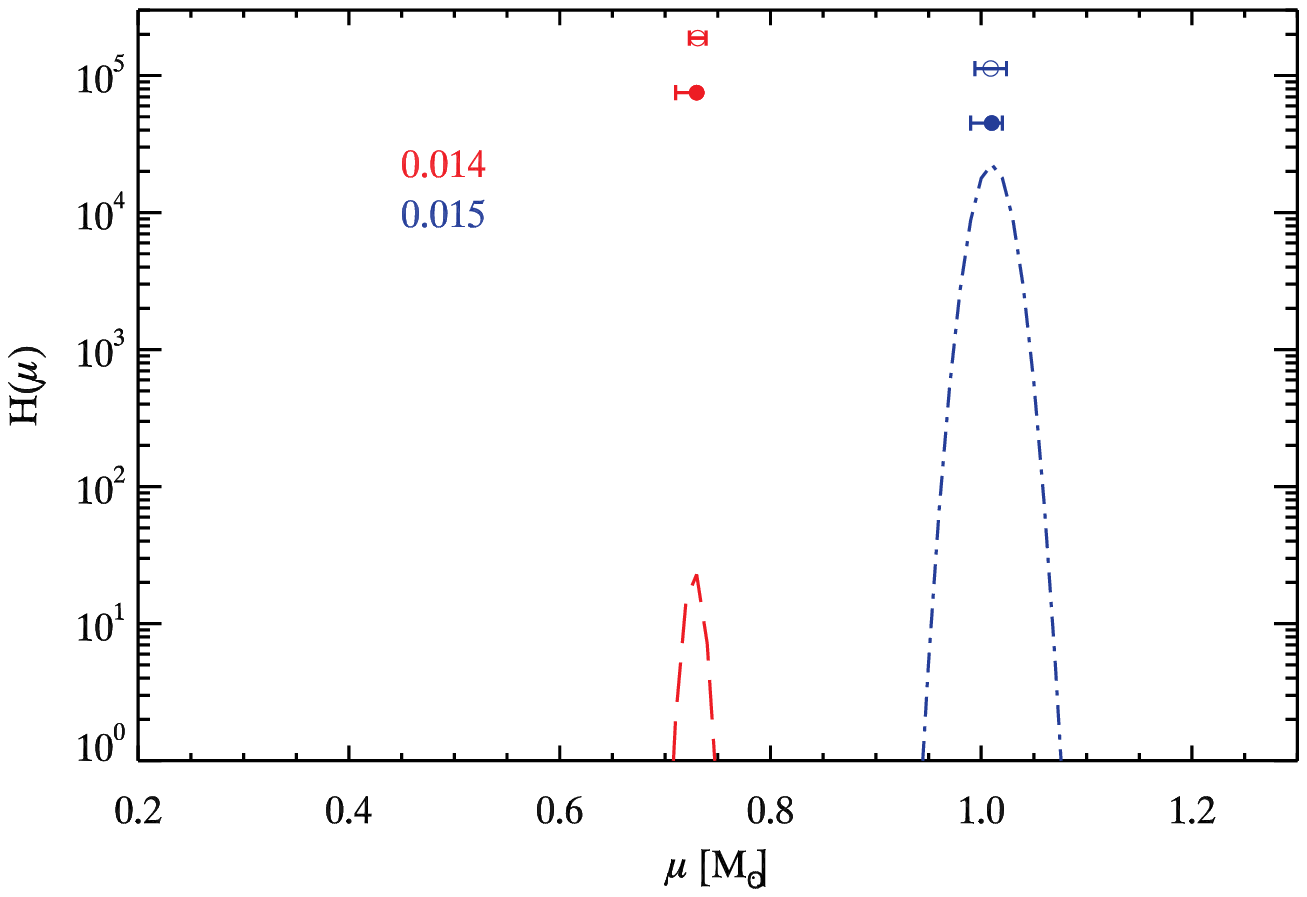}
	\includegraphics[width=0.48\hsize,height=5cm]{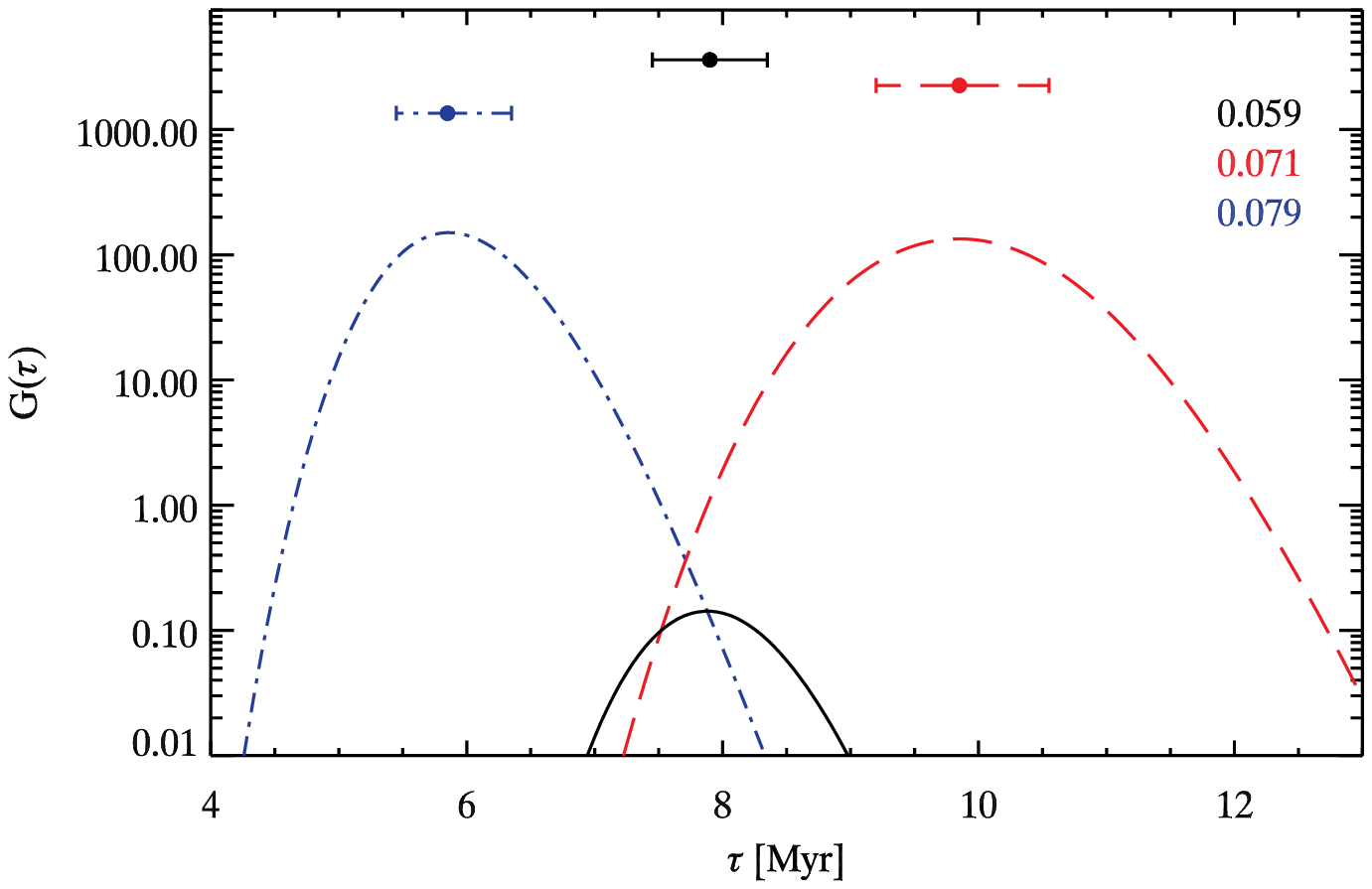}
	\includegraphics[width=0.48\hsize,height=5cm]{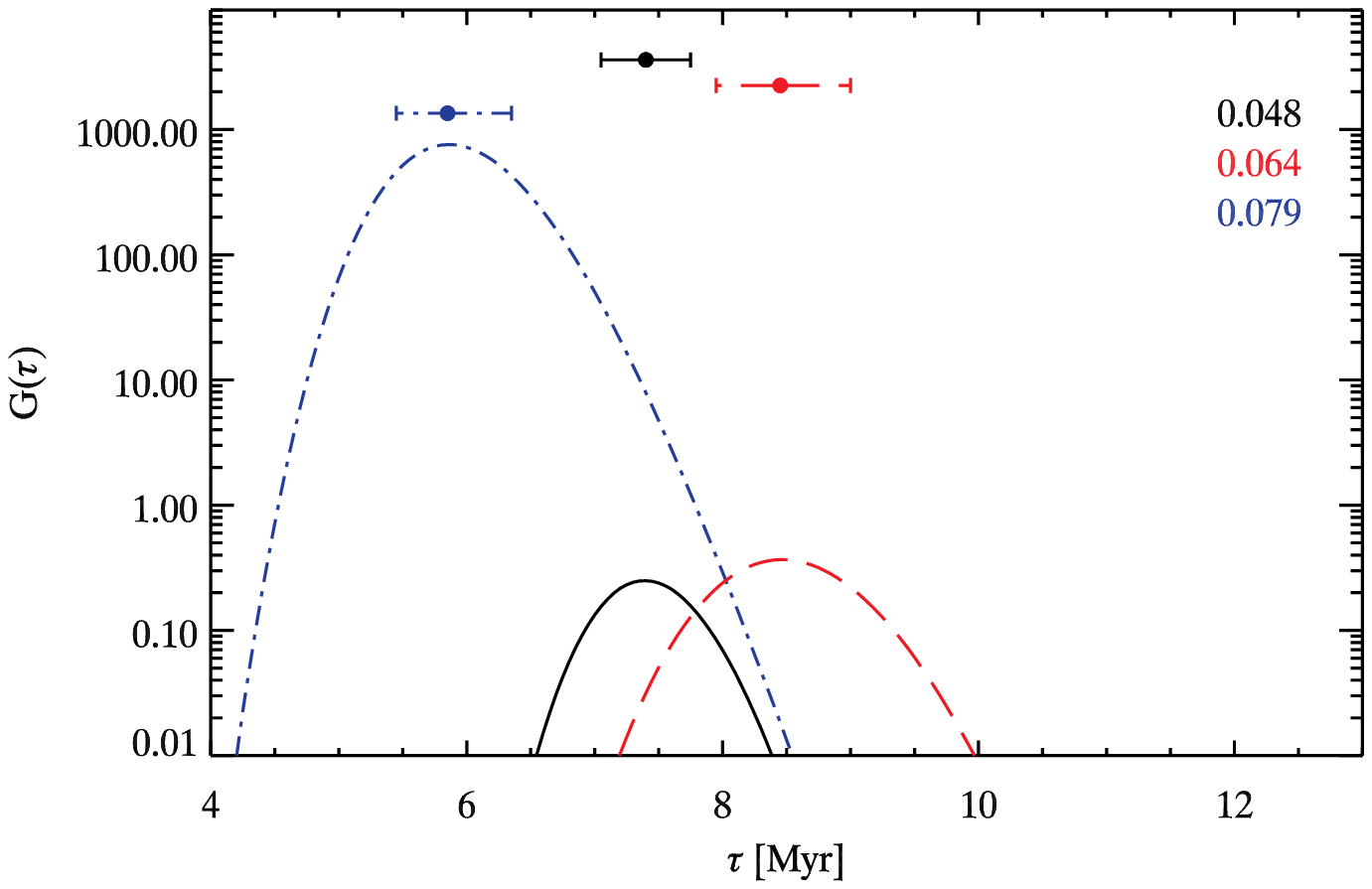}
	\caption{V 1174 Ori components mass and age distribution functions from comparison with the standard set of models (see Fig.~\ref{fig:RSCha_GH} for a description).}
 \label{fig:V1174Ori}
\end{figure*}
\clearpage
\begin{figure*}
 \centering
	\includegraphics[width=0.48\hsize,height=5cm]{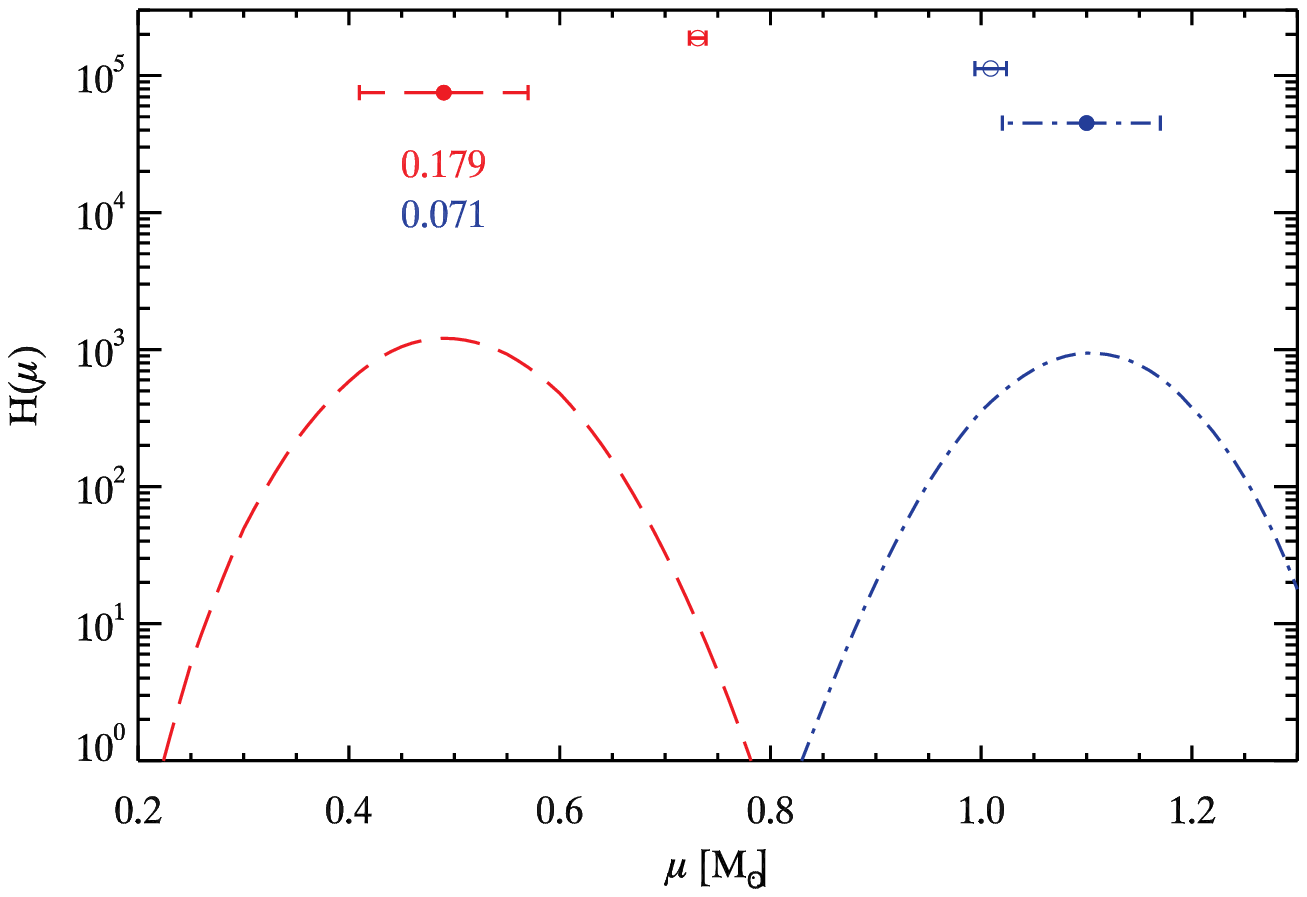}
	\includegraphics[width=0.48\hsize,height=5cm]{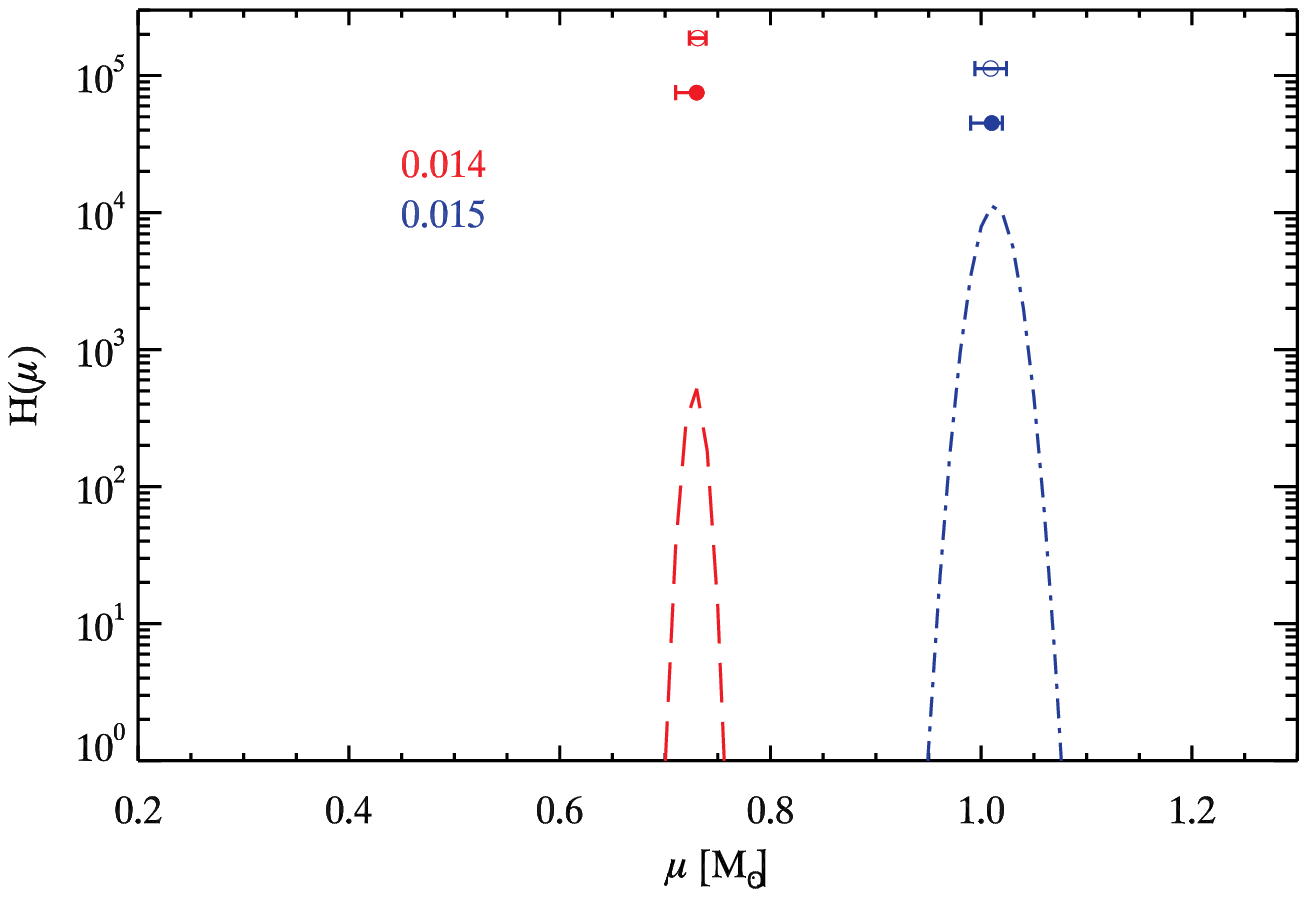}
	\includegraphics[width=0.48\hsize,height=5cm]{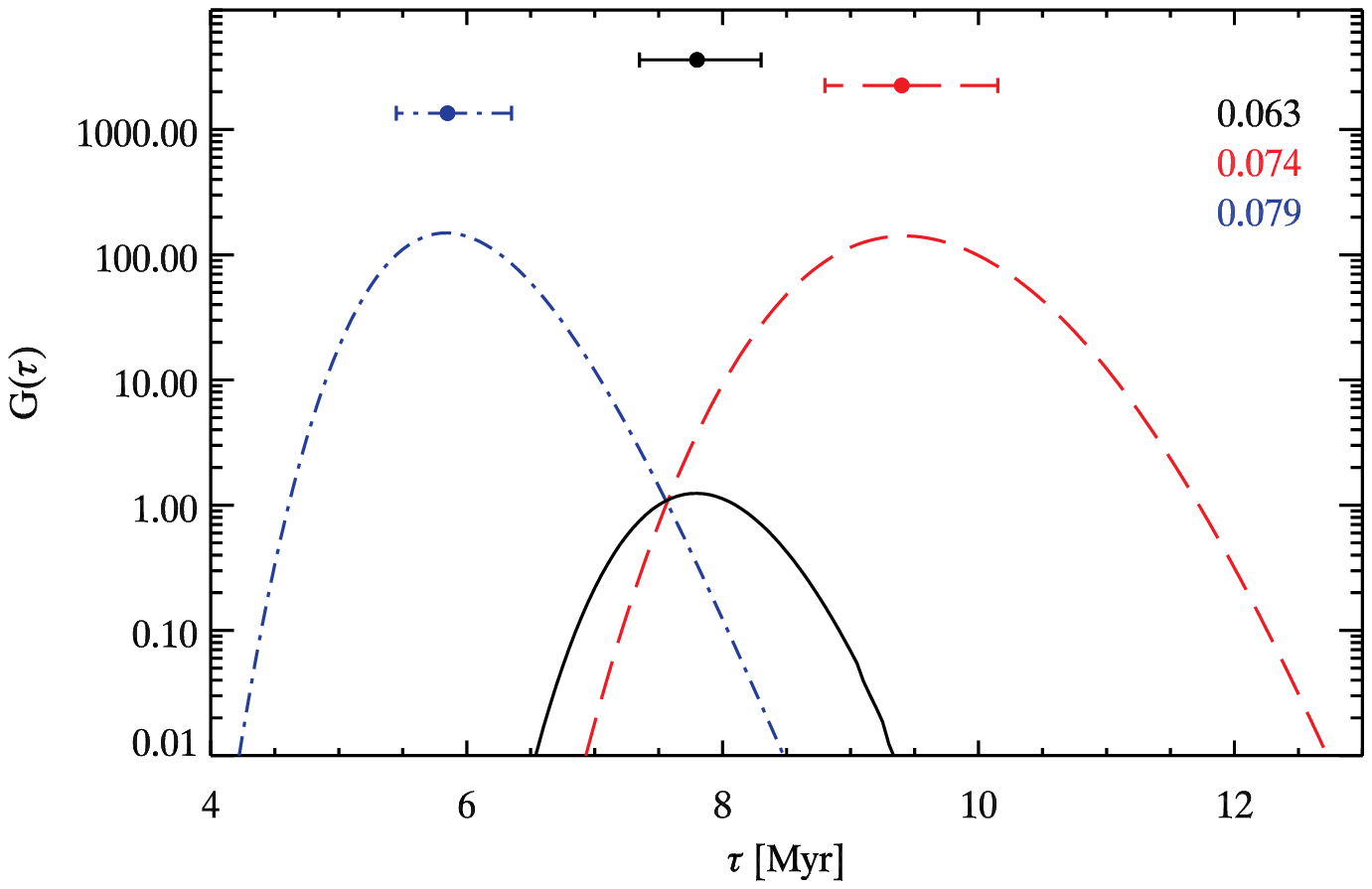}
	\includegraphics[width=0.48\hsize,height=5cm]{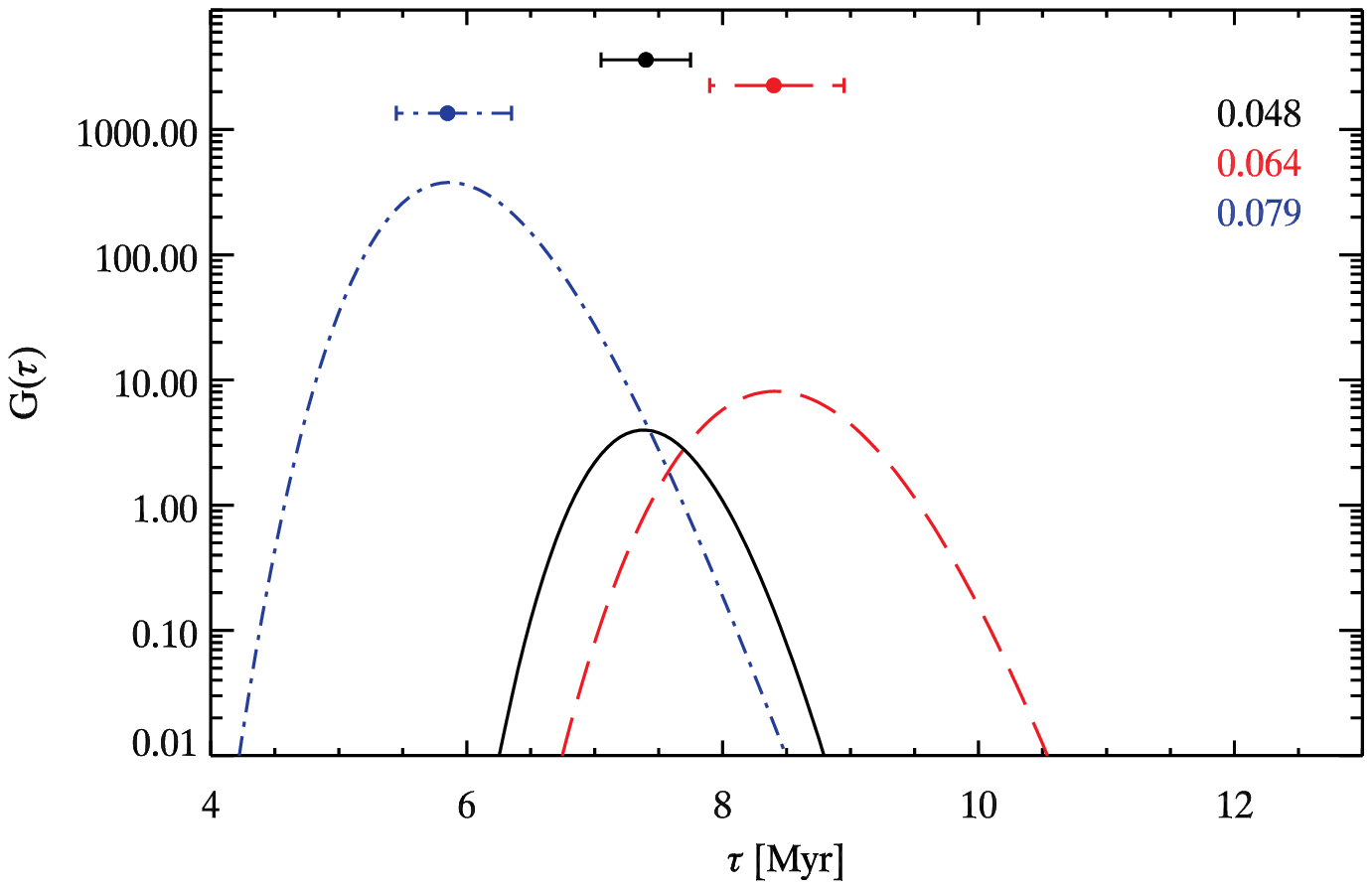}
	\caption{V 1174 Ori components mass and age distribution functions from comparison with the standard set of models (see Fig.~\ref{fig:RSCha_GH} for a description). In this case the temperature of the primary has been artificially raised by 100 K and the temperature of the secondary has been raised accordingly in order to keep the temperature ratio constant.}
 \label{fig:V1174Ori_mod}
\end{figure*}

\begin{figure*}
 \centering
	\includegraphics[width=0.48\hsize,height=5cm]{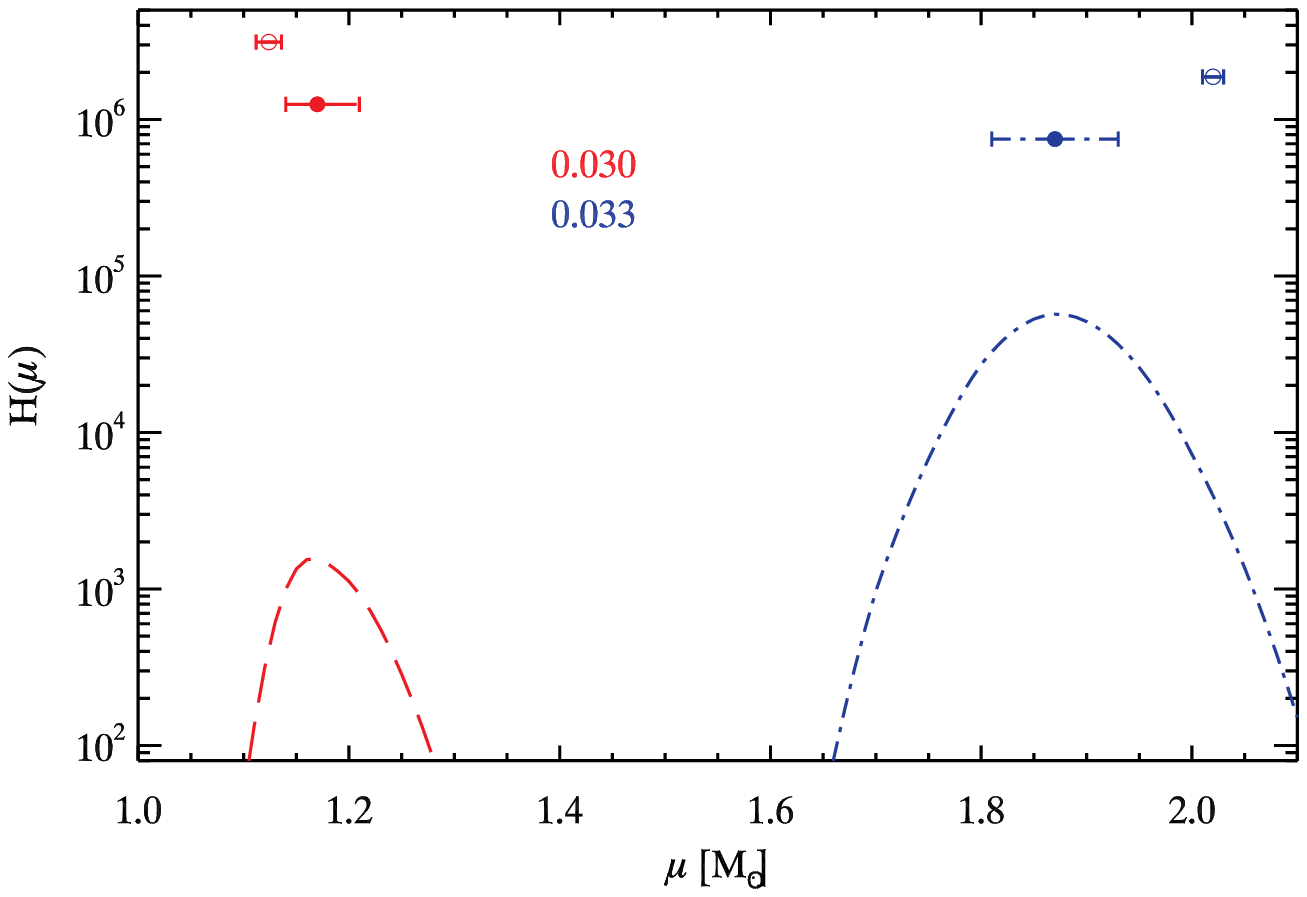}
	\includegraphics[width=0.48\hsize,height=5cm]{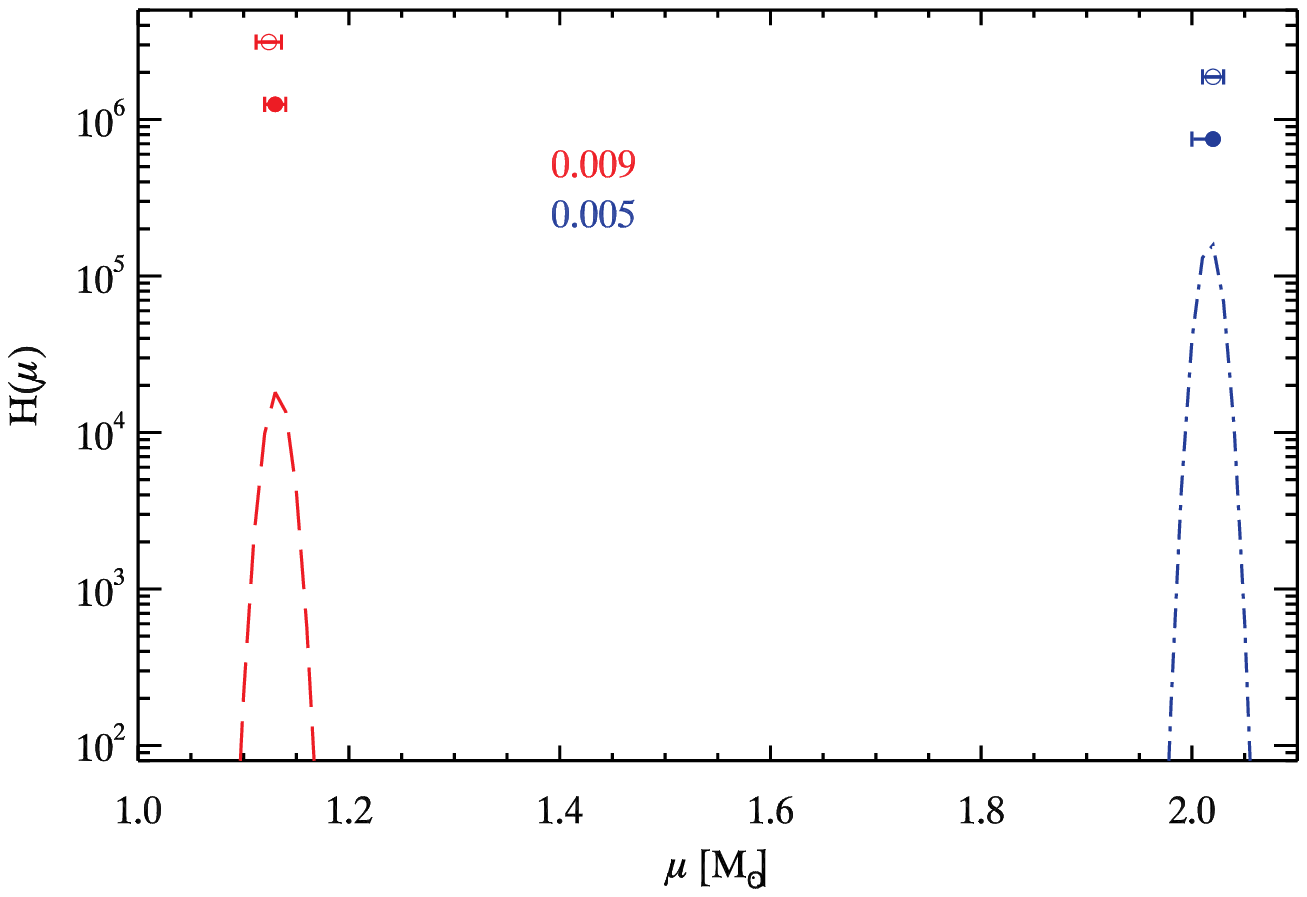}
	\includegraphics[width=0.48\hsize,height=5cm]{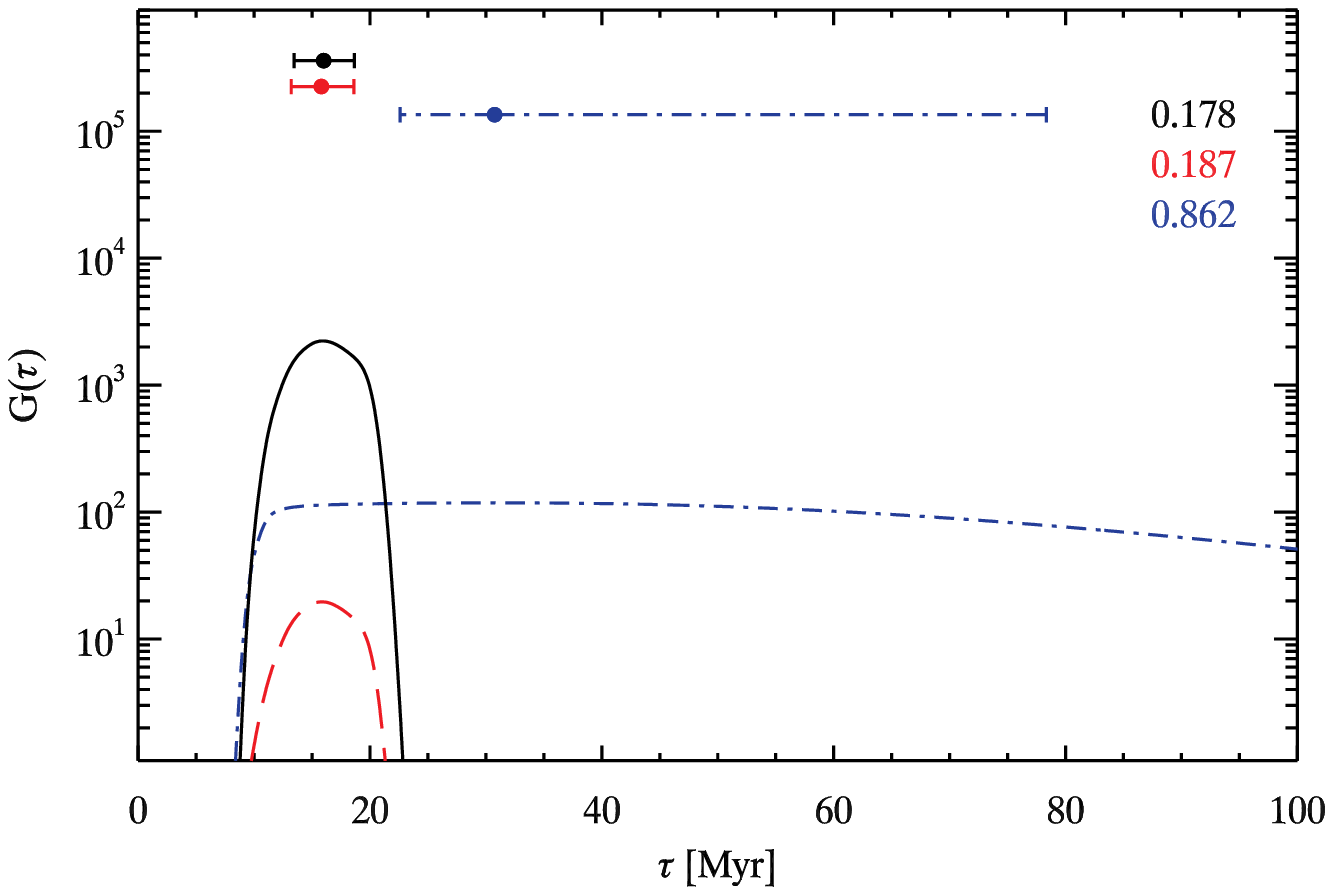}
	\includegraphics[width=0.48\hsize,height=5cm]{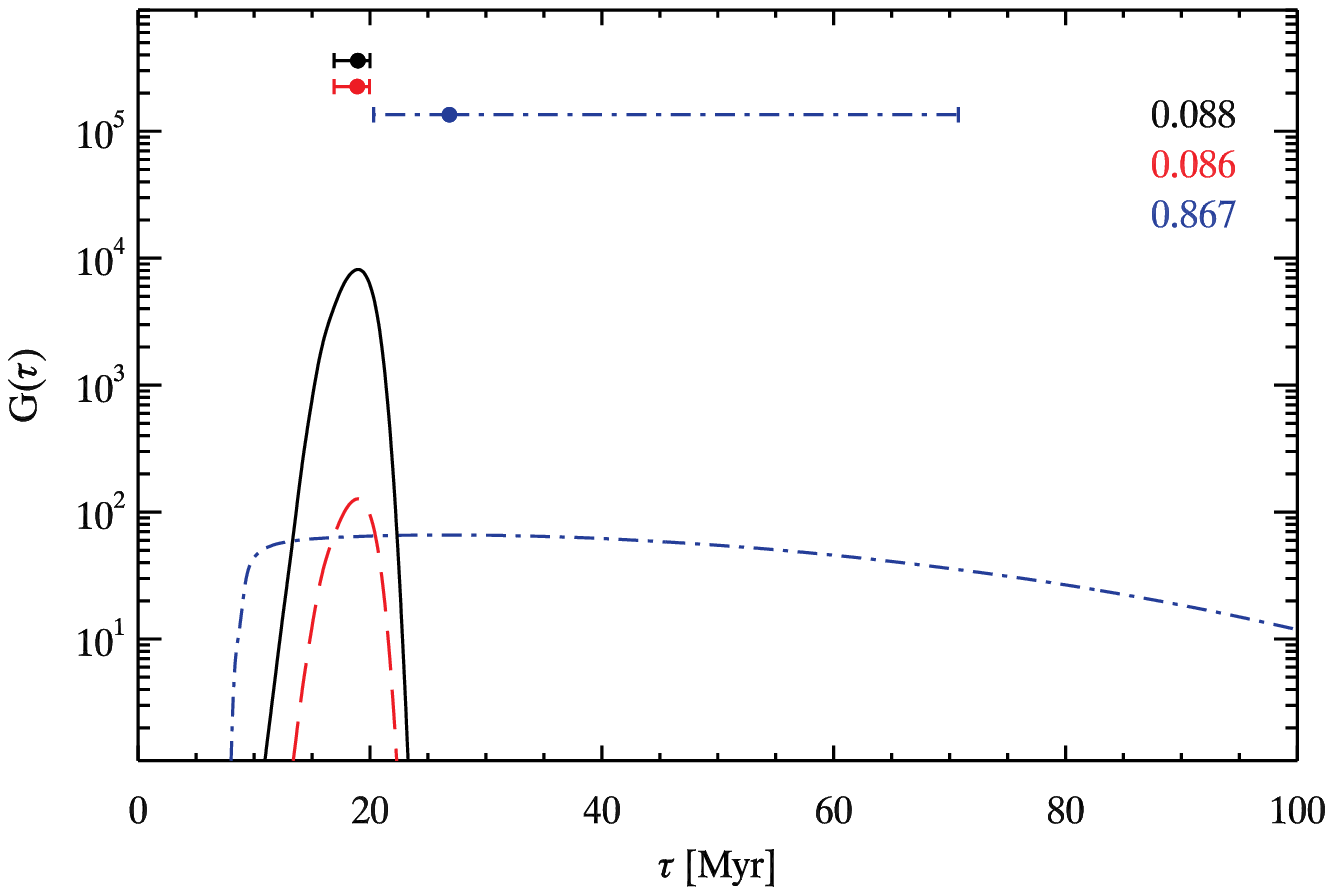}
	\caption{EK Cep components mass and age distribution functions from comparison with the standard set of models (see Fig.~\ref{fig:RSCha_GH} for a description).}
 \label{fig:EKCep}
\end{figure*}
\clearpage
\begin{figure*}
 \centering
	\includegraphics[width=0.48\hsize,height=5cm]{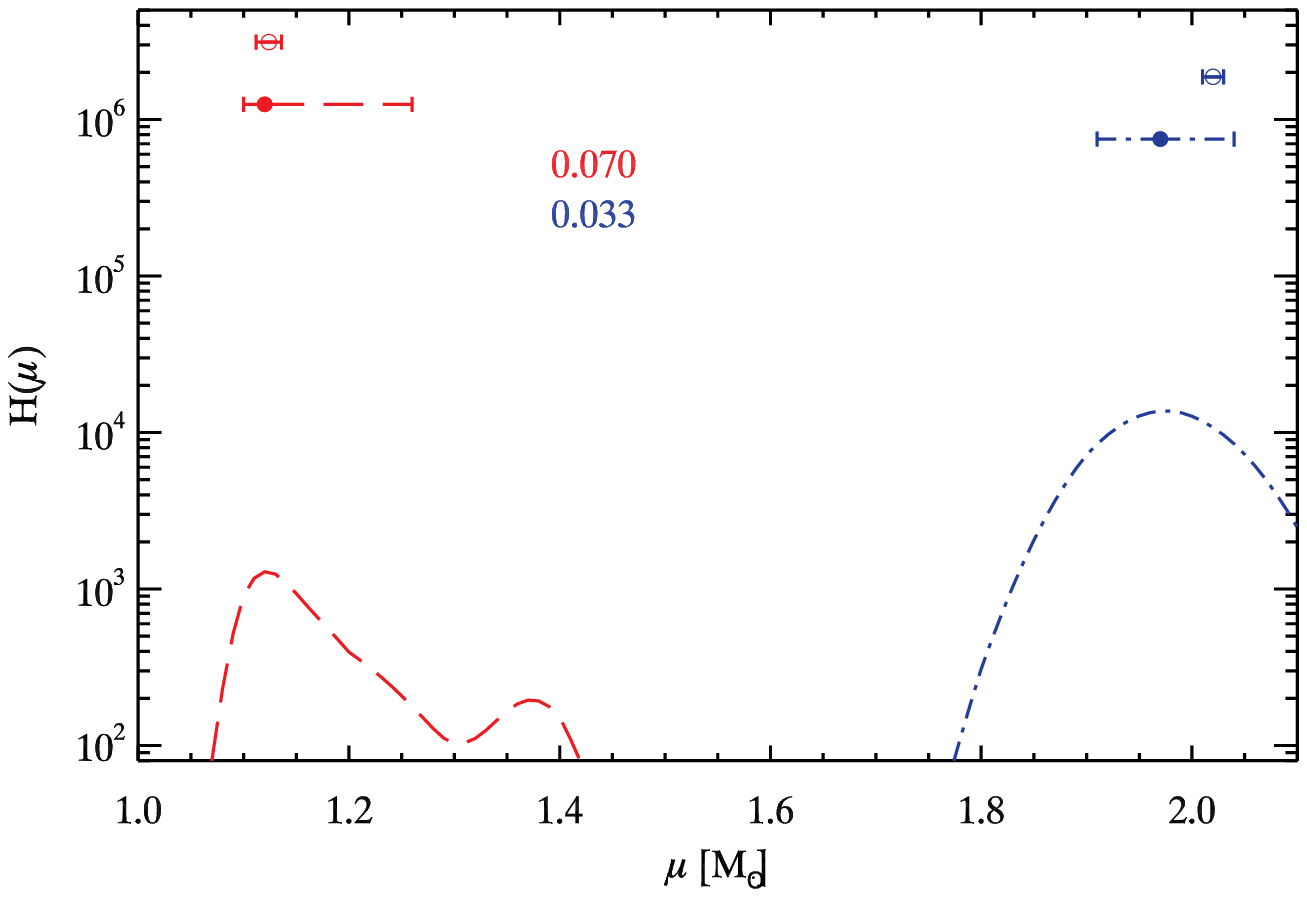}
	\includegraphics[width=0.48\hsize,height=5cm]{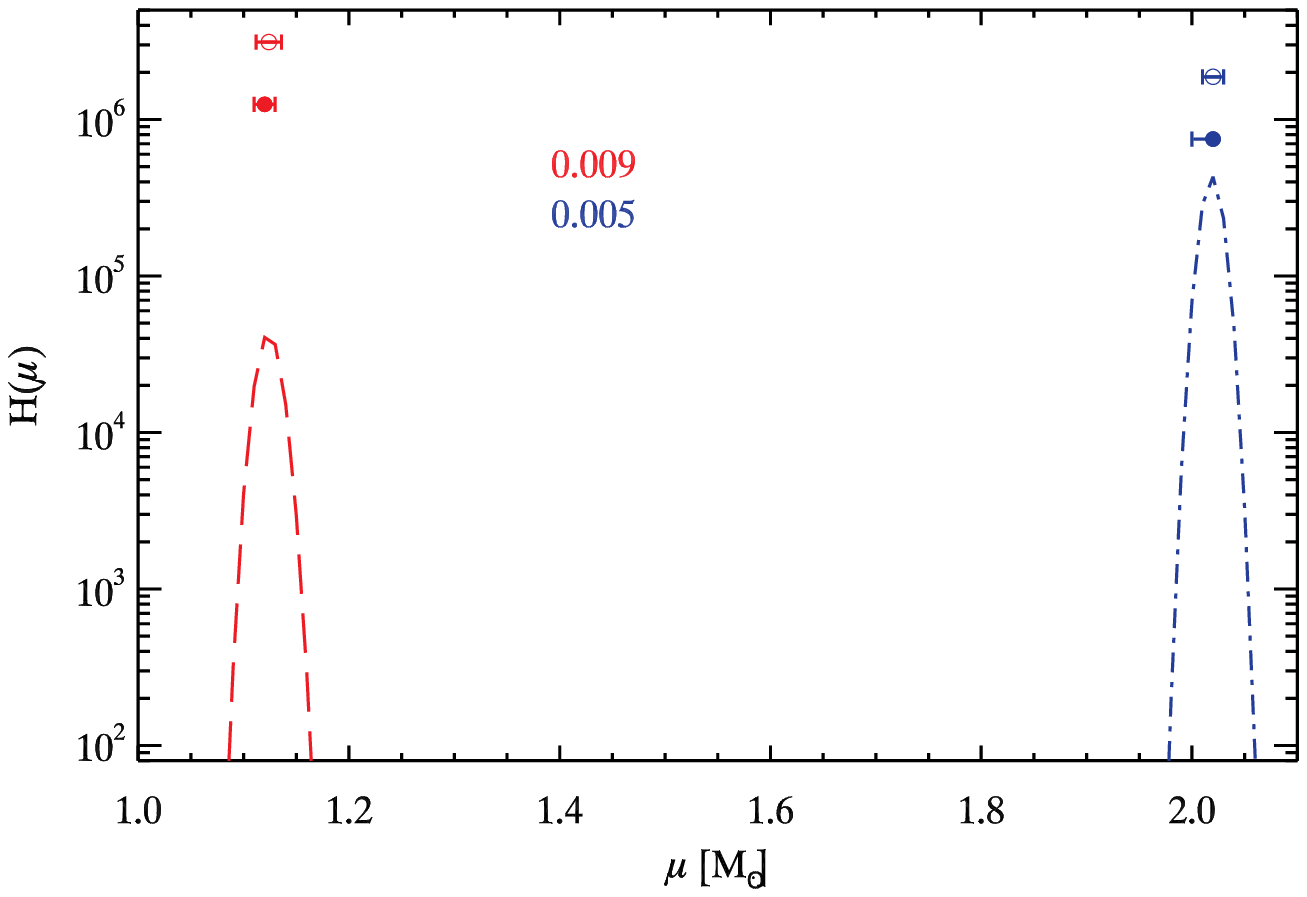}
	\includegraphics[width=0.48\hsize,height=5cm]{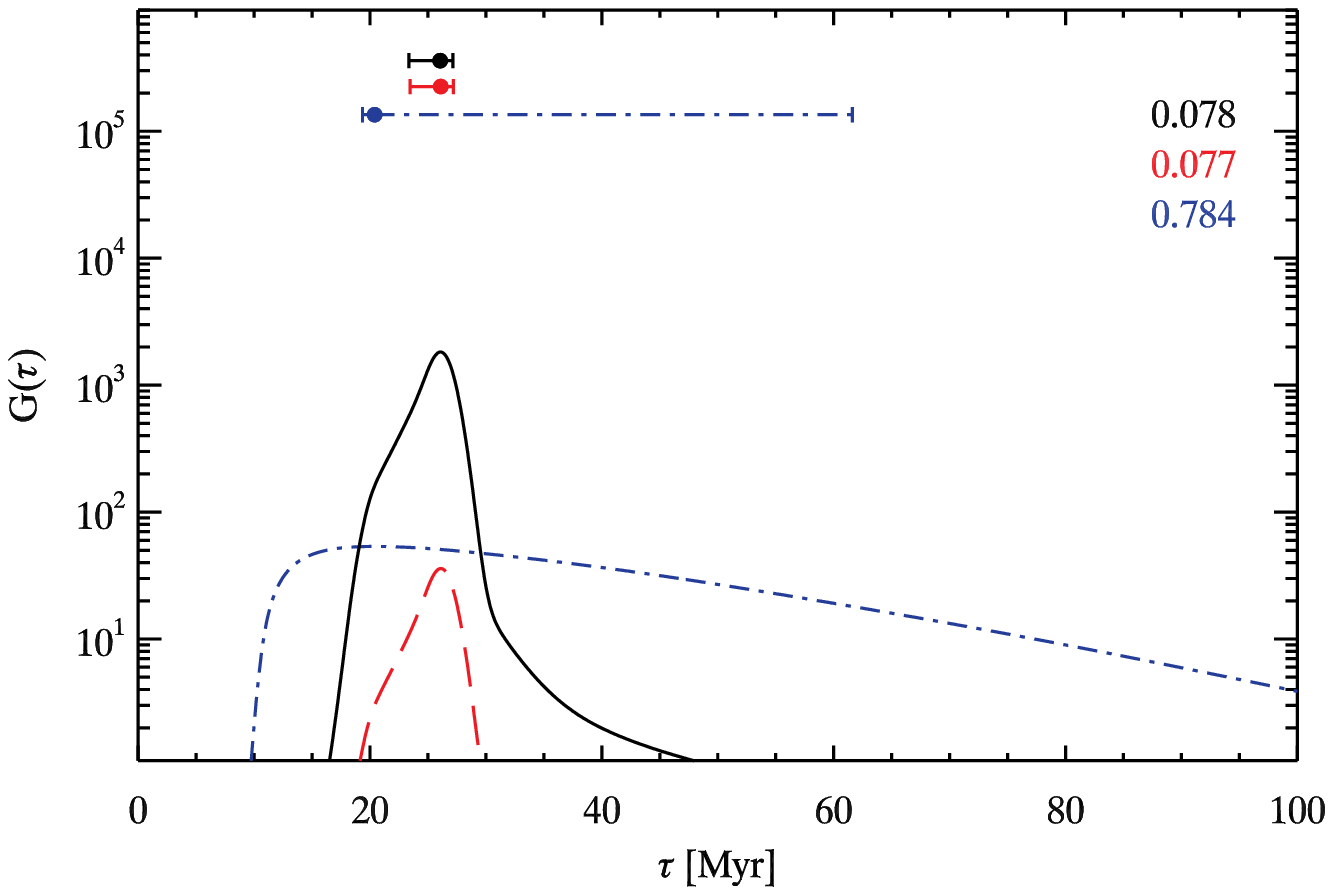}
	\includegraphics[width=0.48\hsize,height=5cm]{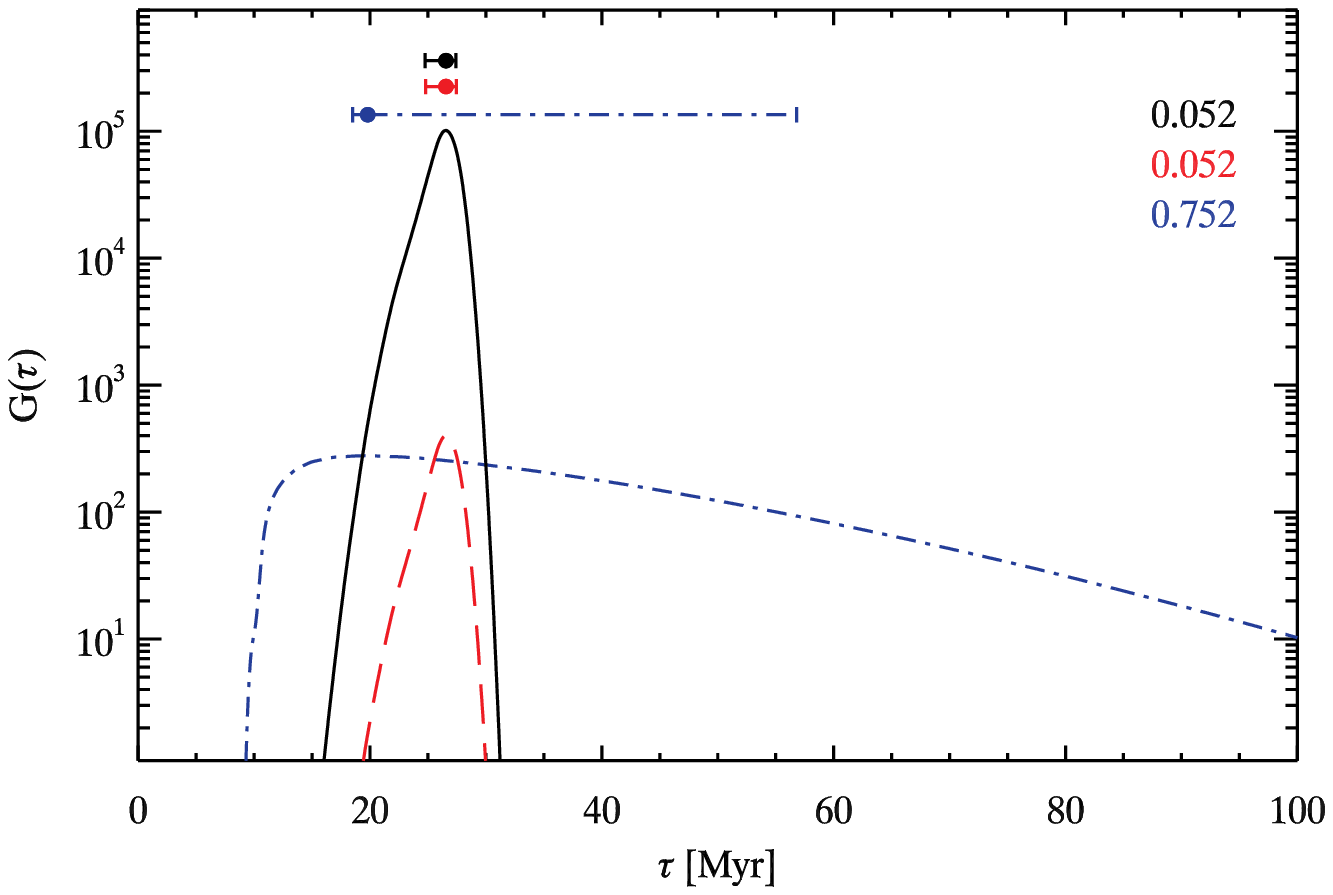}
	\caption{EK Cep components mass and age distribution functions from comparison with the set of models with $\alpha=1.20, Y_{\rmn{P}} = 0.23$ and $\Delta Y / \Delta Z =2$ (see Fig.~\ref{fig:RSCha_GH} for a description). The stellar $Z$ values used for this comparison have been calculated using the observed [Fe/H] and $(Z/X)_{\sun}$ by \citeauthor{1998SSRv...85..161G} \citeyear{1998SSRv...85..161G}.}
 \label{fig:EKCepNS}
\end{figure*}

\begin{figure*}
 \centering
	\includegraphics[width=0.48\hsize,height=5cm]{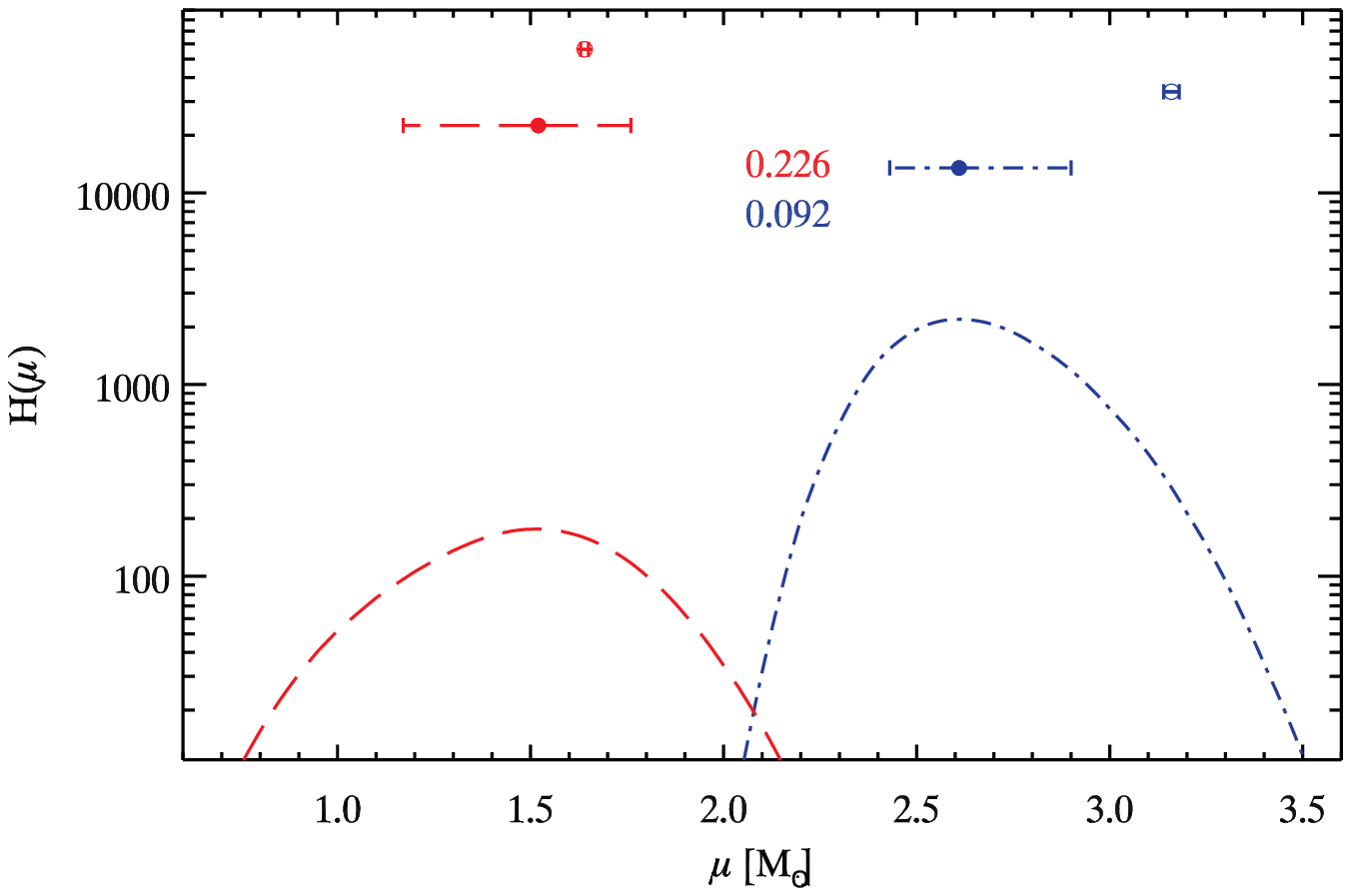}
	\includegraphics[width=0.48\hsize,height=5cm]{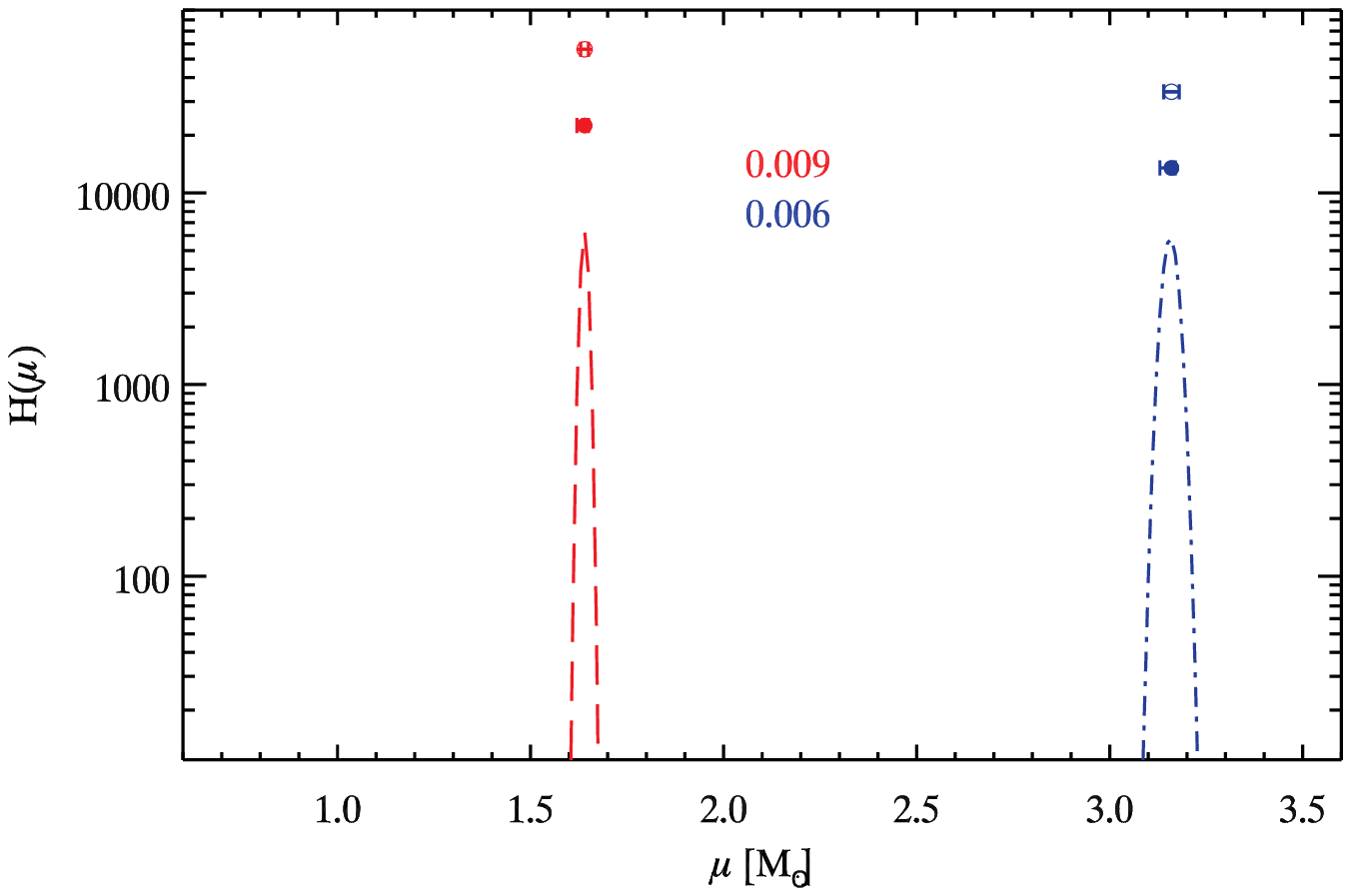}
	\includegraphics[width=0.48\hsize,height=5cm]{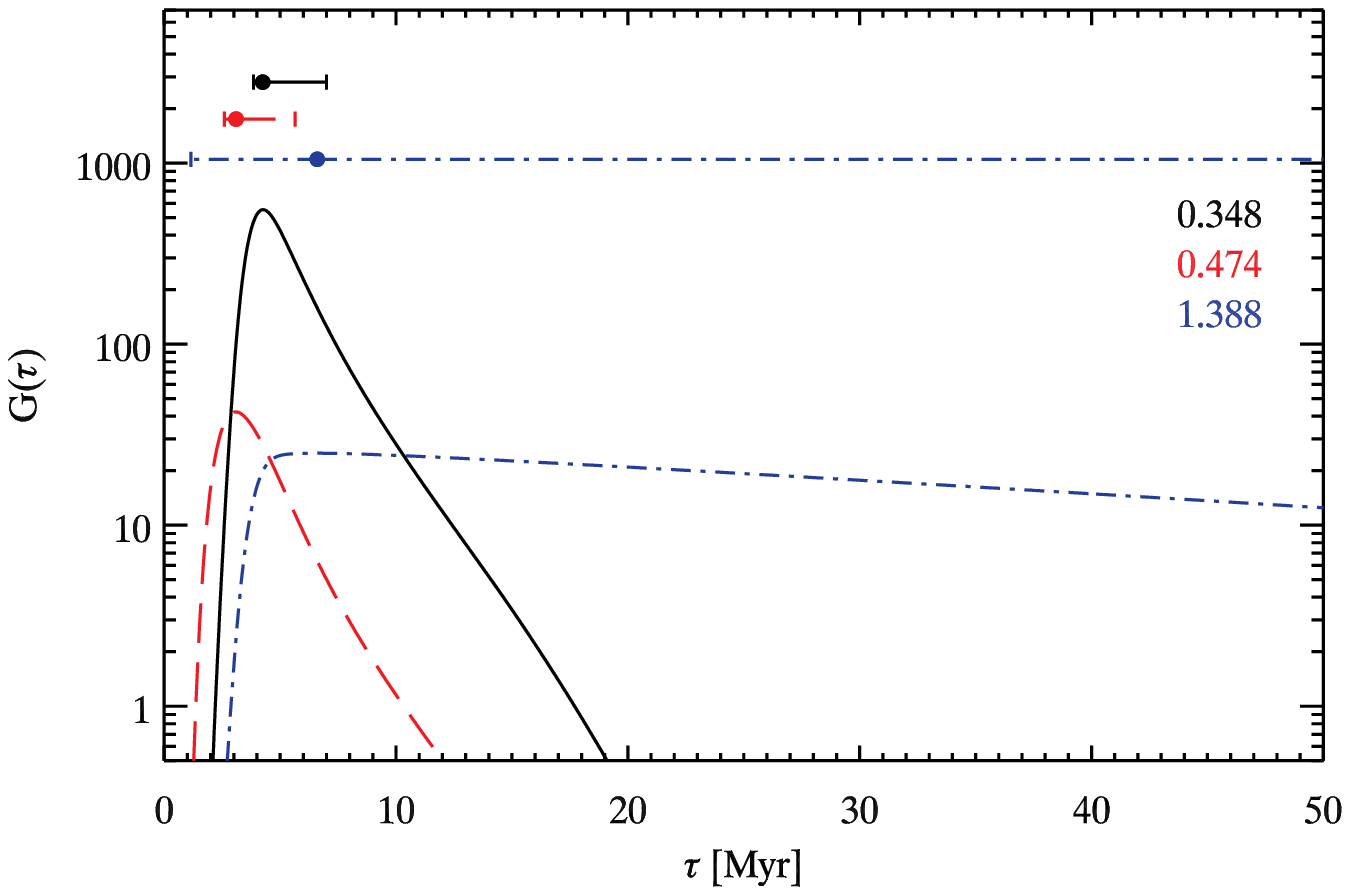}
	\includegraphics[width=0.48\hsize,height=5cm]{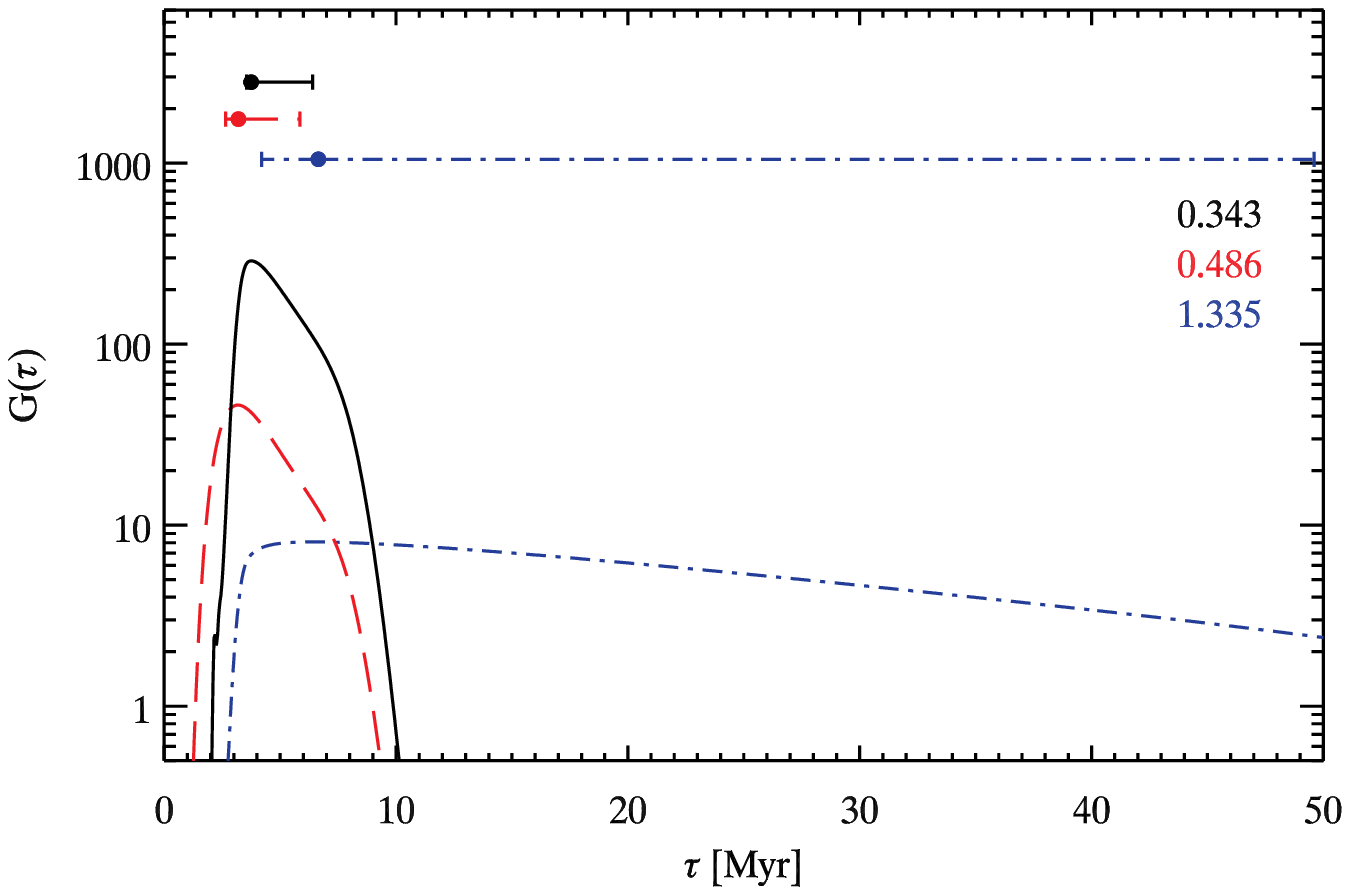}
	\caption{TYCrA components mass and age distribution functions from comparison with the standard set of models (see Fig.~\ref{fig:RSCha_GH} for a description).}
 \label{fig:TYCrA}
\end{figure*}
\clearpage
\begin{figure*}
 \centering
	\includegraphics[width=0.48\hsize,height=5cm]{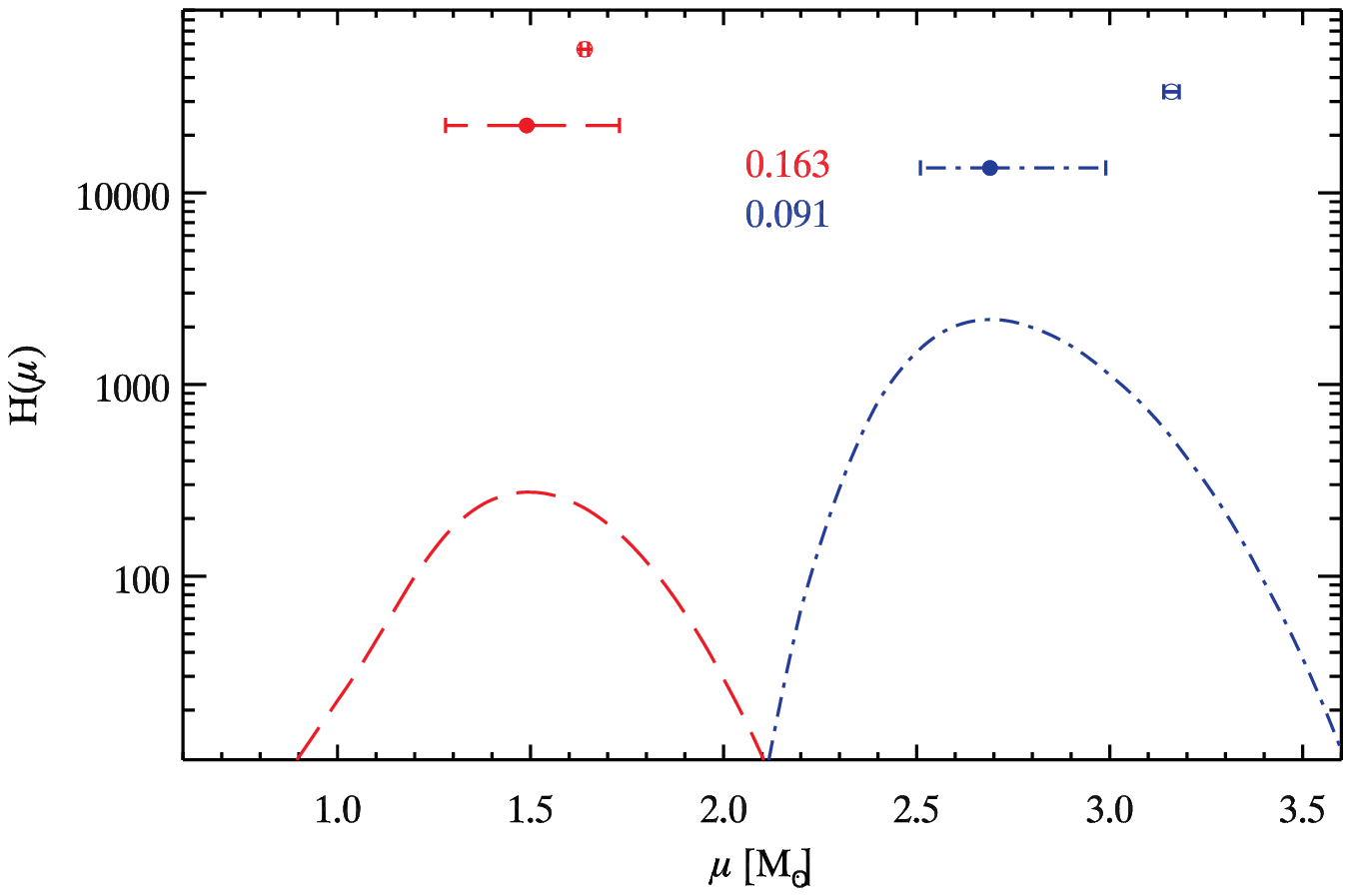}
	\includegraphics[width=0.48\hsize,height=5cm]{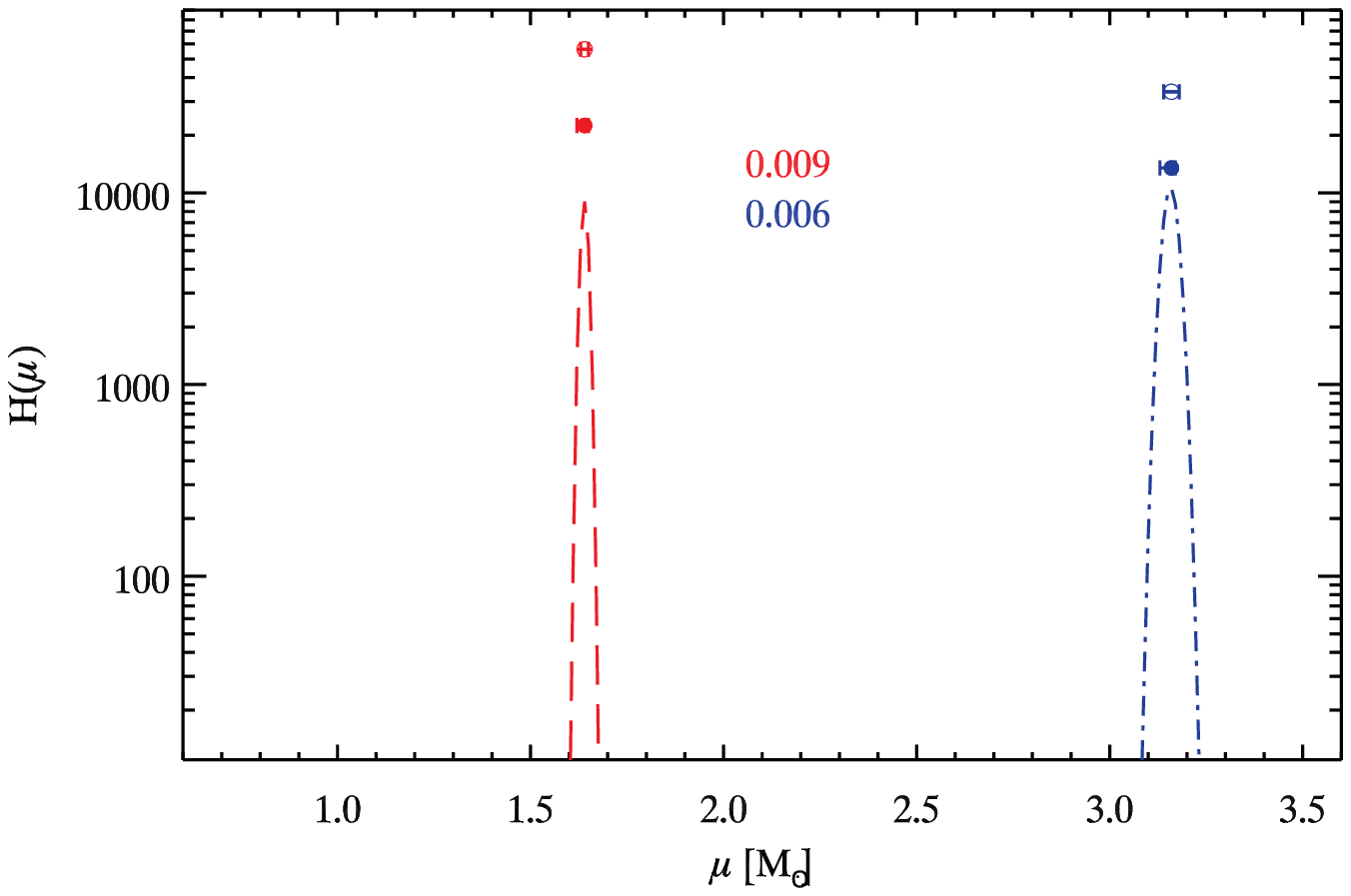}
	\includegraphics[width=0.48\hsize,height=5cm]{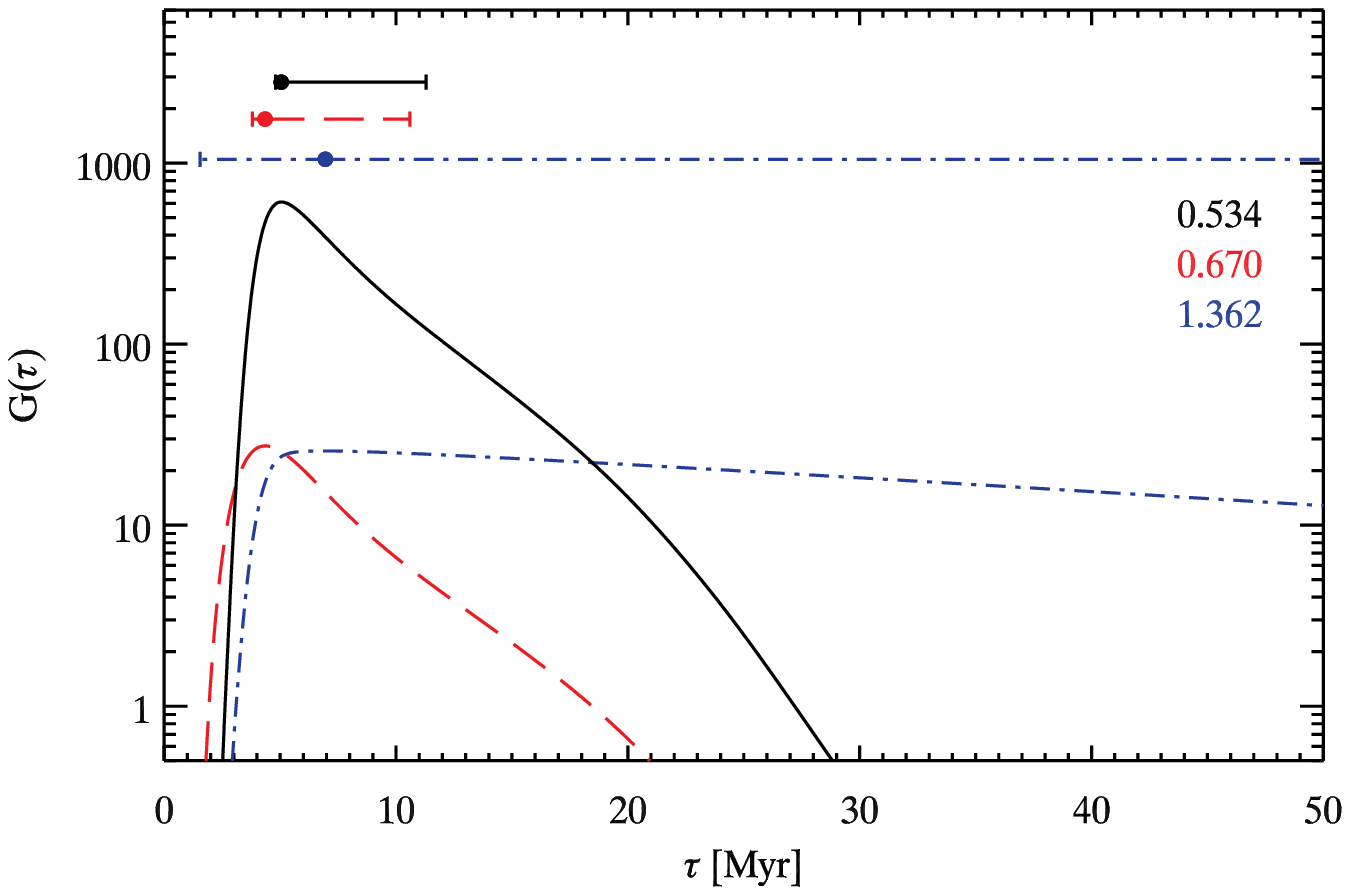}
	\includegraphics[width=0.48\hsize,height=5cm]{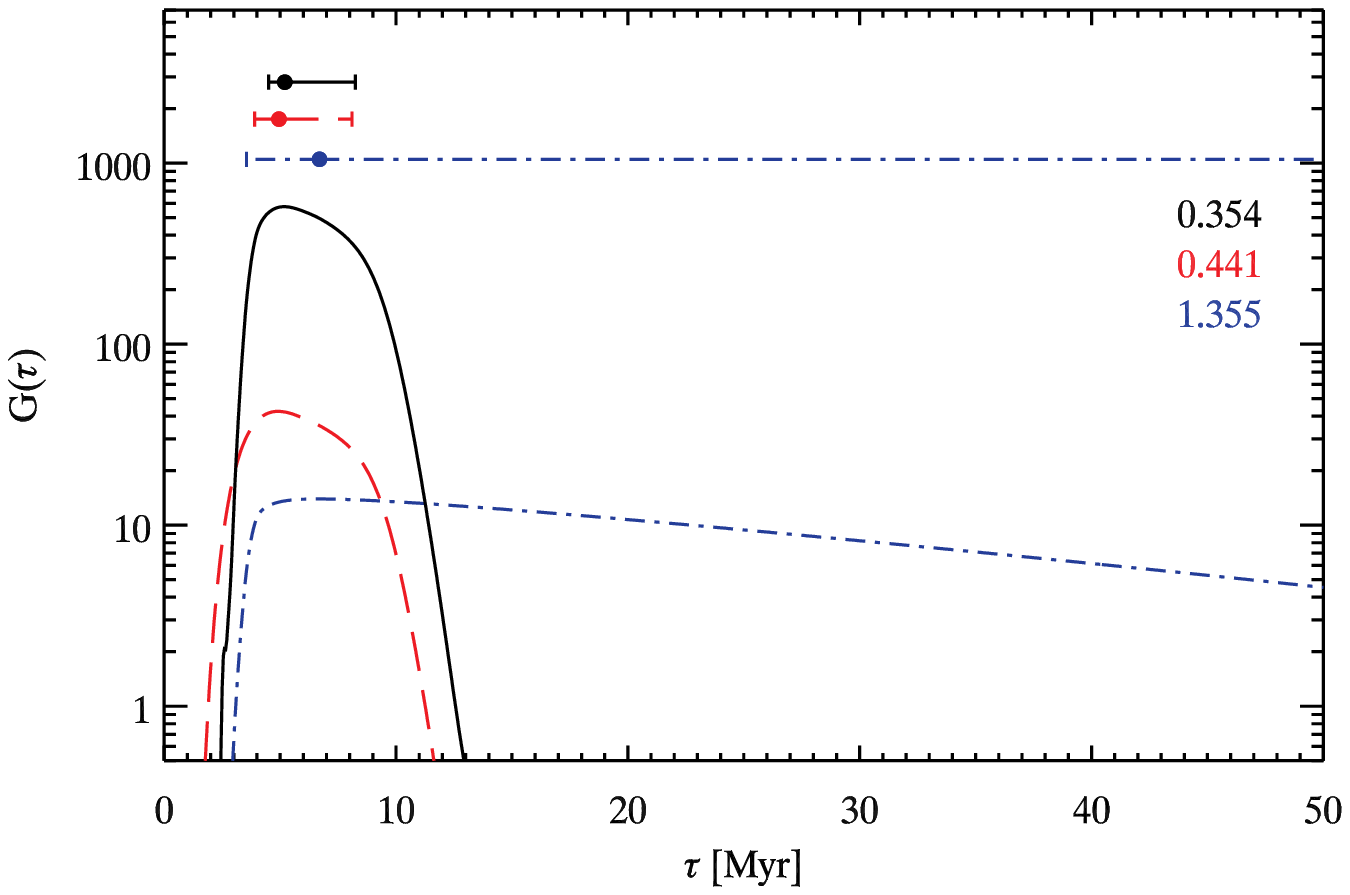}
	\caption{TY CrA components mass and age distribution functions from comparison with the set of models with $\alpha=1.20, Y_{\rmn{P}} = 0.23$ and $\Delta Y / \Delta Z =2$ (see Fig.~\ref{fig:RSCha_GH} for a description).}
 \label{fig:TYCrANS}
\end{figure*}

\begin{figure*}
 \centering
	\includegraphics[width=0.48\hsize,height=5cm]{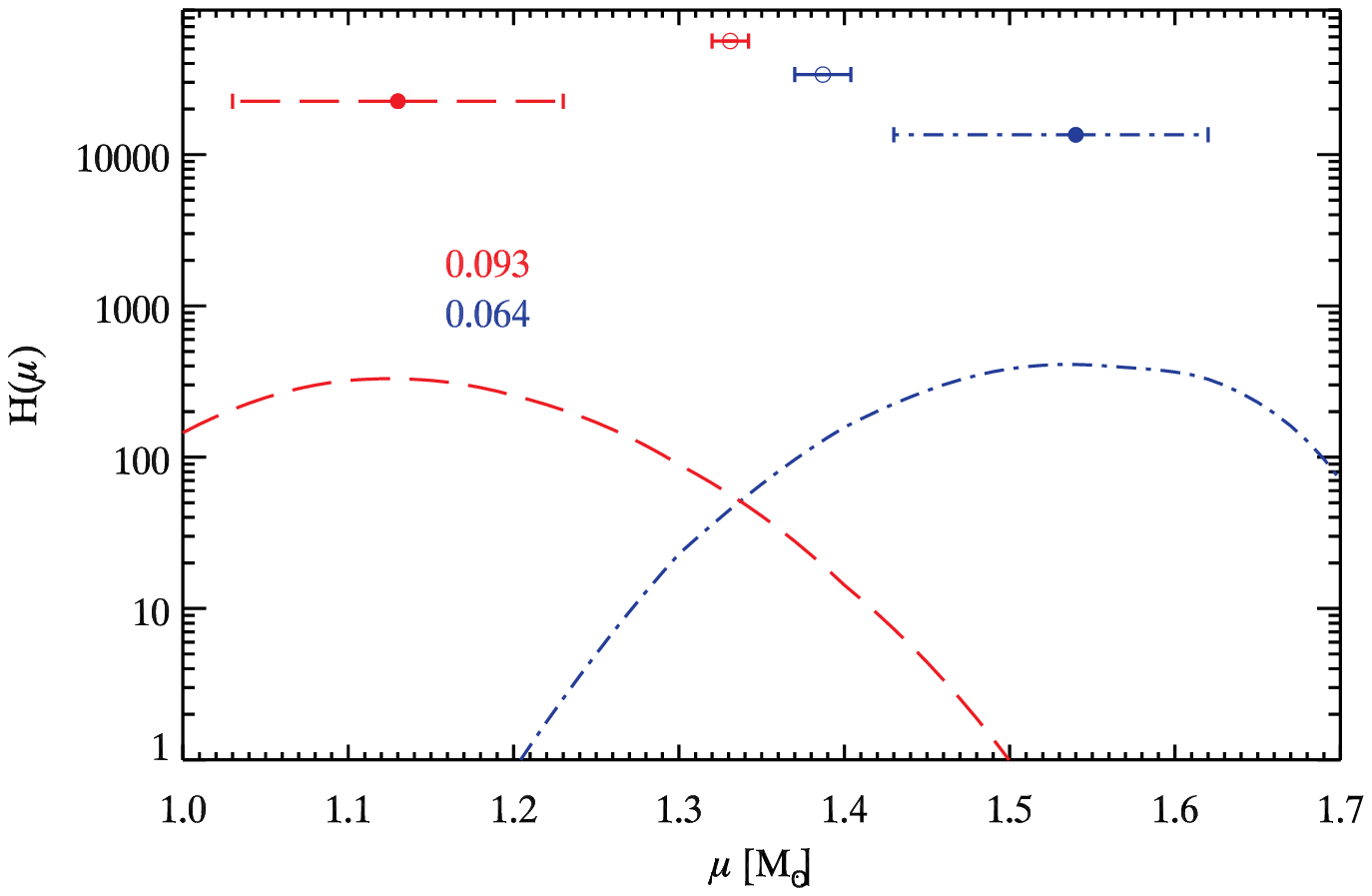}
	\includegraphics[width=0.48\hsize,height=5cm]{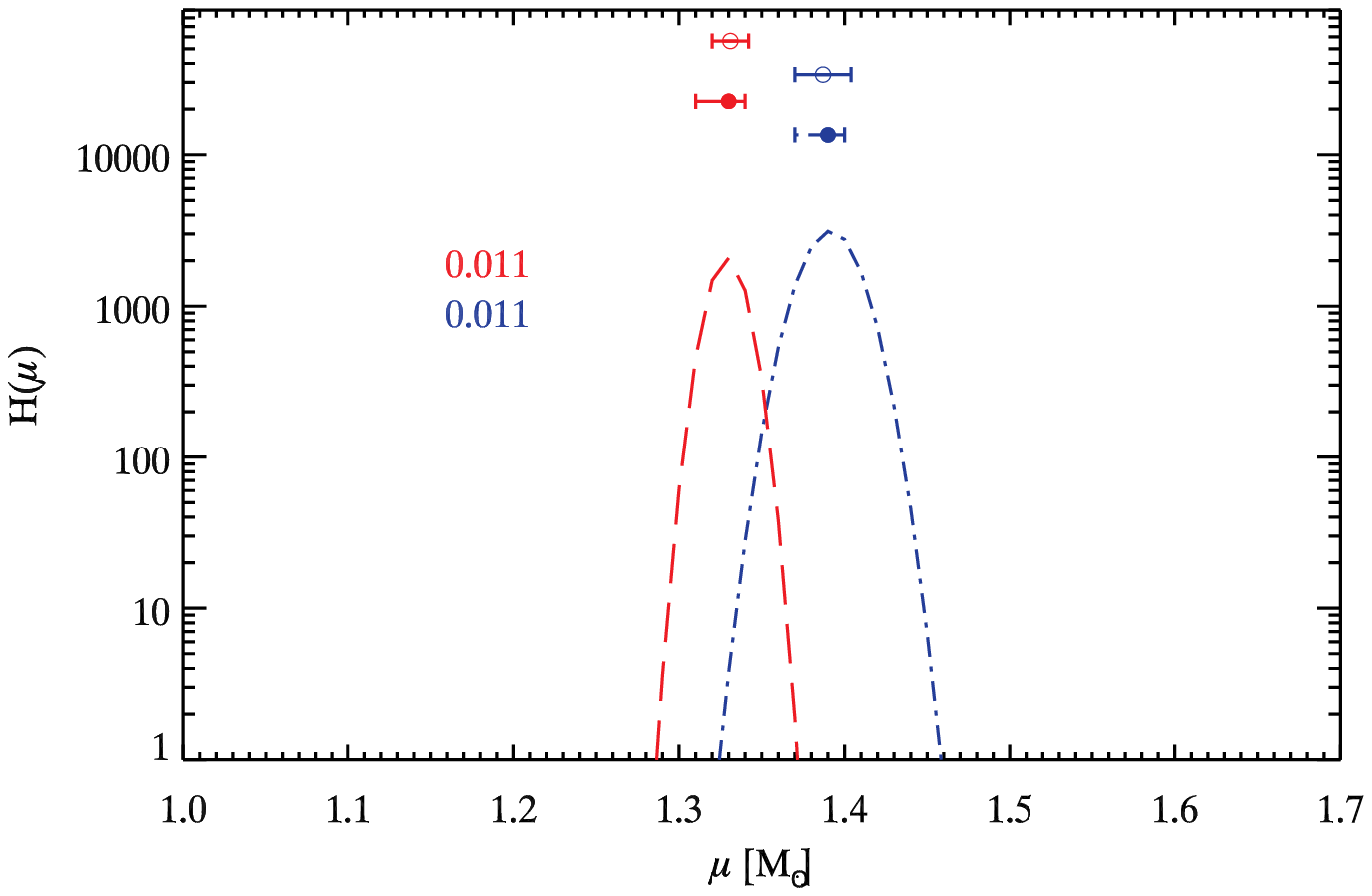}
	\includegraphics[width=0.48\hsize,height=5cm]{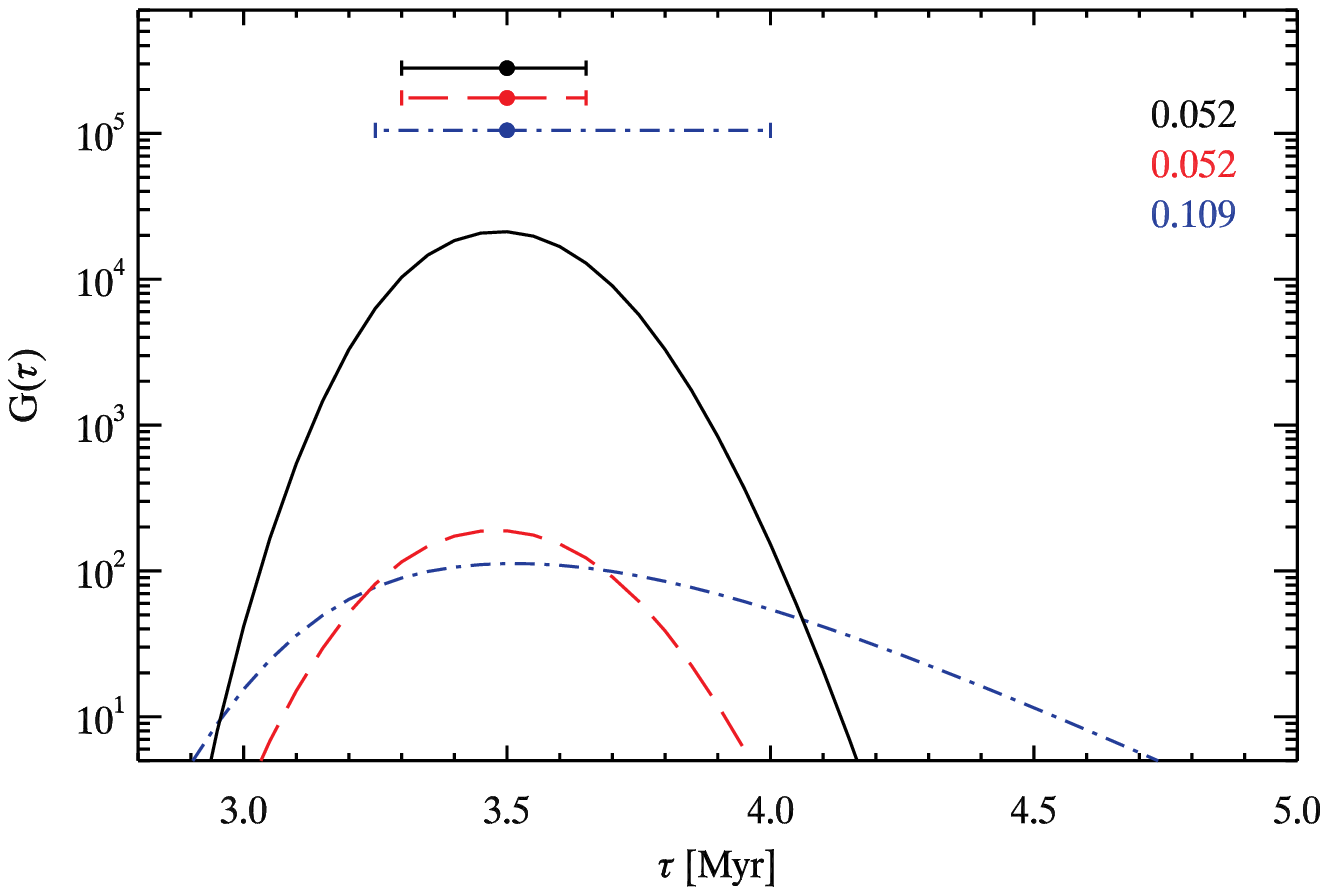}
	\includegraphics[width=0.48\hsize,height=5cm]{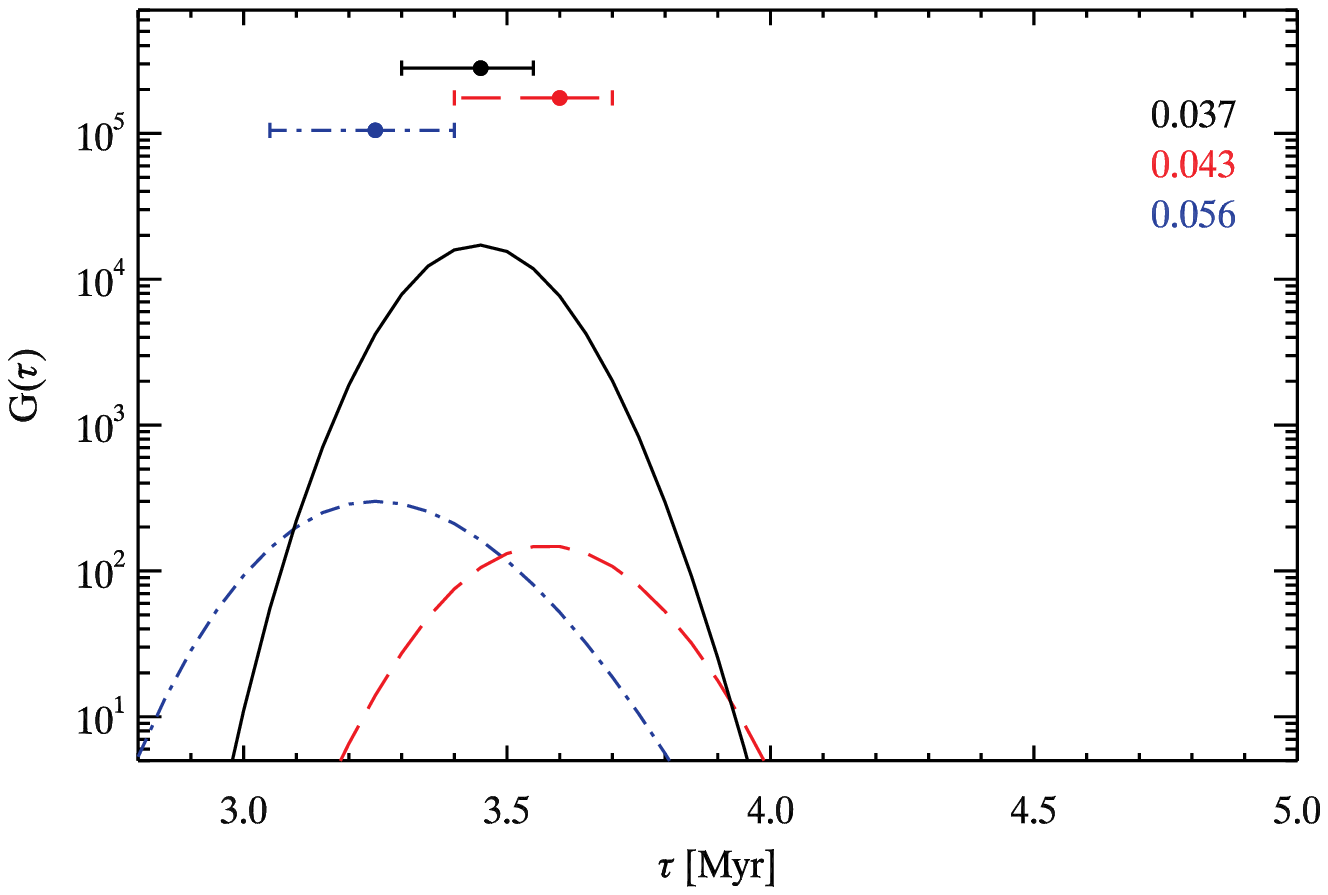}
	\caption{ASAS J052821+0338.5 components mass and age distribution functions from comparison with the standard set of models (see Fig.~\ref{fig:RSCha_GH} for a description).}
 \label{fig:ASAS}
\end{figure*}

\clearpage

\begin{figure*}
 \centering
	\includegraphics[width=0.48\hsize,height=5cm]{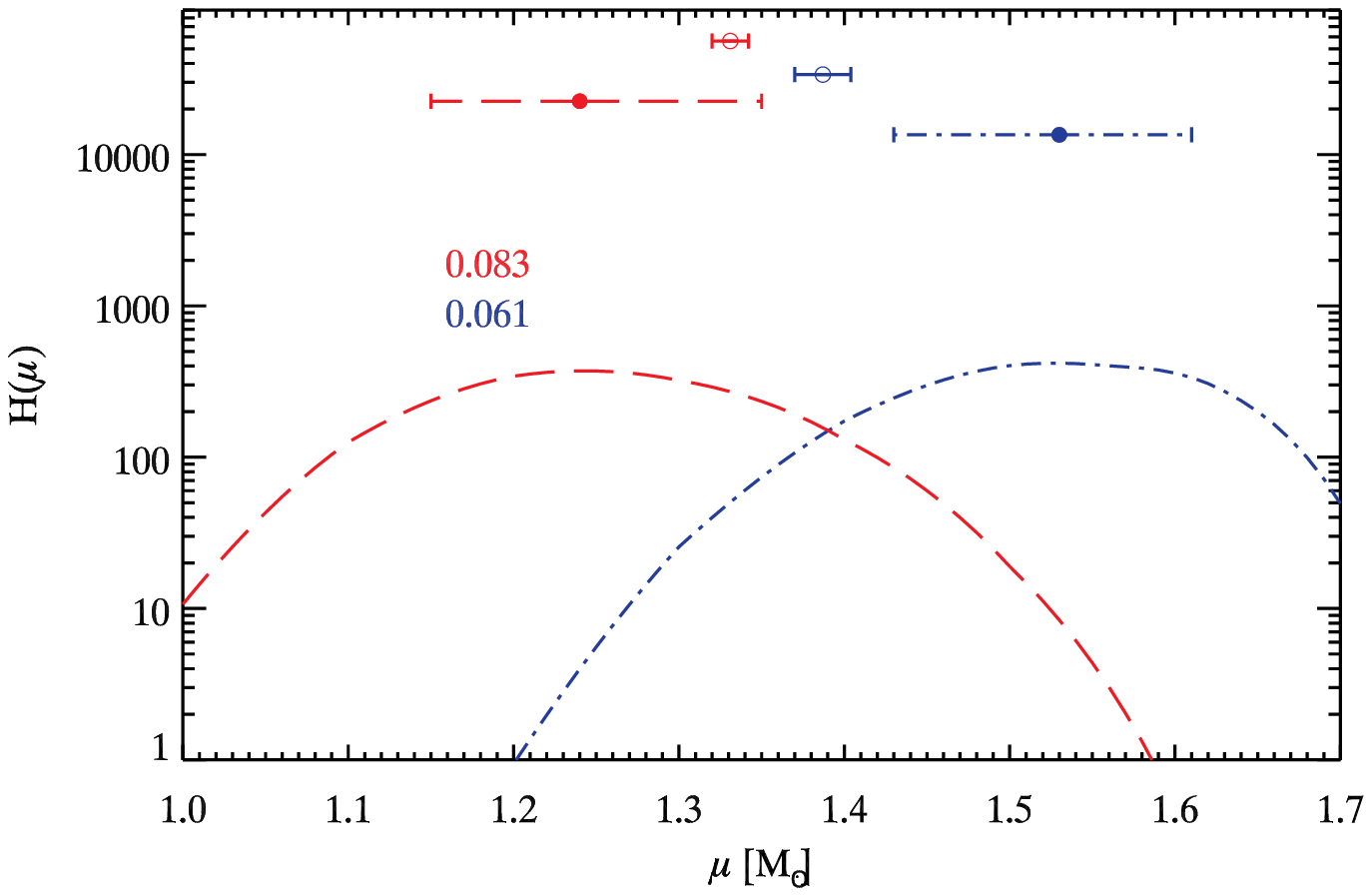}
	\includegraphics[width=0.48\hsize,height=5cm]{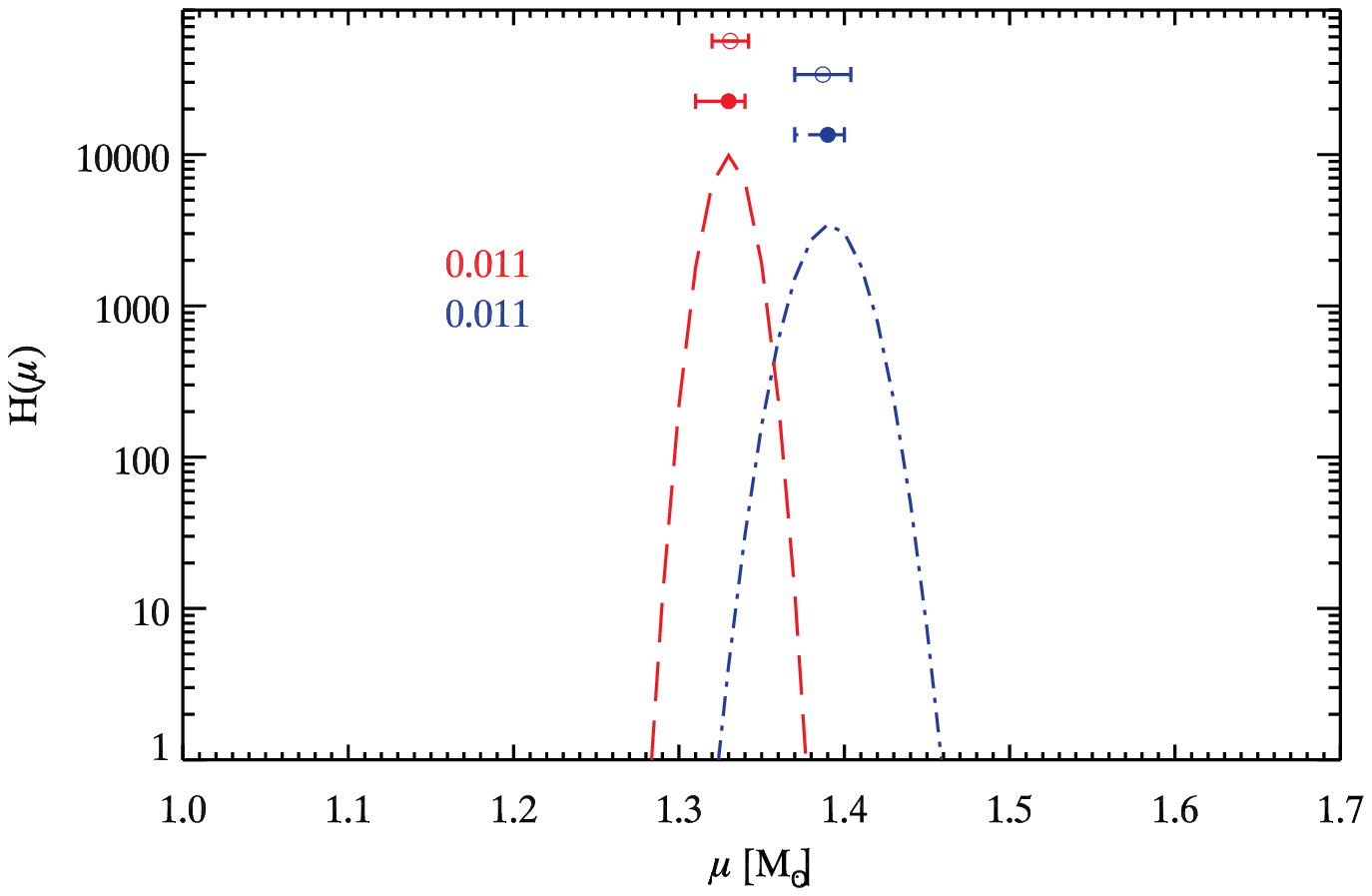}
	\includegraphics[width=0.48\hsize,height=5cm]{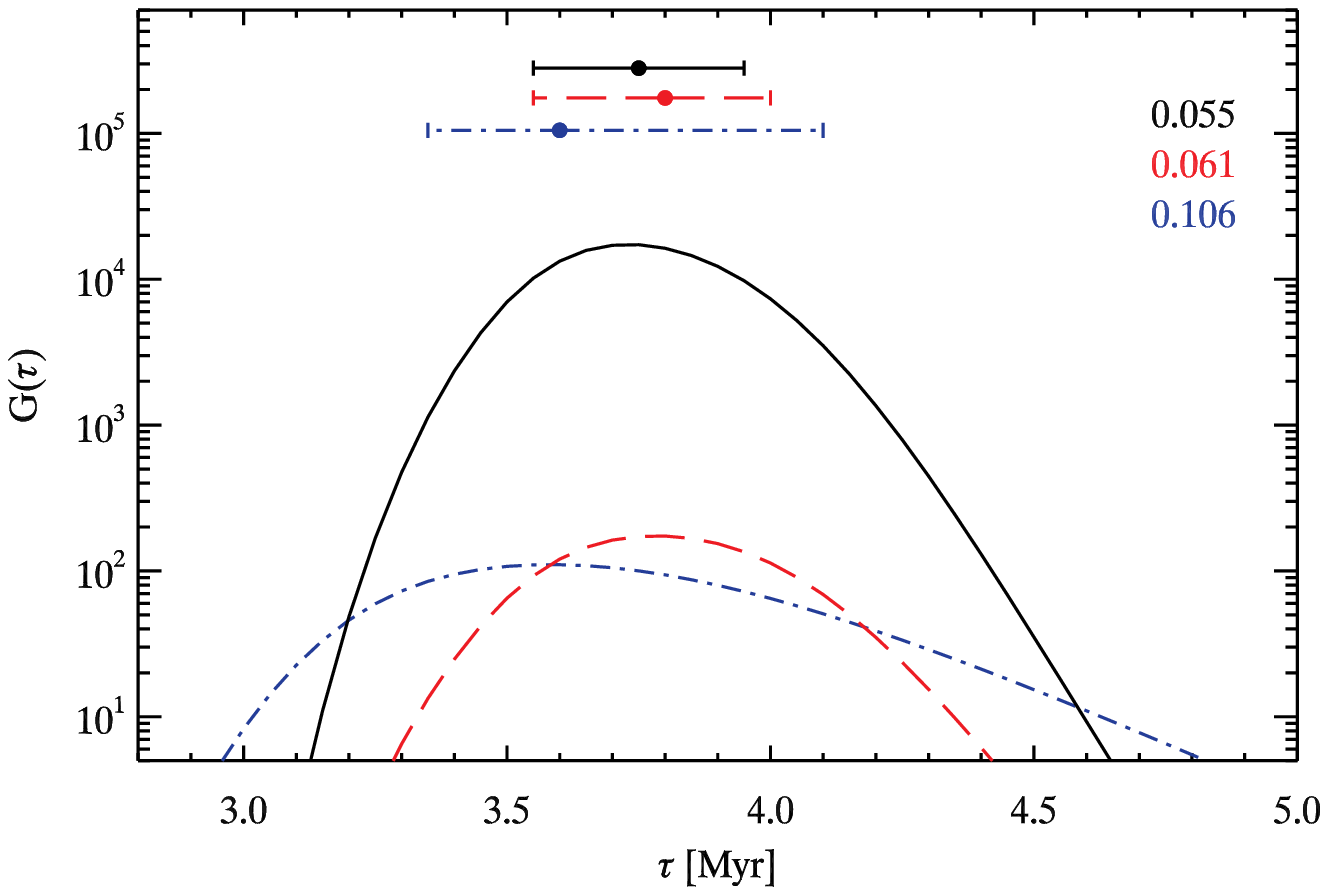}
	\includegraphics[width=0.48\hsize,height=5cm]{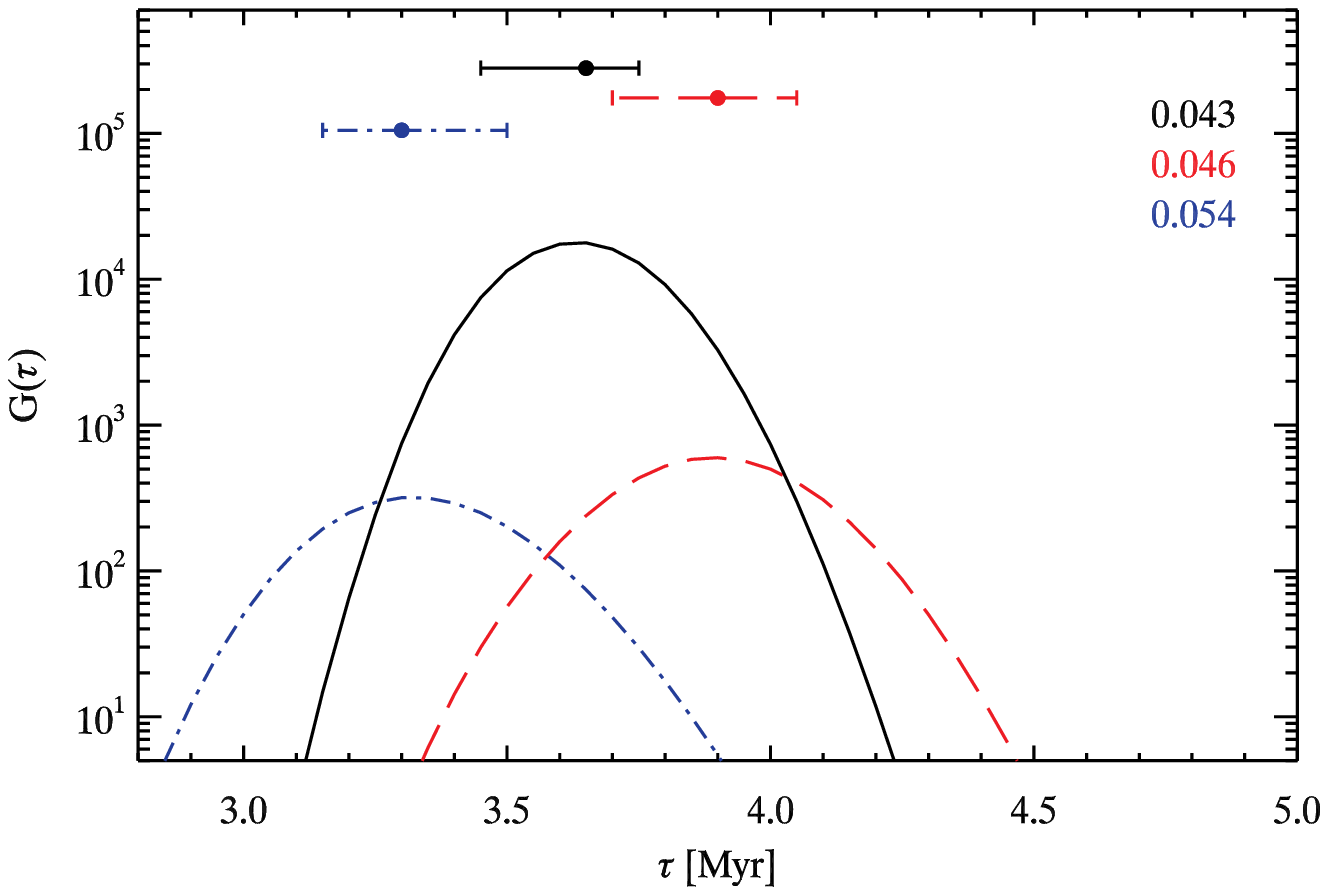}
	\caption{ASAS J052821+0338.5 components mass and age distribution functions from comparison with the standard set of models. In this case we used the data from Table 1 of \citeauthor{2008A&A...481..747S} \citeyear{2008A&A...481..747S} when no spots are included in the light-curve modeling (see Fig.~\ref{fig:RSCha_GH} for a description).}
 \label{fig:ASAS_nospot}
\end{figure*}

\begin{figure*}
 \centering
	\includegraphics[width=0.48\hsize,height=5cm]{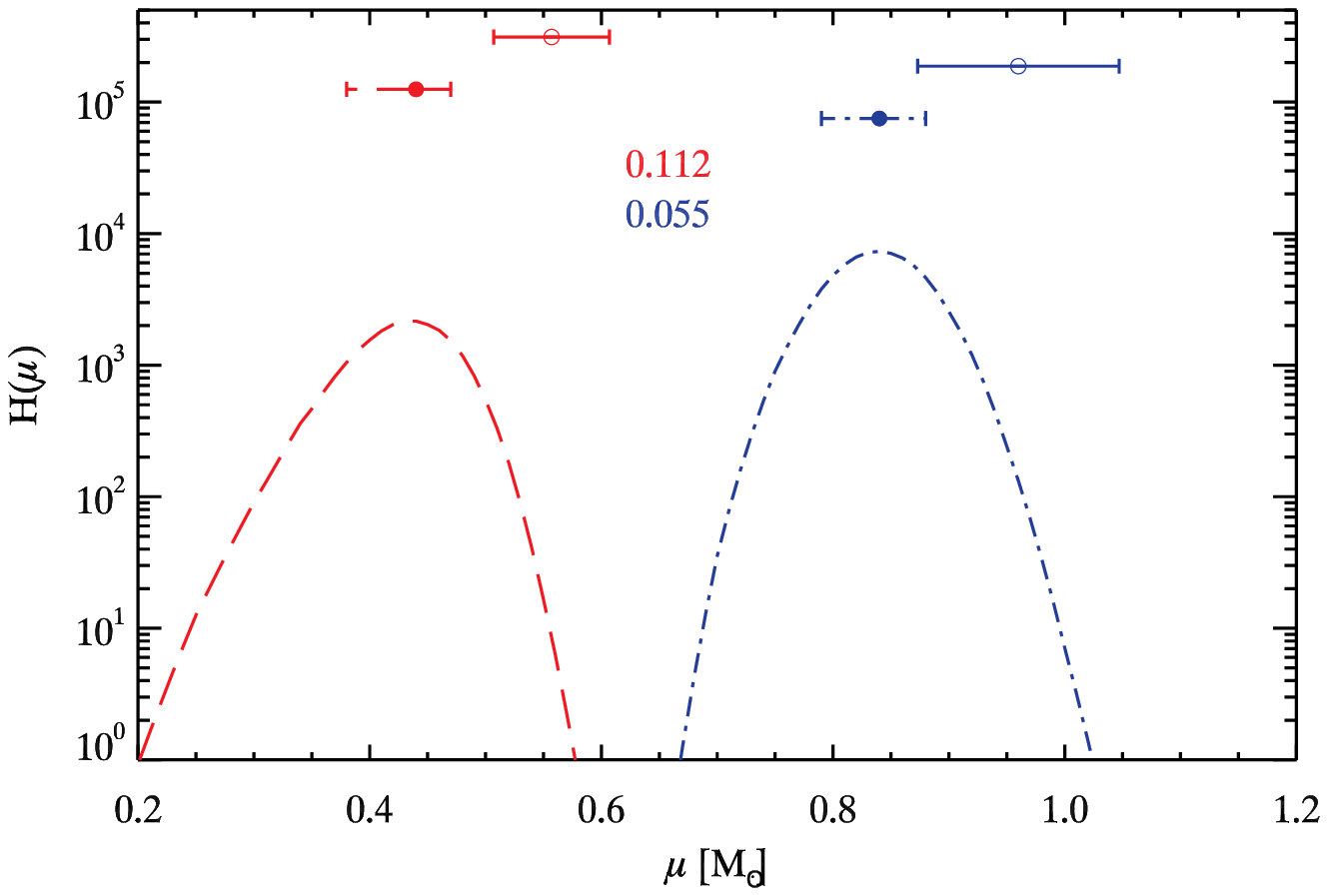}
	\includegraphics[width=0.48\hsize,height=5cm]{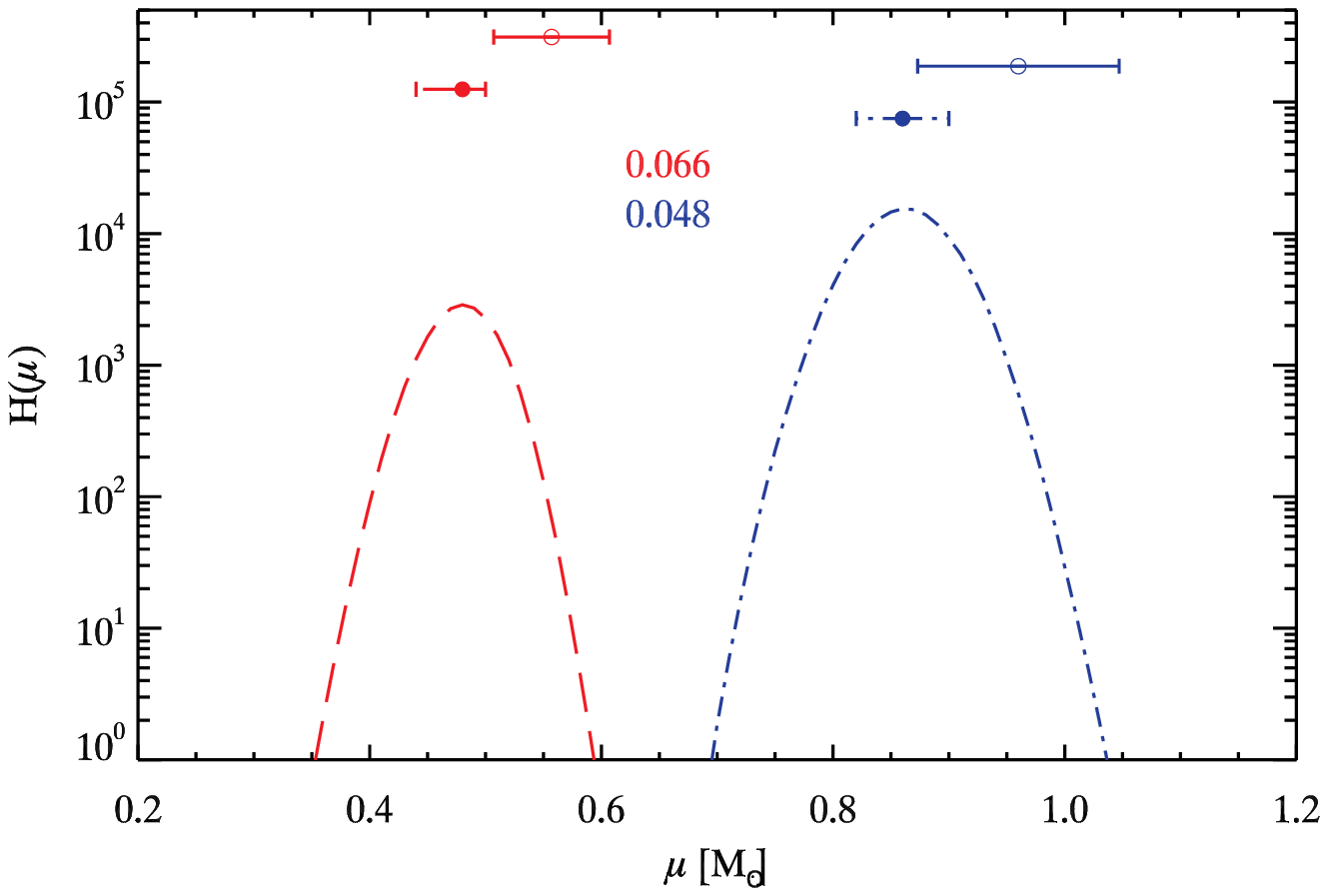}
	\includegraphics[width=0.48\hsize,height=5cm]{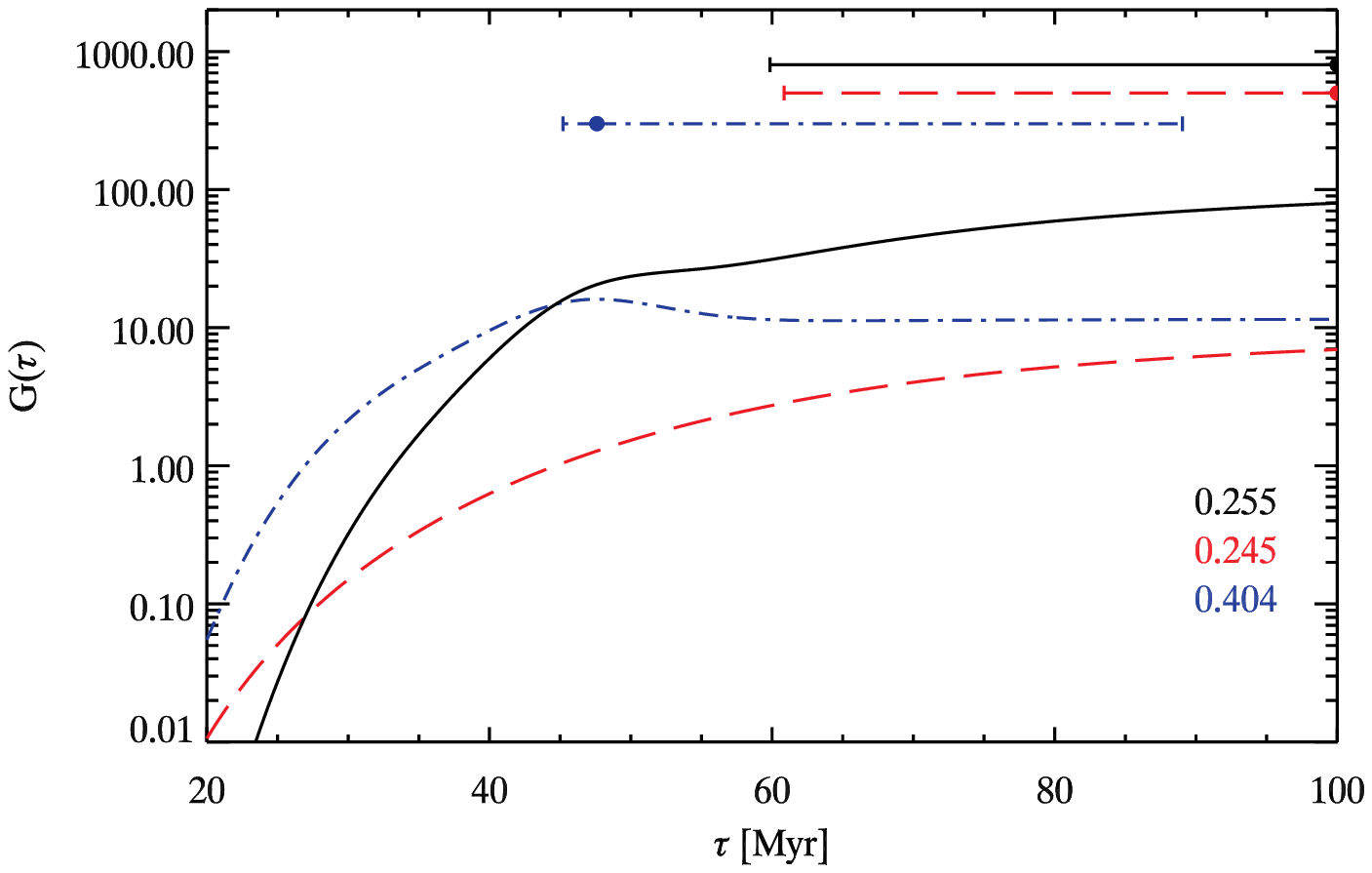}
	\includegraphics[width=0.48\hsize,height=5cm]{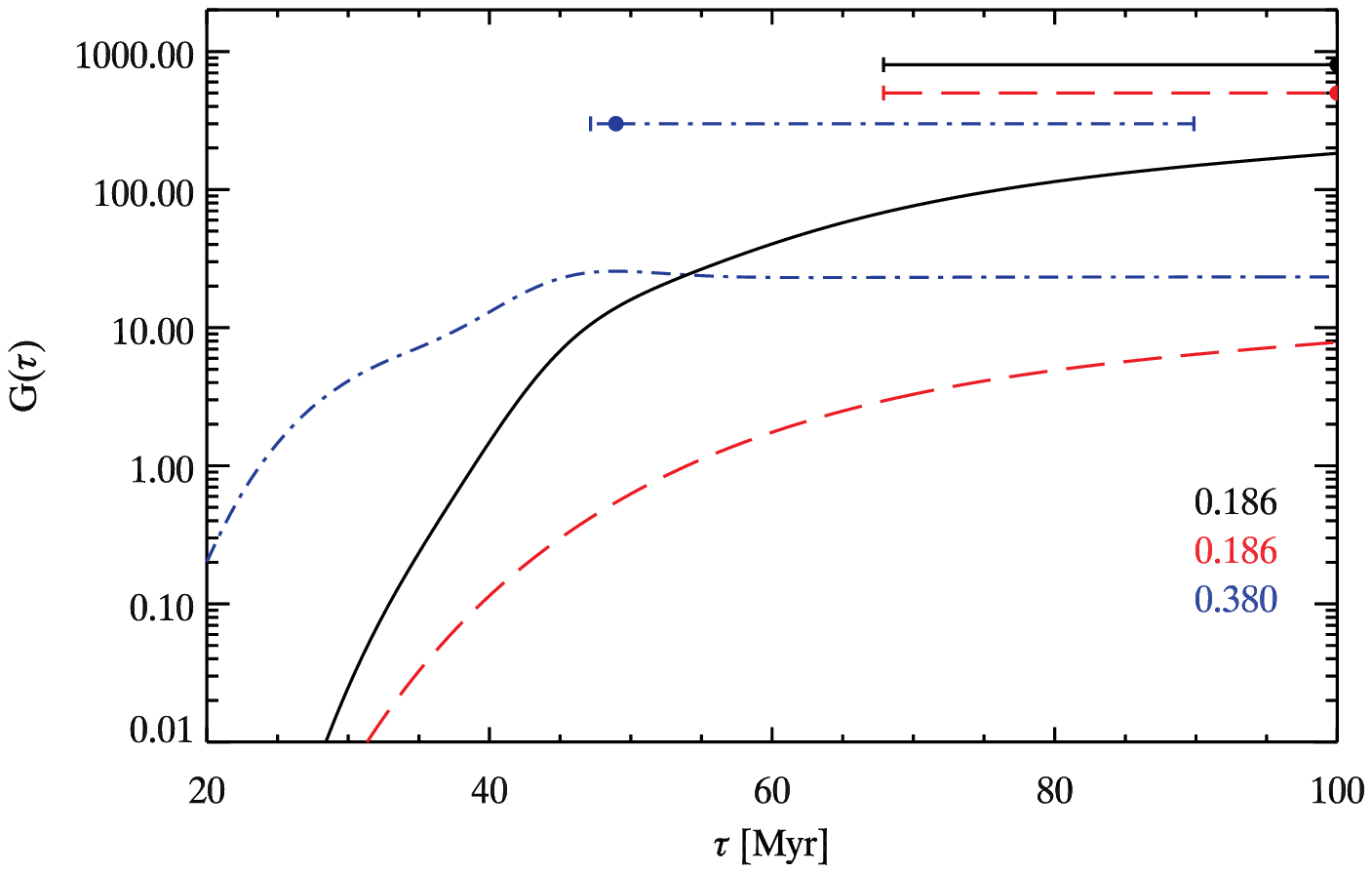}
	\caption{HD 113449 components mass and age distribution functions from comparison with the standard set of models (see Fig.~\ref{fig:RSCha_GH} for a description).}
 \label{fig:HD113449}
\end{figure*}
\clearpage
\begin{figure*}
 \centering
	\includegraphics[width=0.48\hsize,height=5cm]{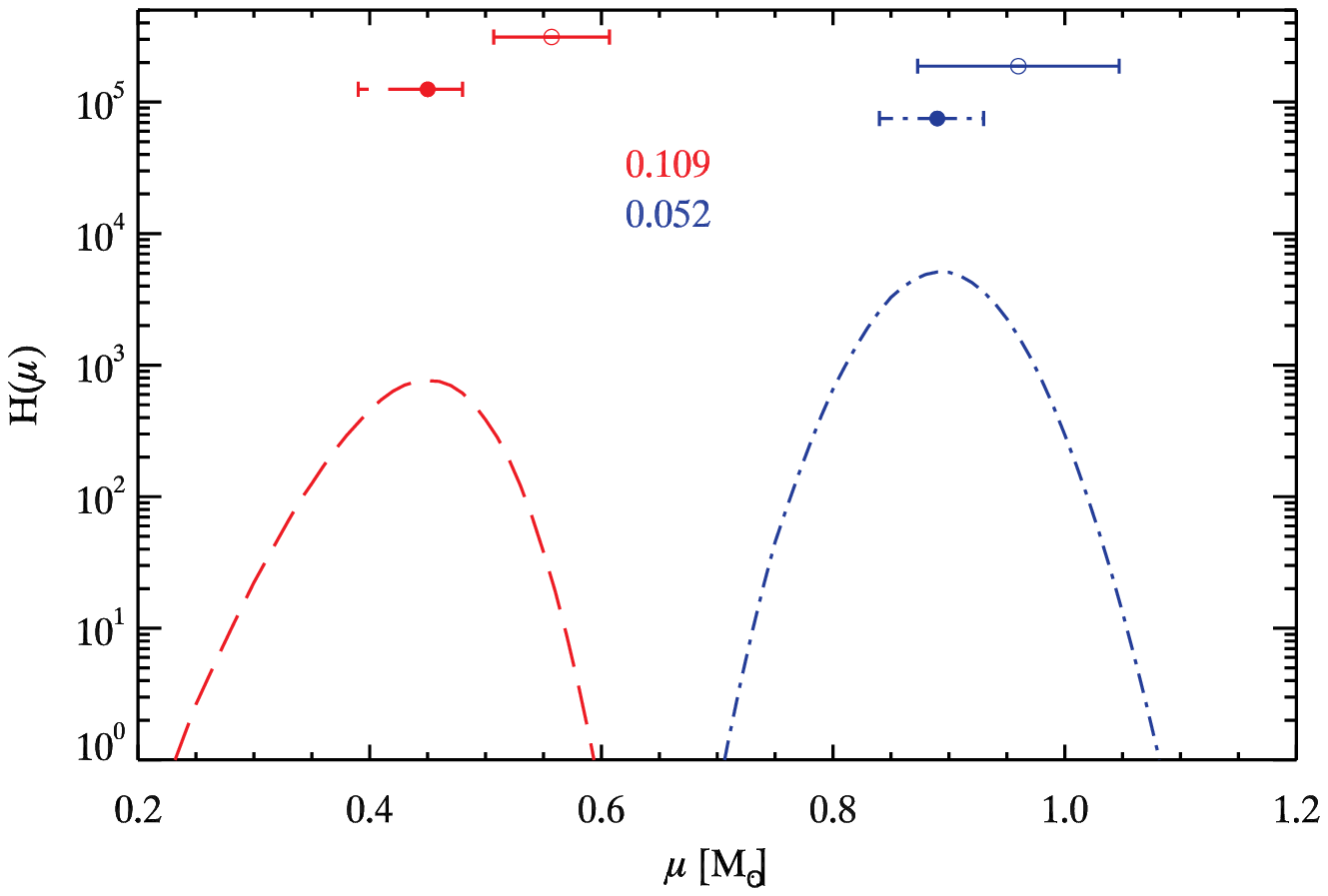}
	\includegraphics[width=0.48\hsize,height=5cm]{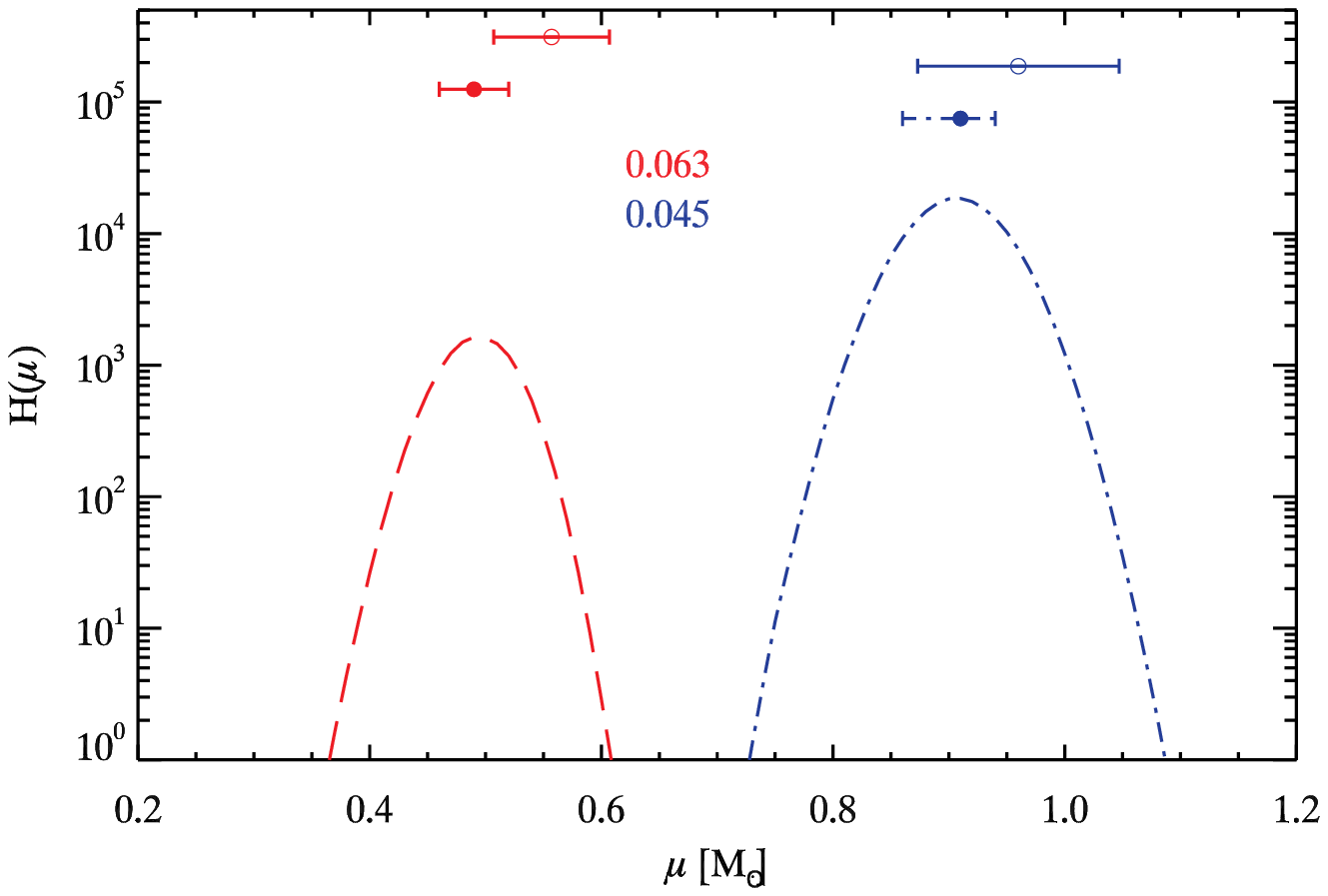}
	\includegraphics[width=0.48\hsize,height=5cm]{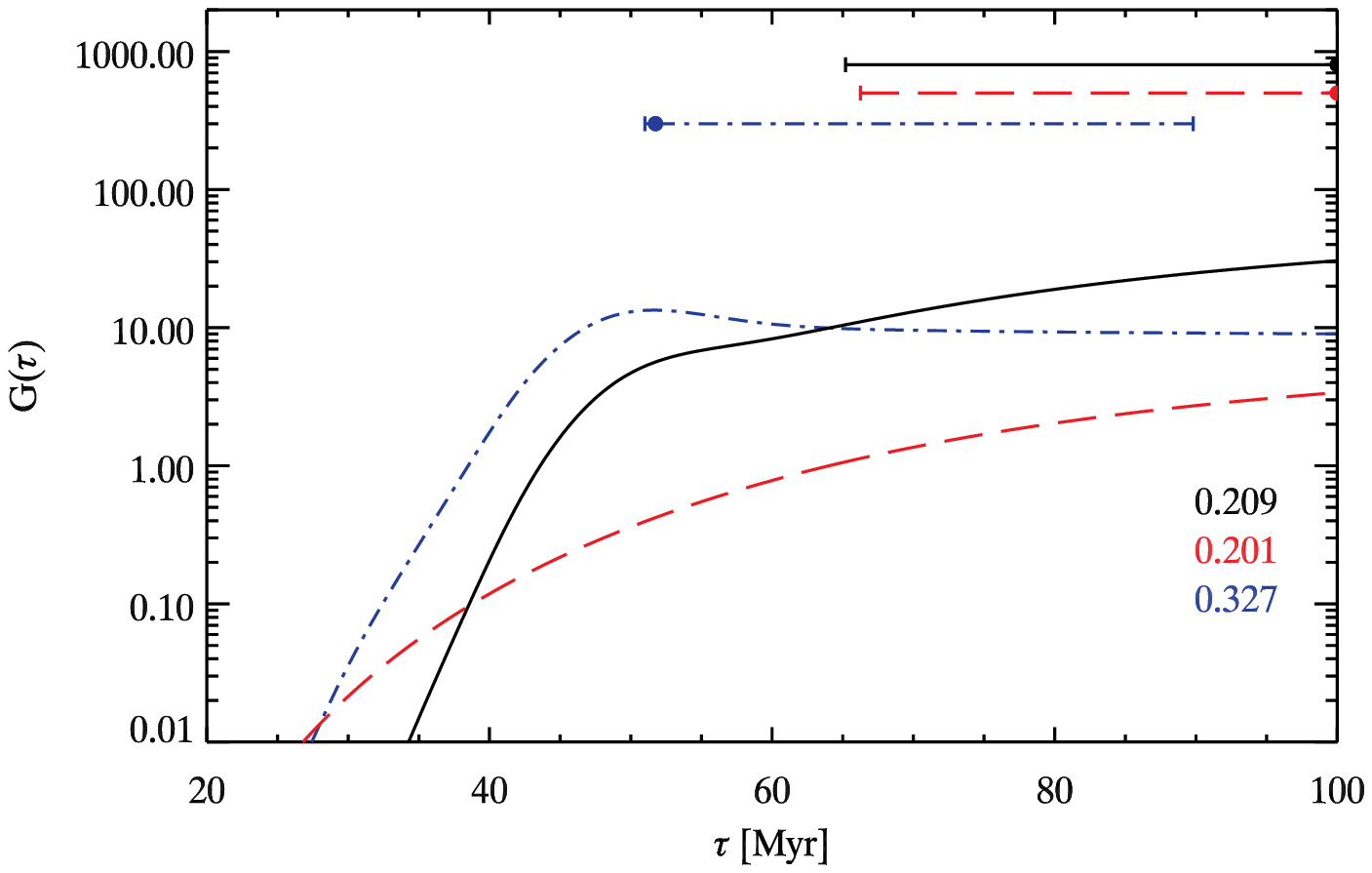}
	\includegraphics[width=0.48\hsize,height=5cm]{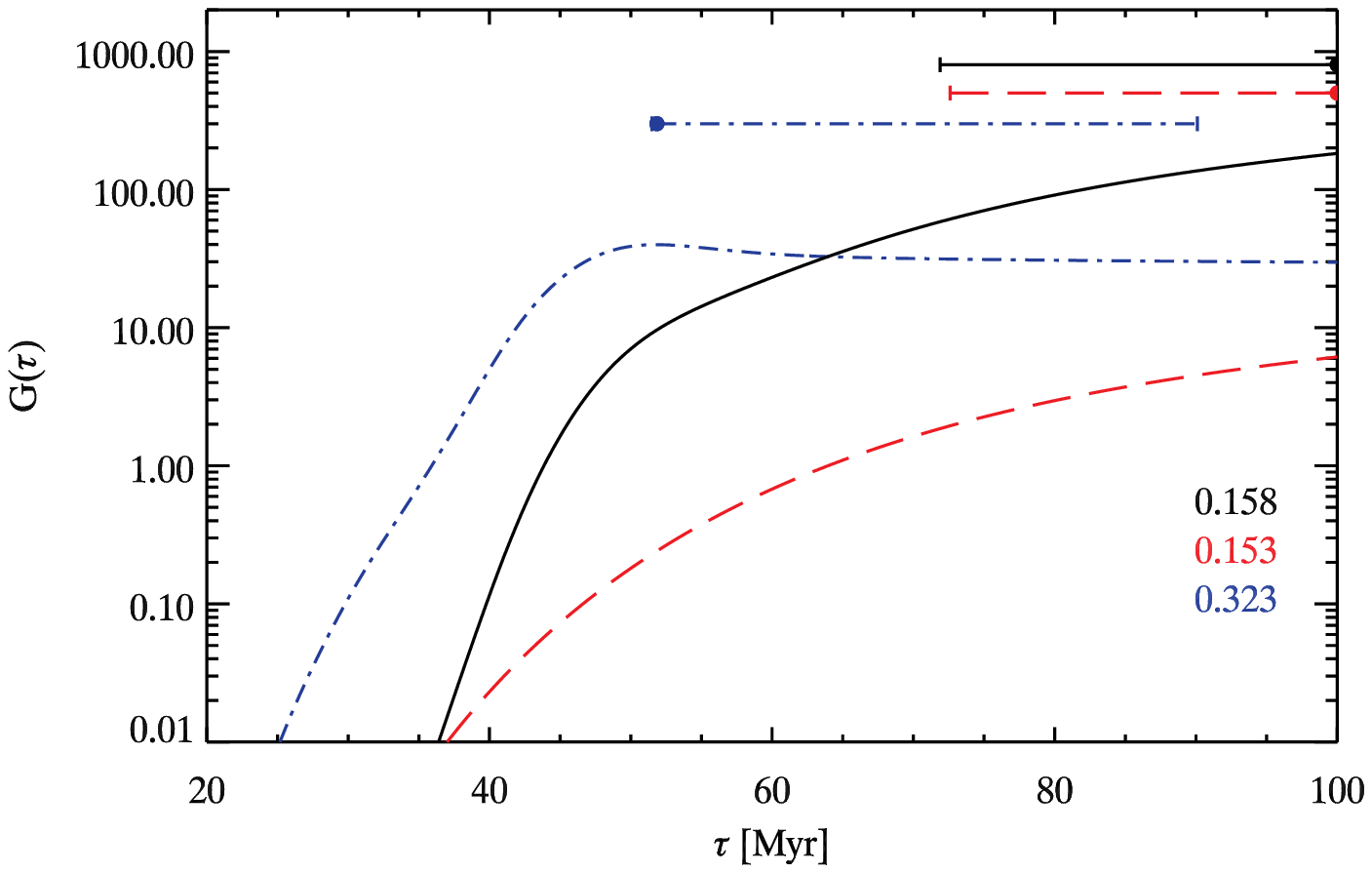}
	\caption{HD 113449 components mass and age distribution functions from comparison with the set of models with $\alpha=1.20, Y_{\rmn{P}} = 0.23$ and $\Delta Y / \Delta Z =2$ (see Fig.~\ref{fig:RSCha_GH} for a description). The stellar $Z$ values used for this comparison have been calculated using the observed [Fe/H] and $(Z/X)_{\sun}$ by \citeauthor{1998SSRv...85..161G} \citeyear{1998SSRv...85..161G}.}
 \label{fig:HD113449NS}
\end{figure*}

\begin{figure*}
 \centering
	\includegraphics[width=0.48\hsize,height=5cm]{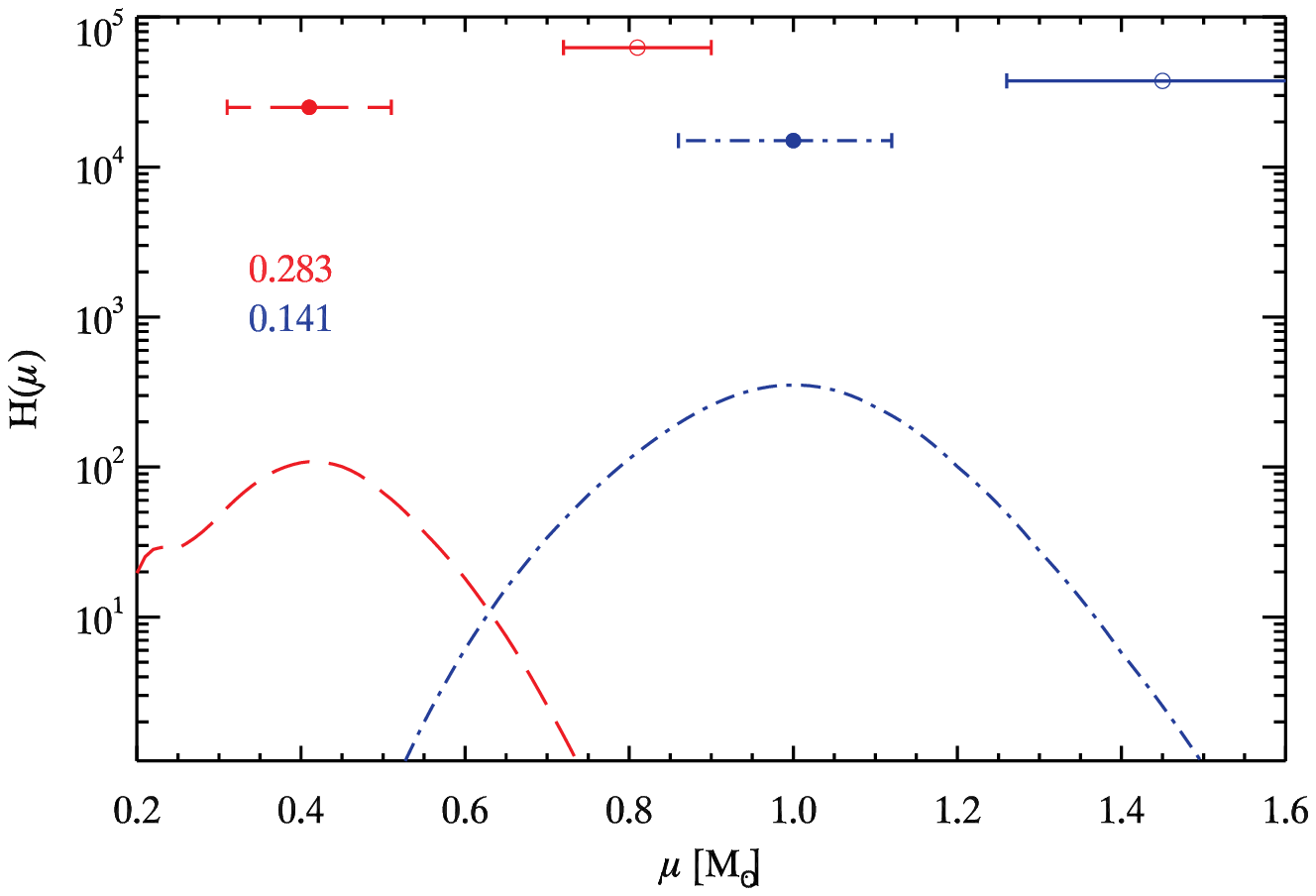}
	\includegraphics[width=0.48\hsize,height=5cm]{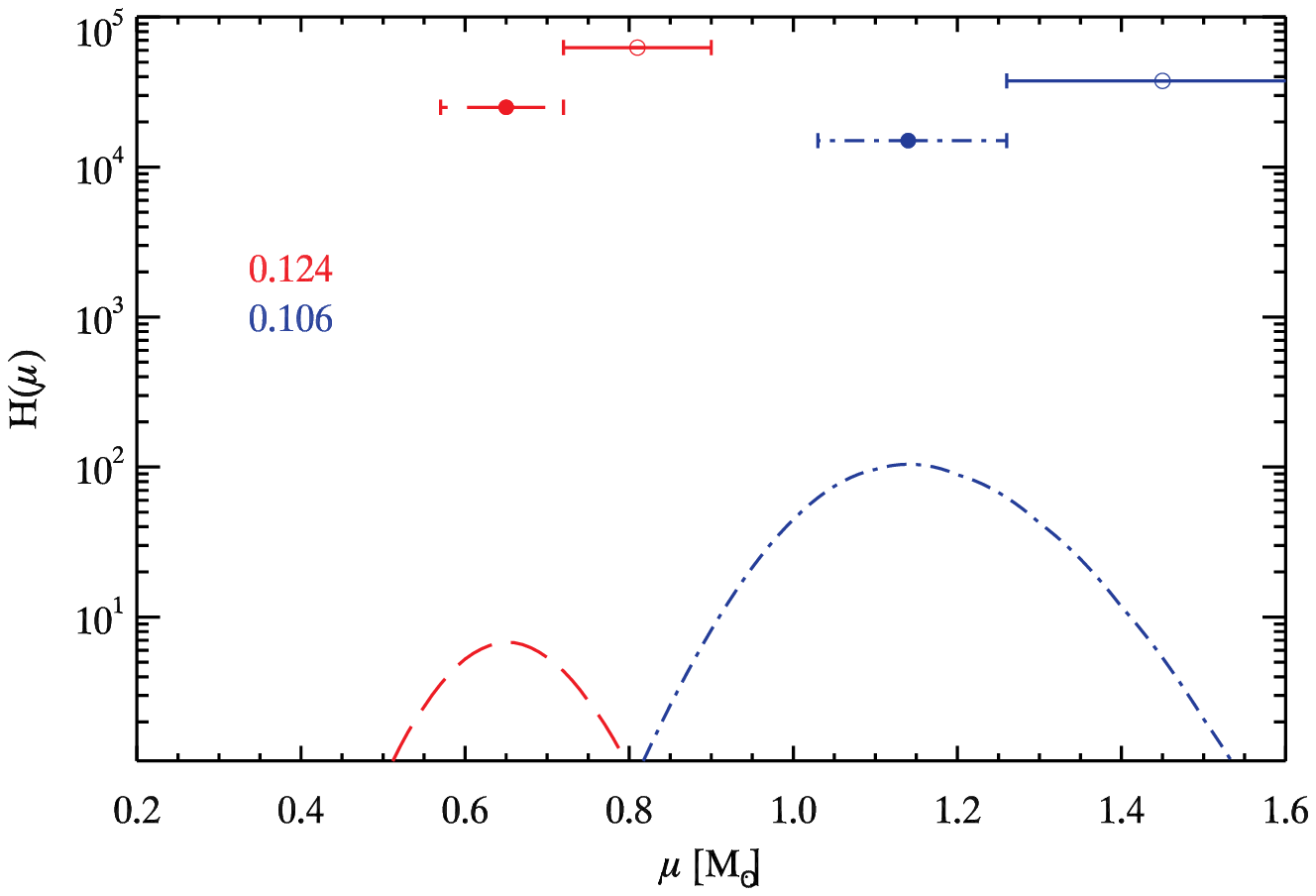}
	\includegraphics[width=0.48\hsize,height=5cm]{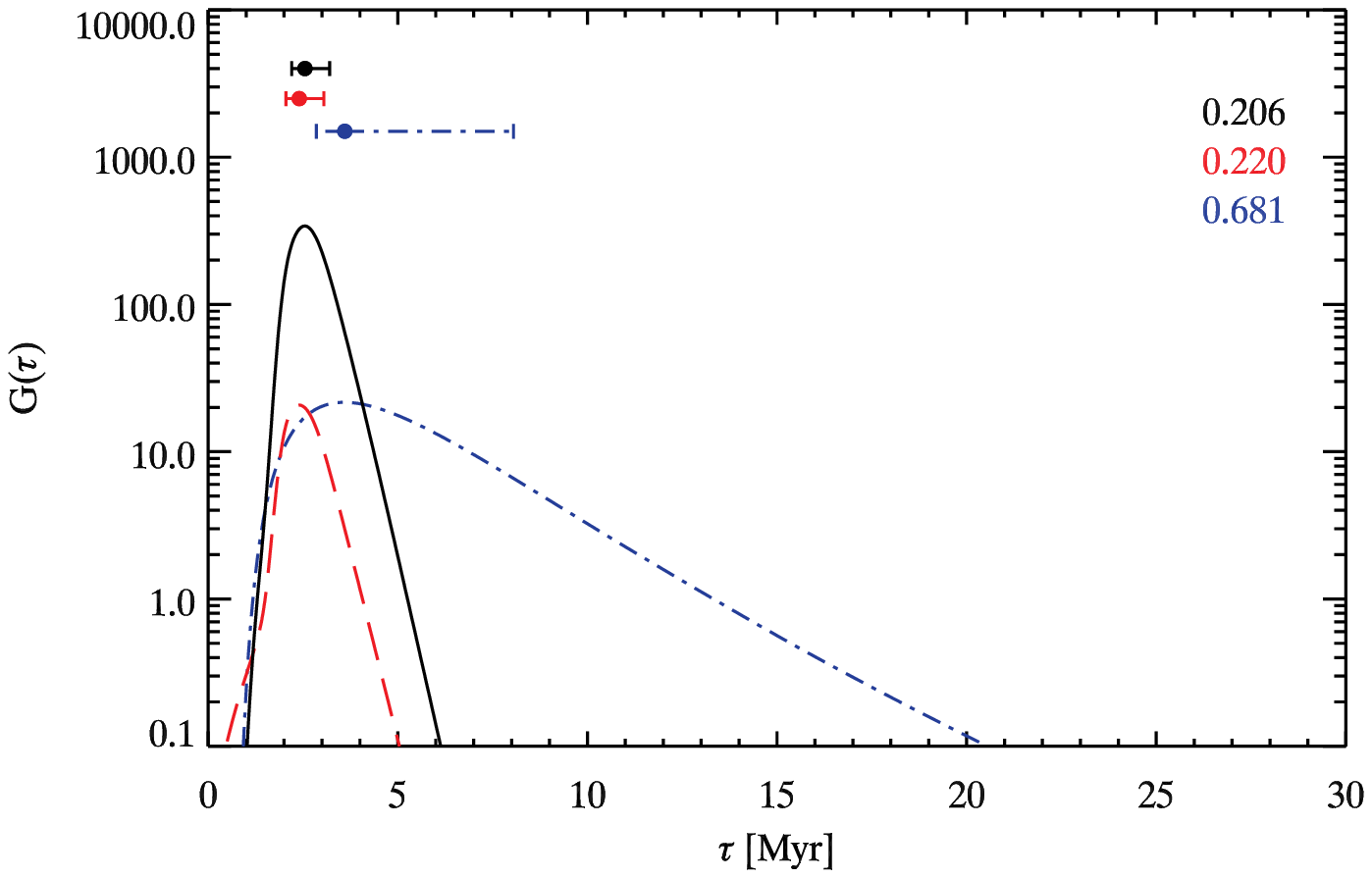}
	\includegraphics[width=0.48\hsize,height=5cm]{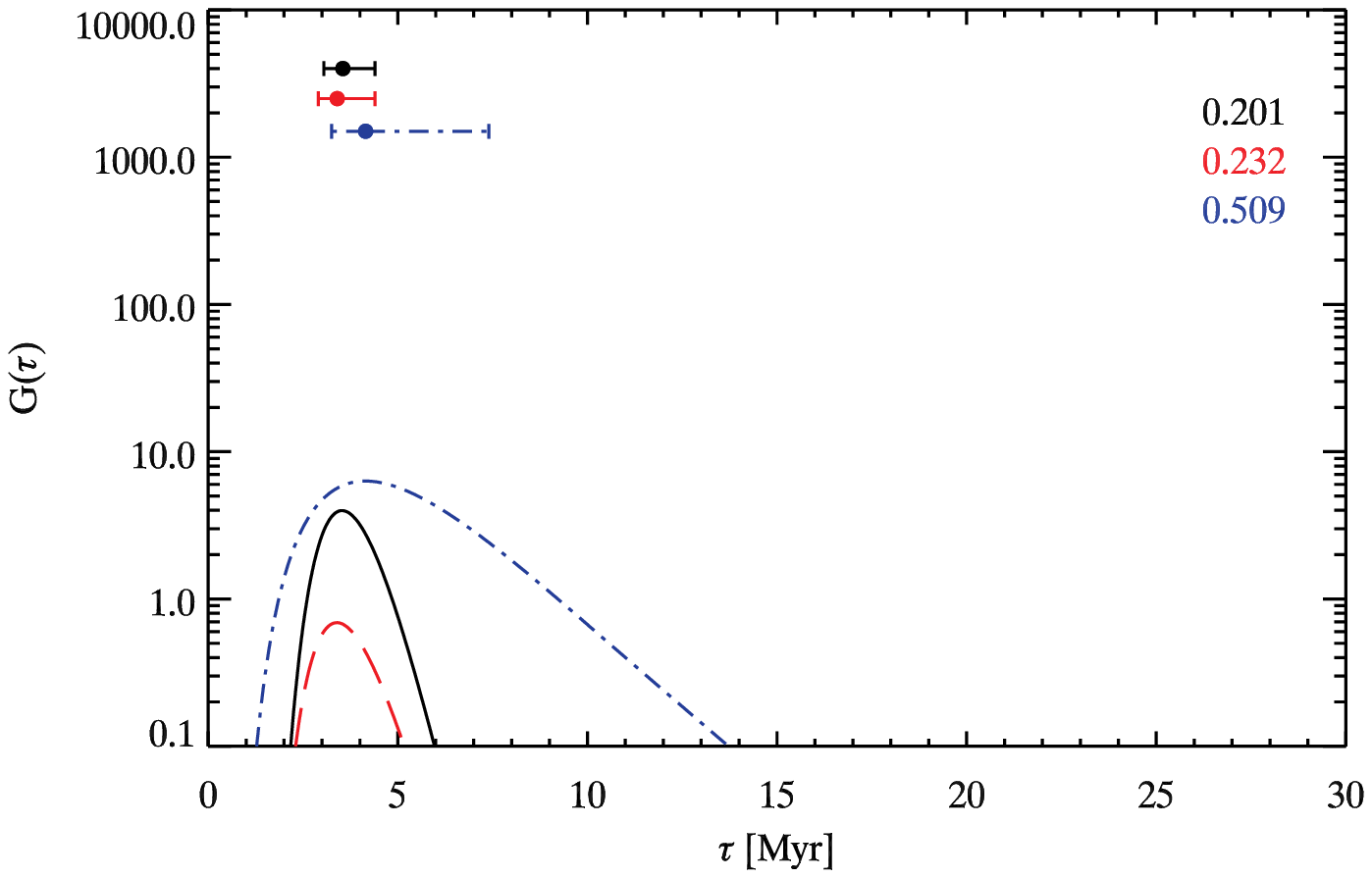}
	\caption{NTT 045251+3016 components mass and age distribution functions from comparison with the standard set of models (see Fig.~\ref{fig:RSCha_GH} for a description).}
 \label{fig:NTT}
\end{figure*}
\clearpage
\begin{figure*}
 \centering
	\includegraphics[width=0.48\hsize,height=5cm]{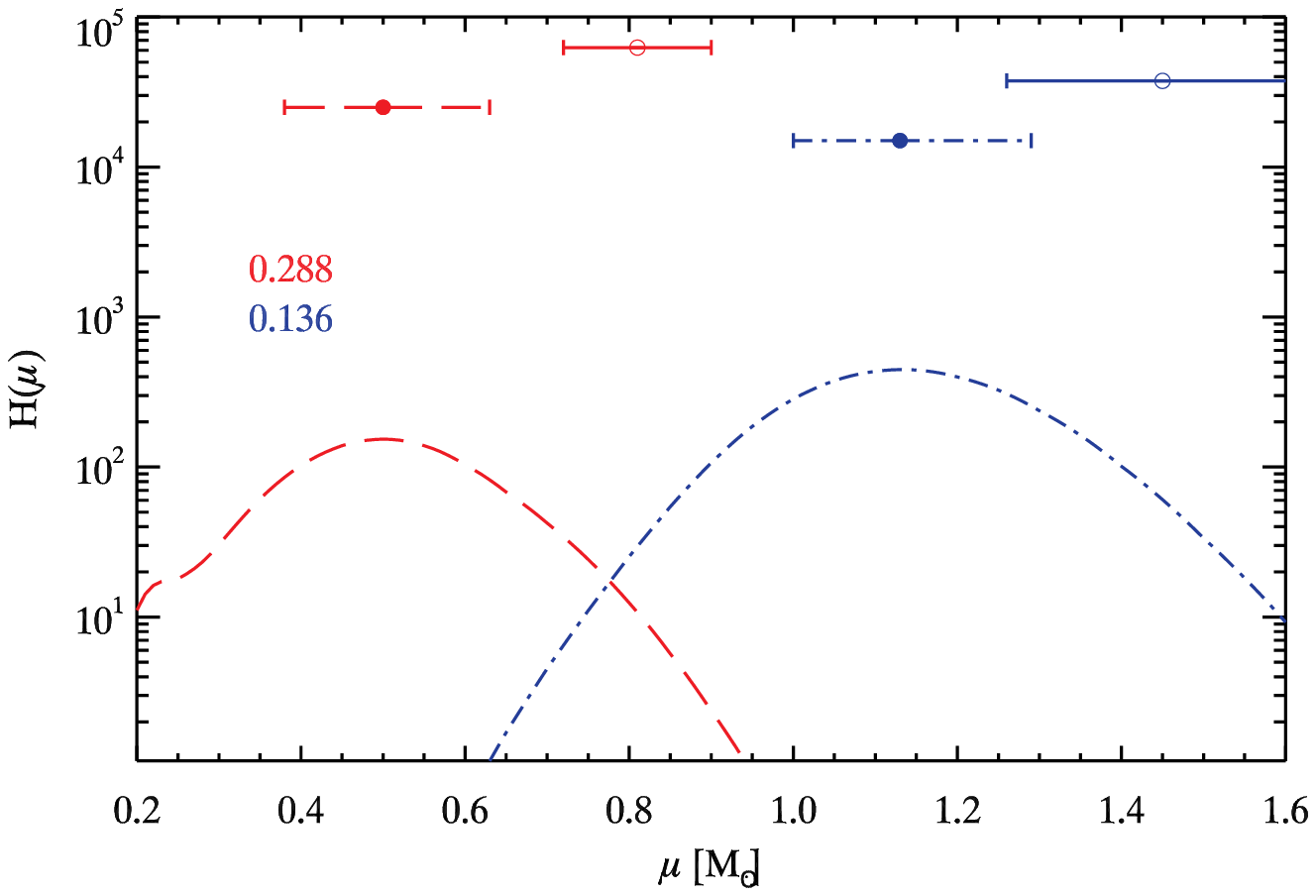}
	\includegraphics[width=0.48\hsize,height=5cm]{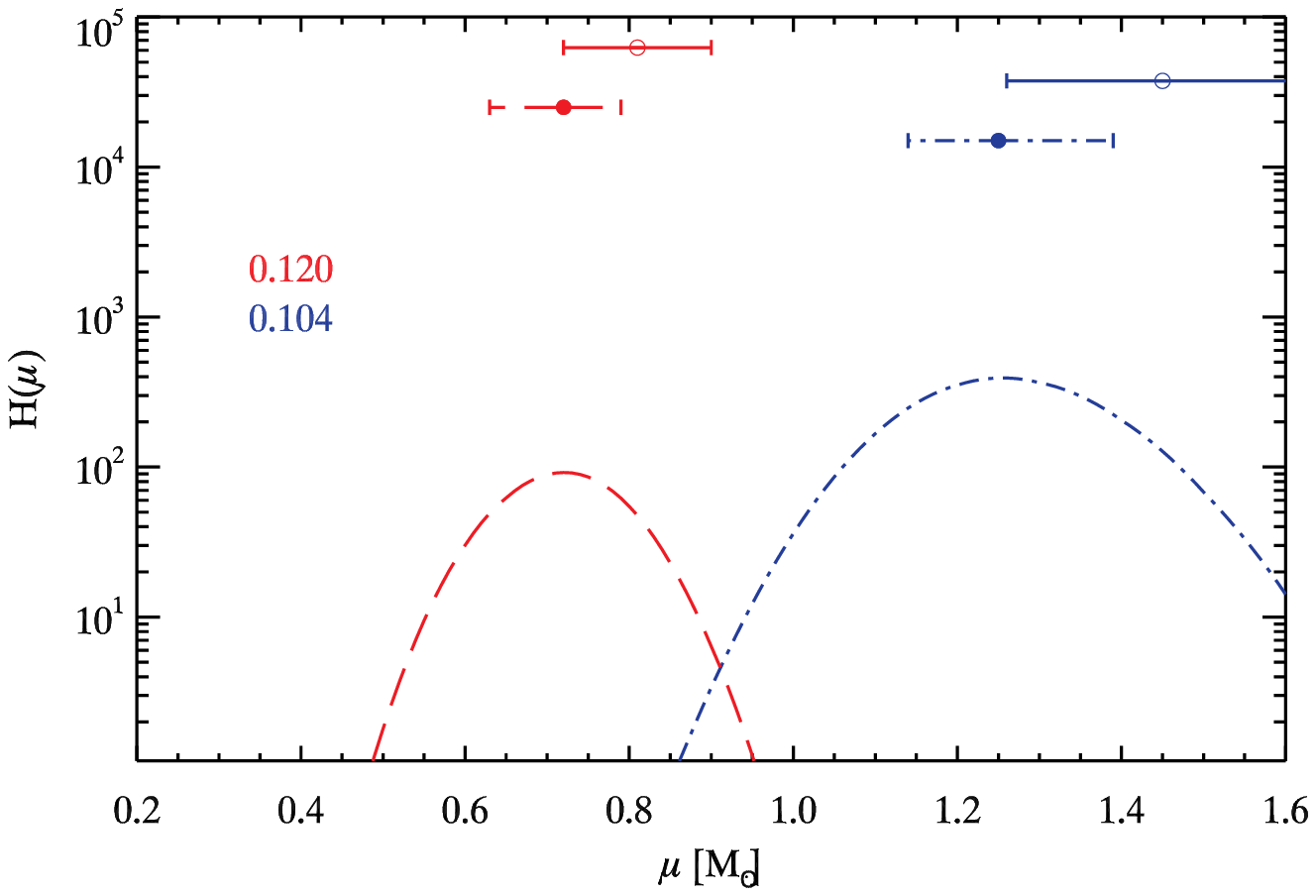}
	\includegraphics[width=0.48\hsize,height=5cm]{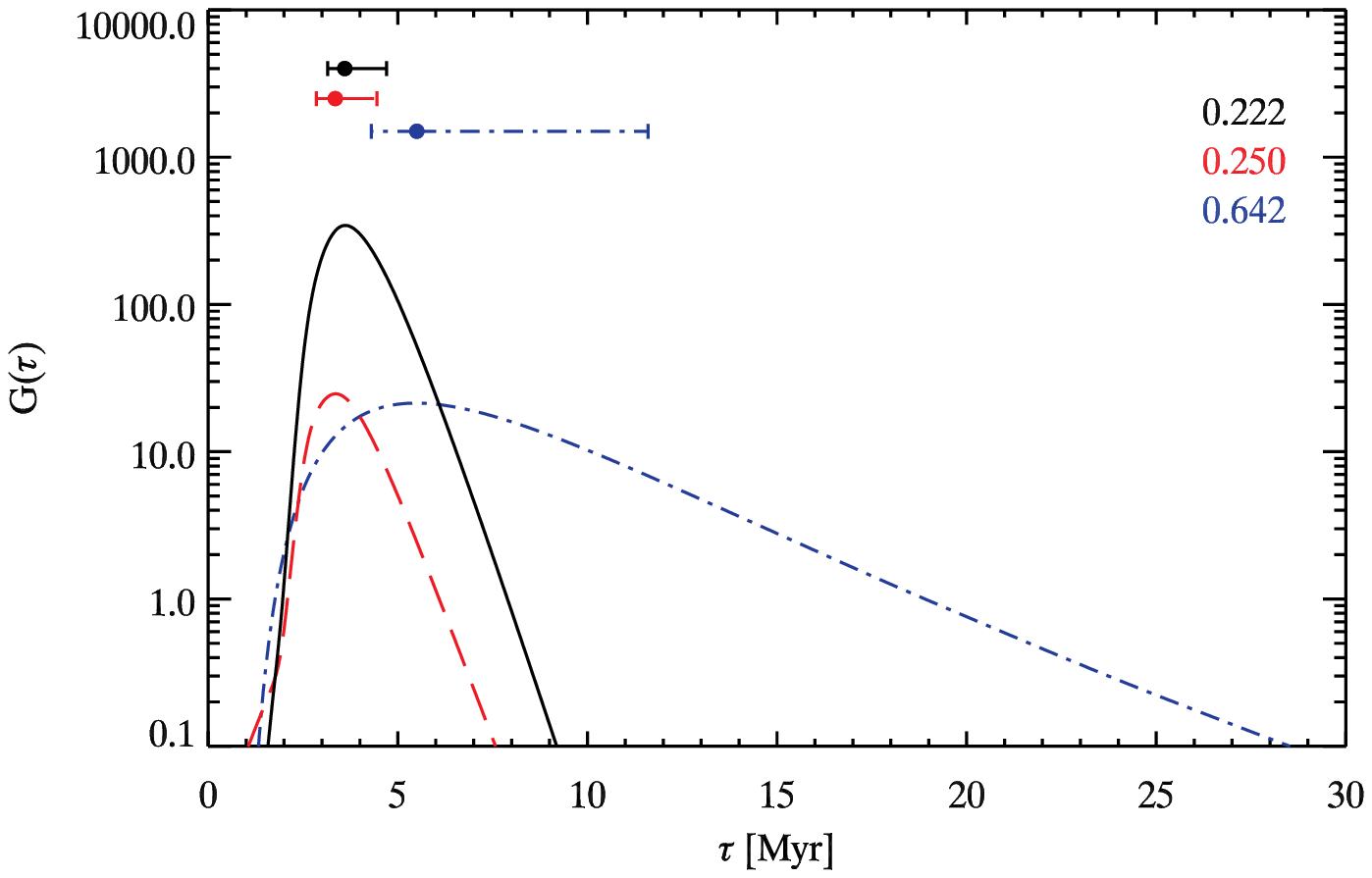}
	\includegraphics[width=0.48\hsize,height=5cm]{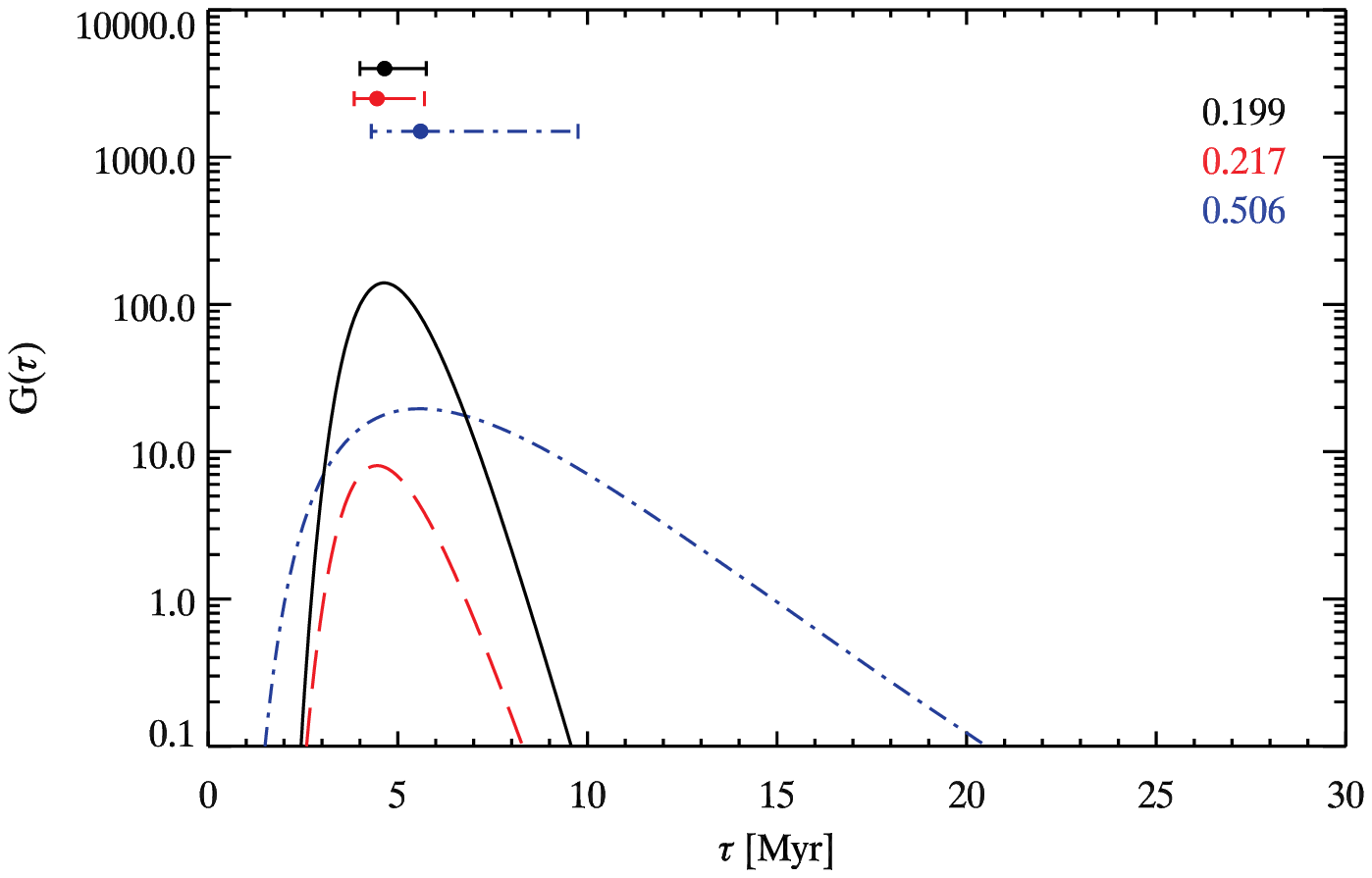}
	\caption{NTT 045251+3016 components mass and age distribution functions from comparison with the set of models with $\alpha=1.20, Y_{\rmn{P}} = 0.23$ and $\Delta Y / \Delta Z =2$ (see Fig.~\ref{fig:RSCha_GH} for a description).}
 \label{fig:NTTNS}
\end{figure*}

\begin{figure*}
 \centering
	\includegraphics[width=0.48\hsize,height=5cm]{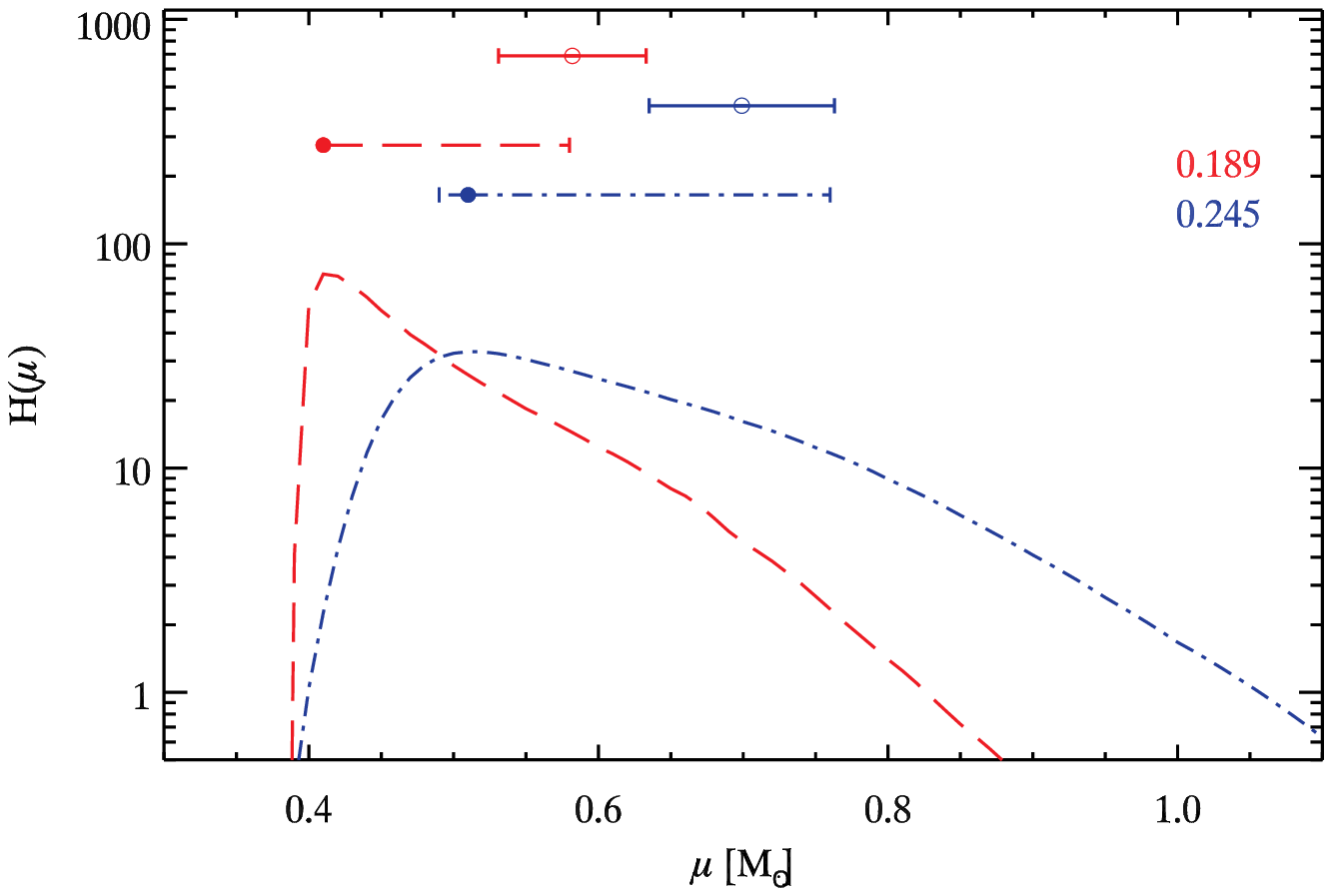}
	\includegraphics[width=0.48\hsize,height=5cm]{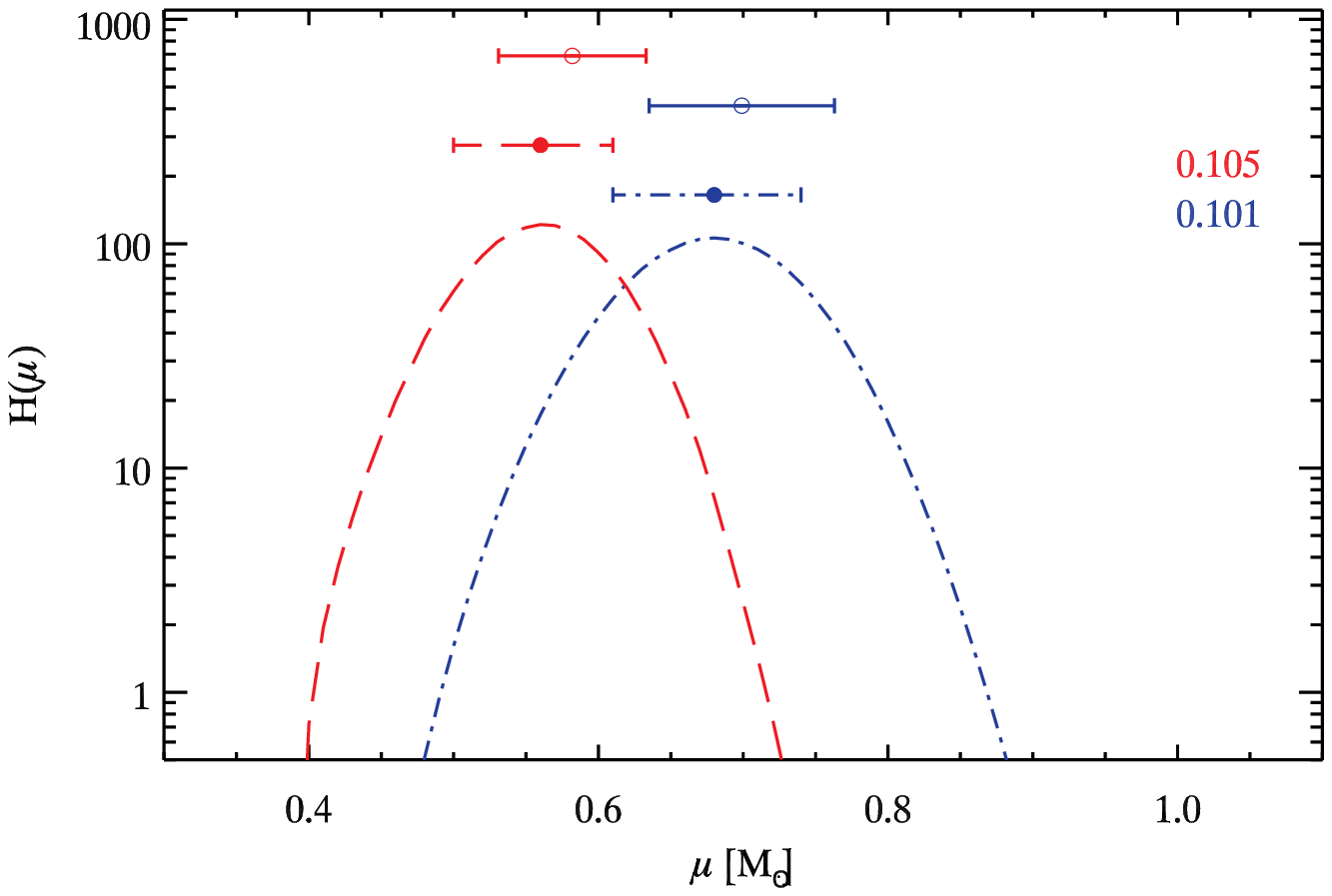}
	\includegraphics[width=0.48\hsize,height=5cm]{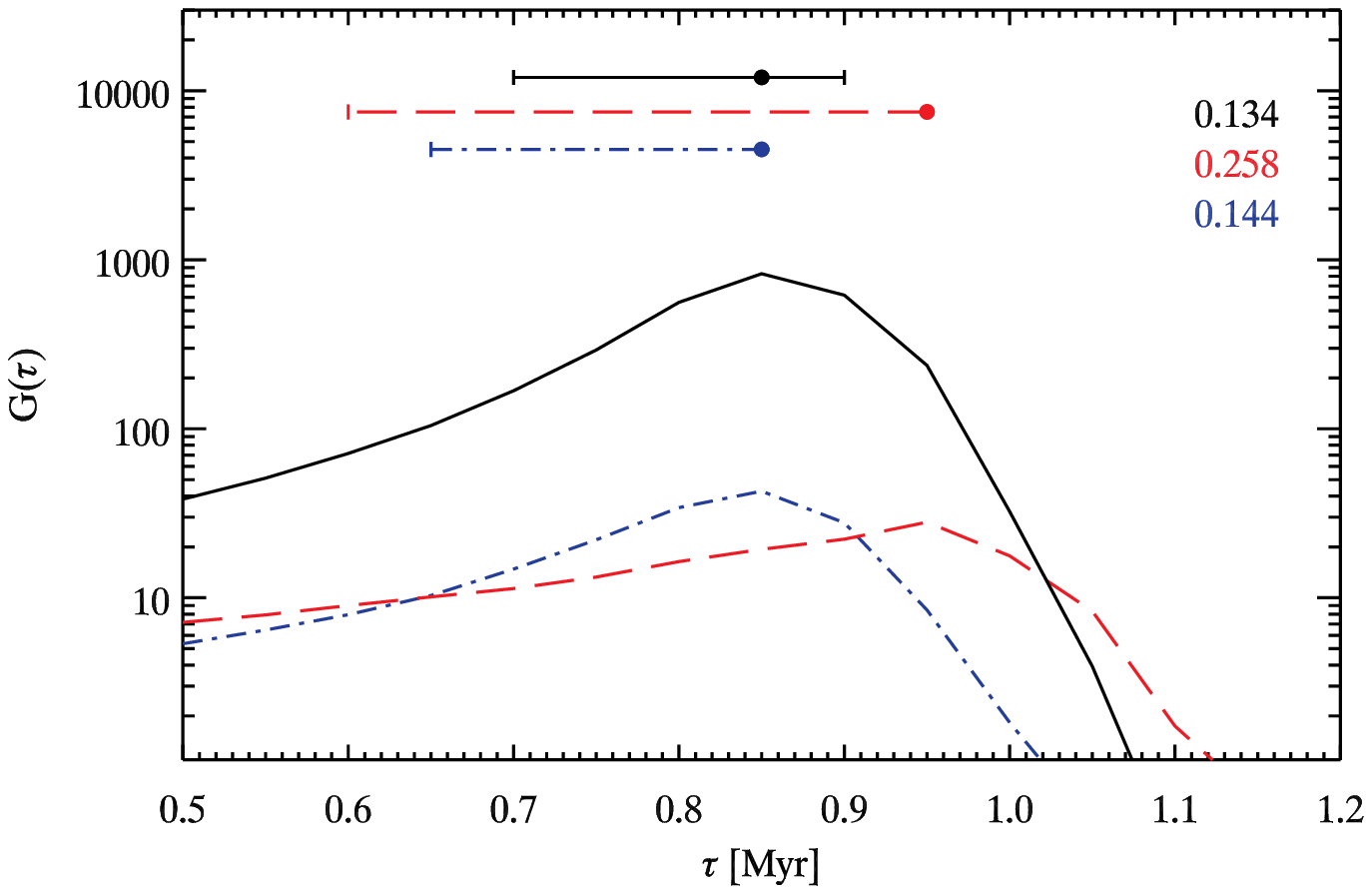}
	\includegraphics[width=0.48\hsize,height=5cm]{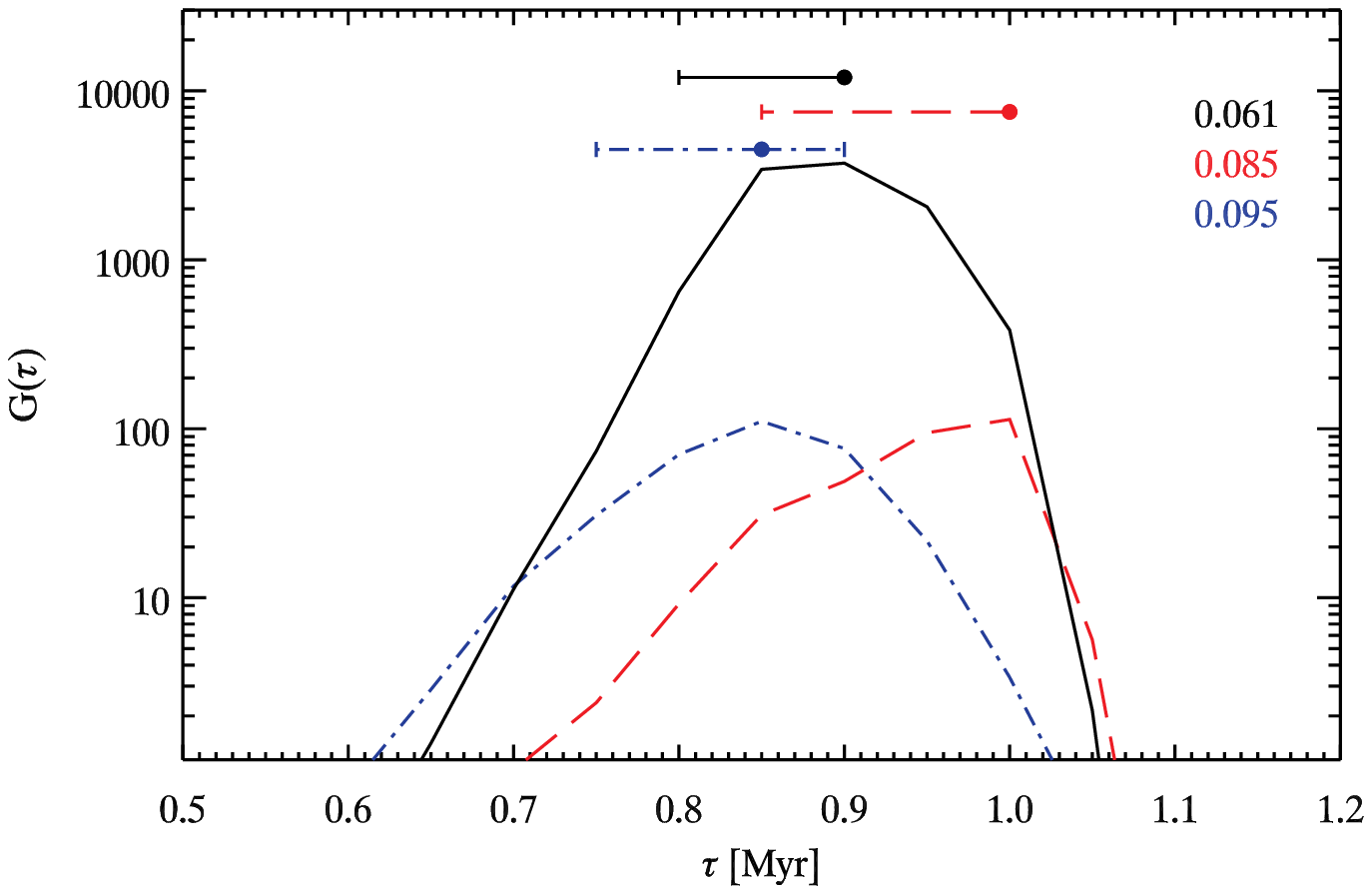}
	\caption{HD 98800 B components mass and age distribution functions from comparison with the standard set of models (see Fig.~\ref{fig:RSCha_GH} for a description).}
 \label{fig:HD98800}
\label{lastpage} 
\end{figure*}


\end{document}